\documentclass[twocolumn, floatfix, prb, aps, showpacs, longbibliography]{revtex4-2}
\usepackage{graphicx, amssymb, color, amsmath}
\usepackage{overpic} 
\usepackage{subcaption}
\usepackage{nicefrac}
\usepackage{multirow, array, booktabs}
\usepackage{float}
\usepackage{romannum}
\usepackage{mathtools}
\usepackage{bbm}
\usepackage{bm}
\usepackage{mathrsfs}
\usepackage{IEEEtrantools}

\usepackage[titletoc, title]{appendix}
\usepackage[colorlinks, bookmarks=true, citecolor=blue, linkcolor=red, urlcolor=blue]{hyperref}
\usepackage{orcidlink}
\AtBeginDocument{\pagenumbering{arabic}}
\begin{document}

\newcommand{\addPRD}[1]{{\bf {\color{red}{#1}}}}
\newcommand{\addACB}[1]{{\bf {\color{blue}{#1}}}}

\title{Static structure factor and the dispersion of the Girvin-MacDonald-Platzman density mode for fractional quantum Hall fluids on the Haldane sphere}
 
\author{Rakesh K. Dora\orcidlink{0009-0009-0043-2982}}
\email{prakeshdora@imsc.res.in}
\affiliation{Institute of Mathematical Sciences, CIT Campus, Chennai, 600113, India}
\affiliation{Homi Bhabha National Institute, Training School Complex, Anushaktinagar, Mumbai 400094, India}

\author{Ajit C. Balram\orcidlink{0000-0002-8087-6015}}
\email{cb.ajit@gmail.com}
\affiliation{Institute of Mathematical Sciences, CIT Campus, Chennai 600113, India}
\affiliation{Homi Bhabha National Institute, Training School Complex, Anushaktinagar, Mumbai 400094, India} 

\date{\today}

\begin{abstract}
We study the neutral excitations in the bulk of the fractional quantum Hall (FQH) fluids generated by acting with the Girvin-MacDonald-Platzman (GMP) density operator on the uniform ground state. Creating these density modulations atop the ground state costs energy, since any density fluctuation in the FQH system has a gap stemming from underlying interparticle interactions. We calculate the GMP density-mode dispersion for many bosonic and fermionic FQH states on the Haldane sphere using the ground state static structure factor computed on the same geometry. Previously, this computation was carried out on the plane. Analogous to the GMP algebra of the lowest Landau level (LLL) projected density operators in the plane, we derive the algebra for the LLL-projected density operators on the sphere, which facilitates the computation of the density-mode dispersion. Contrary to previous results on the plane, we find that, in the long-wavelength limit, the GMP mode accurately describes the dynamics of the primary Jain states. 
\end{abstract}

\maketitle

\section{Introduction}

The fractional quantum Hall (FQH) effect~\cite{Tsui82} is a fascinating quantum phase of matter realized in a two-dimensional system of electrons placed in a perpendicular magnetic field. Although each FQH phase originates from the Coulomb interactions between the electrons, the particulars of a given FQH phase are characterized by its unique and intricate topological and geometric structure. The FQH ground state is an incompressible topologically ordered fluid~\cite{Wen95} evidenced by the fact that its elementary-charged excitations have fractional charges~\cite{Laughlin83} and exhibit an anyonic character, in that they carry fractional braid statistics~\cite{Halperin84, Arovas84, Nakamura20, Bartolomei20}. The gapped neutral excitation of the FQH state, which is the subject of this paper, encodes an emergent quantum geometry~\cite{Haldane11}.

Nearly four decades ago, Girvin, Macdonald, and Platzman (GMP)~\cite{Girvin85, Girvin86} put forth a description of the neutral excitation in the bulk of a FQH fluid. Drawing inspiration from the construction of neutral excitations in superfluid He$^{4}$~\cite{Bijl40, Feynman53, Feynman54a, Feynman54}, GMP proposed that the neutral excitations of an FQH liquid can be similarly constructed using the single-mode approximation (SMA), i.e., as a collective density-wave excitation on top of the ground state. More recently, a geometrical description of FQH states was put forth by Haldane~\cite{Haldane11, Haldane11a}, which has sparked renewed interest in density-wave excitations. Specifically, the fluctuation of the geometric degrees of freedom related to the shape deformations of the correlation hole around an electron attached to its guiding center coordinates corresponds to the long-wavelength limit of density fluctuations in many FQH states~\cite{Haldane11a}. Consequently, probing the long-wavelength limit of the GMP mode (for example, via an anisotropic geometric quench~\cite{Liu18, Liu20}) provides direct evidence of these underlying emergent geometric degrees. The quanta of these geometric fluctuations possess spin-$2$ (exhibiting a quadrupolar character with total orbital angular momentum $L{=}2$ in the spherical geometry) and are called ``chiral FQH graviton" as they have a definite chirality due to the presence of a magnetic field~\cite{Liou19}. Recently, the FQH graviton and its chirality for certain FQH states have been observed in circularly polarized inelastic light scattering measurements~\cite{Liang24} and these experimental results align well with the theoretical predictions of Ref.~\cite{Liou19}.    

In a strong magnetic field, the Landau level (LL) degrees of freedom are frozen and the density-wave excitations reside in a LL. Consequently, for states in the lowest LL (LLL), GMP constructed the neutral excitations by applying the LLL-projected momentum-space density operator to the FQH ground state. Although obtaining the exact FQH ground state is challenging, several accurate trial states can be used to obtain the neutral excitation spectrum. Examples of such trial states include the Laughlin wave function at filling $\nu{=}1/(2m{+}1)$~\cite{Laughlin83}, Jain's composite fermion (CF) wave functions at $\nu{=}n/(2np{\pm} 1)$~\cite{Jain89}, the Moore-Read state at $\nu{=}1/2$~\cite{Moore91}, Read-Rezayi states at $\nu{=}k/(k{+}2)$~\cite{Read99}, and various parton states~\cite{Jain89b} at rational fillings. Here, $m,n,k$, and $p$ are positive integers. While the aforementioned FQH states can be constructed in various geometries, such as disk, torus, or sphere, we primarily use spherical geometry in this paper. We are interested in bulk excitations, which are best studied in compact geometries with no boundaries such as the sphere or torus. Furthermore, unlike in the torus geometry, ground states on the sphere are nondegenerate making it the ideal geometry for our purposes.

For repulsive interactions like Coulomb, the GMP ansatz generates a gapped neutral collective mode. Notably, the dispersion of the GMP mode is nonmonotonic, exhibiting a minimum at a finite wavevector, referred to as the magnetoroton minimum, analogous to the roton minimum in superfluid He$^{4}$. The Coulomb GMP gap for the $\nu{=}1/3$ Laughlin state agrees well with the neutral gap obtained from the exact diagonalization up to the roton minimum; beyond that point, the GMP mode overestimates the gap~\cite{Girvin85, Girvin86, Platzman94, He94, Balram24}. This is because, at large momentum, the GMP mode enters the continuum while in the actual spectrum, there are several other low-lying modes~\cite{Majumder09, Balram24}. Thus, the GMP mode ceases to be a sharp excitation in this regime, rendering the SMA invalid, and providing only the average energy of the neutral excitations as opposed to describing a single mode~\cite{Balram24}. For the non-Laughlin primary Jain states $[n{>}1,~p{=}1$ in $n/(2pn{\pm}1)]$, the GMP mode does not quantitatively capture the neutral excitation~\cite{Platzman94, He94, Balram24}, except in the long-wavelength limit.

To obtain the GMP gap, the following two inputs are needed: (i) knowledge of the density-density correlation, i.e., the static structure factor, in the ground state since the GMP state is constructed by applying the density operator on the ground state, and (ii) the algebra of the LLL-projected density operators. This algebra, known as the GMP-algebra was worked out on the plane by GMP and was shown to be closed~\cite{Girvin85, Girvin86}. Thus, previously, the GMP mode dispersion was computed in the planar geometry. In this paper, we derive the GMP algebra for the LLL-projected density operators on the sphere and show it to be closed. However, unlike in planar geometry, where the commutator of two projected density operators is a single projected density operator, on the sphere, the commutator of two projected density operators is a linear sum of projected density operators. 

For completeness, we mention other approaches for constructing neutral excitations of a FQH fluid. These include the CF exciton (CFE) within the CF theory~\cite{Jain89} and the Jack-polynomial approach~\cite{Rodriguez12b, Yang12b, Yang13a}. In the CF theory, neutral excitations are constructed naturally by creating excitons of CFs over the filled IQH state of CFs. For FQH states in the Jain sequence, CFEs provide an accurate description of the neutral excitations seen in exact diagonalization~\cite{Jain97, Jain97b, Balram24} as well as with experimentally observed results~\cite{Pinczuk93, Mellor95, book_Pinczuk96, Davies97, Zeitler99, Kang00}. For specific FQH states, such as $\nu{=}1/3$ Laughlin~\cite{Laughlin83} and $\nu{=}1/2$ Moore-Read~\cite{Moore91}, by leveraging the underlying Jack polynomial structure of their ground states~\cite{Bernevig08}, the neutral mode was constructed in Ref.~\cite{Yang12b}, which also agrees very well with the exact diagonalization results. However, unlike the CFE, the Jack polynomial approach is limited to small system sizes. A notable feature of the GMP mode is that it only needs the ground state wave function as input whereas the other approaches, such as the Jack and CFE construction, can only be carried out when the ground state has a particular structure. This makes the GMP construction more widely applicable than the other approaches we mentioned. Interestingly, for the Laughlin states, different ways of constructing the neutral excitation, i.e, following the GMP ansatz, CFE construction, or Jack-polynomial approach, all become equivalent in the long-wavelength limit~\cite{Kamilla96b, Kamilla96c, Yang12b, Pu24}. 

The authors of Ref.~\cite{Scarola00} found that the long-wavelength planar Coulomb GMP gap of non-Laughlin primary Jain states, like 2/5 and 3/7, obtained from the ground state structure factor of large systems, deviates significantly from the corresponding CFE gap. Contrastingly, numerical results on small systems from the recent Ref.~\cite{Balram24} suggest that for the 2/5 and 3/7 Jain states, the energy of the CFE and GMP states at $L{=}2$ on the sphere (which provides the long-wavelength gap) lies very close to each other. To resolve this discrepancy, we compute the GMP gap on the sphere semi-analytically for large systems by directly using the structure factor computed on the same geometry and find that the thermodynamic extrapolated $L{=}2$ Coulomb GMP gap is indeed very close to the corresponding CFE gap, in agreement with the small system results of Ref.~\cite{Balram24}. We show that the issue with the computation in Ref.~\cite{Scarola00} was that $S_{4}$, the coefficient of $(q\ell)^4$ in the small-momentum $q$ expansion of the structure factor, was set to an incorrect value. By imposing the correct constraints on the long-wavelength limit of the structure factor, we show the $q{\to}0$ GMP and CFE gaps agree with each other even while using the method of Ref.~\cite{Scarola00}, thereby resolving the discrepancy.

Finally, for the secondary Jain states $[n{>}1,~p{=}2$ in $n/(2pn{\pm}1)]$, the GMP mode is not the lowest-energy neutral excitation even in the long-wavelength limit~\cite{Balram24}. In this limit, the GMP state, referred to as the GMP graviton, splits into two gravitons~\cite{Balram21d, Nguyen22, Wang22, Balram24}, which have been interpreted as arising from a low-lying CFE graviton (that does provide an accurate description of the actual lowest-energy graviton seen in exact diagonalization) and a high-energy parton graviton~\cite{Balram21d, Balram24}. 

The rest of the article is organized as follows. In Sec.~\ref{sec: interacting_particles_sphere_B}, we present the mathematical preliminaries defining the density operator and structure factor on a sphere in Secs.~\ref{ssec: density_operators} and~\ref{ssec: structure_factor}, respectively, followed by the interaction Hamiltonian in Sec.~\ref{ssec: interaction_Hamiltonian} and Haldane pair-pseudopotentials in Sec.~\ref{ssec: pair_pesudopotential}. In Sec.~\ref{sec: algebra_projected_density_operator}, we present one of our main results on the algebra of projected density operators on the sphere. In Secs.~\ref{sec: magnetoroton_dipersion} and~\ref{sec: CFE_gap}, we derive the GMP gap equation on a sphere and outline the computation of the CFE gap, respectively. In Sec.~\ref{sec: results}, we validate the algebra of projected density operators and compare the GMP and CFE gaps. We revisit the planar GMP gap computation of Ref.~\cite{Scarola00} for the primary Jain states in Sec.~\ref{sec: planar_GMP_gap} and suggest suitable modifications to it that bring the GMP and CFE gaps in agreement with each other in the long-wavelength limit consistent with small system exact diagonalization results of Ref.~\cite{Balram24}. We close the article in Sec.~\ref{sec: discussions} by summarizing our results and discussing the scope of its potential extensions. Several technical details are presented in Appendices~\ref{app: projected_unprojected_Sq},~\ref{app: Derivation_not_normal_ordered_interaction_Hamiltonia}, \ref{app: derivation_GMP_algebra_coefficients} and~\ref{app: Oscillator strength_sphere}. In Appendices~\ref{app: ground_state_energies} and \ref{app: GMP_gap_fermionic_MR_Pf_SLL}, we provide the ground state energies of various FQH states for different interparticle interactions, calculated using our formalism. In Appendices~\ref{app: GMP_gap_short_range_interactions} and~\ref{app: GMP_gap_bosonic_FQH_states}, we present the GMP gaps for short-range interactions and results for bosonic states, respectively. In Appendix~\ref{app: fit_gr_Sq}, we provide results for the fits of the numerically computed pair-correlation function and compare the structure factor computed from it with the numerically obtained structure factor data.

\section{Interacting particles on a sphere in a magnetic field}
\label{sec: interacting_particles_sphere_B}
Throughout the paper, unless otherwise specified our primary system of consideration consists of $N$ spin-polarized particles moving on the surface of a Haldane sphere subjected to an outward radial magnetic field of flux-strength $N_{\phi}{=}2Qhc/e$ ($Q{\geq}0$)~\cite{Haldane83}. The magnetic field is produced by a magnetic monopole of strength $Q$ placed at the sphere's center. Owing to the Dirac quantization condition, $Q$ is an integer or half-integer. The radius $R$ of the sphere is related to $Q$ as $R{=}\sqrt{Q}\ell$, where $\ell{=}\sqrt{\hbar/(eB)}$ is the magnetic length at magnetic field $B$, and $e$ is the absolute value of the electric charge of the particles. The single-particle spectrum of the system comprises a discrete set of levels, called LLs, which are labeled by the orbital angular momentum quantum number $l{=}Q$,{~}$Q{+}1$,~${\cdots}$. The corresponding LL eigenstates are the monopole spherical harmonics $Y^{Q}_{l,m}(\boldsymbol{\Omega})$, where the $z$-component of the orbital angular momentum, i.e., azimuthal quantum number $m$ ranges from ${-}l$ to $l$ in steps of one. Here, $\boldsymbol{\Omega}{=}(\theta,\phi)$ is the coordinate of a particle on the sphere with $\theta$ and $\phi$ its polar and azimuthal angles. On the sphere, incompressible quantum Hall states occur when $2Q{=}\nu^{-1}N{-}\mathcal{S}$, where $\mathcal{S}$ is the Wen-Zee ``shift" ~\cite{Wen92} that characterizes the nature of the state. Furthermore, these incompressible quantum Hall states are uniform, i.e., they have total orbital angular momentum $L{=}0$ (therefore, $L_{z}$, quantum number of the $\hat{L}_{z}$ operator, which is the $z$-component of $\hat{L}$, is also zero).  

We are interested in the intra-LL density-wave excitation created over a FQH ground state. Next, we will introduce density operators and their projection to the LLL. We note that $l{=}Q$ represents the LLL while in general, for a LL indexed by $n{=}0,1,2,{\cdots}$, $l{=}Q{+}n$.

\subsection{Density operator}
\label{ssec: density_operators}

In the first quantized notation, the density operator on the sphere is defined as
\begin{equation}
\label{eq: first_quantized_density}
\rho({\boldsymbol{\Omega}})=\sum_{i}\delta(\boldsymbol{\Omega}-\boldsymbol{\Omega}_{i}),
\end{equation}
where $\boldsymbol{\Omega}_{i}$ is the coordinate of the $i$th particle. The density operator in the angular momentum-space, which is used to create a density-wave, can be obtained through the following analog of the Fourier transformation~\cite{He94}:
\begin{equation}
    \rho_{L,M}=\int d\boldsymbol{\Omega}~Y_{L,M}({\boldsymbol{\Omega}})~\rho({\boldsymbol{\Omega}})=\sum_{i}Y_{L,M}(\boldsymbol{\Omega}_{i}),
    \label{eq: Angular_momentum_space_density_operator}
\end{equation}
where $Y_{L, M}({\boldsymbol{\Omega}})$ is the usual spherical harmonic that matches with the monopole spherical harmonic $Y^{Q}_{L,m}(\boldsymbol{\Omega})$ when $Q{=}0$. The angular momentum index $L$ in Eq.~\eqref{eq: Angular_momentum_space_density_operator} can take any non-negative integer value, while the azimuthal quantum number $M{=}{-}L,{-}L{+}1,{\cdots}, L$ takes integral values. This is in contrast to the LL index $l$, which can either be an integer or half-integer depending on whether $Q$ is an integer or half-integer.  

In second quantization, the density operator can be written in terms of the real-space field operators as $\rho^{\sigma}(\boldsymbol{\Omega}){=}\left[\Psi^{\sigma}(\boldsymbol{\Omega})\right]^{\dagger}\Psi^{\sigma}(\boldsymbol{\Omega})$, where the field operator $[\Psi^{\sigma}]^{\dagger}$ ($\Psi^{\sigma}$) creates (annihilates) a particle at position $\boldsymbol{\Omega}$. The symbol $\sigma {\in} (b,f)$ denotes whether the particles are bosons ($b$) or fermions ($f$). Utilizing the completeness of the LL eigenstates, one can express the field operator $[\Psi^{\sigma}]^{\dagger}$ in terms of the LL creation operators $[\chi^{\sigma}_{l,m}]^{\dagger}$ as follows~\cite{He94}:
\begin{equation}
\label{eq: creation_operator_landau_level_basis}
 \left[\Psi^{\sigma}(\boldsymbol{\Omega})\right]^{\dagger}=\sum_{l,m}[Y^{Q}_{l,m}(\boldsymbol{\Omega})]^{*}~[\chi^{\sigma}_{l,m}]^{\dagger}.   
\end{equation}
The operator $[\chi^{\sigma}_{l,m}]^{\dagger}$ represents the bosonic (fermionic) creation operator for $\sigma{=}b$ ($\sigma{=}f$) and satisfies the usual bosonic (fermionic) commutation (anti-commutation) relations. Using Eq.~\eqref{eq: creation_operator_landau_level_basis}, the density operator can be expressed in terms of the LL creation and annihilation operators as
\begin{equation}
\label{eq: second_quantized_density}
    \rho^{\sigma}(\boldsymbol{\Omega})=\sum_{\substack{l_1, m_1,\\l_2, m_2}} \left[Y_{l_1, m_1}^Q(\boldsymbol{\Omega})\right]^* Y_{l_2, m_2}^Q(\boldsymbol{\Omega})~ [\chi^{\sigma}_{l_1, m_1}]^{\dagger} \chi^{\sigma}_{l_2, m_2}.
\end{equation}
Similarly, the second quantized form of the angular momentum-space density operator is obtained by substituting Eq.~\eqref{eq: second_quantized_density} into Eq.~\eqref{eq: Angular_momentum_space_density_operator} to obtain
\begin{equation}
\label{eq: second_quantized_angular_momentum_density}
    \rho^{\sigma}_{L,M}=\sum_{\substack{l_1, m_1,\\l_2, m_2}} \rho\left(L, M, l_1, m_1, l_2, m_2\right)~ [\chi^{\sigma}_{l_1, m_1}]^{\dagger} \chi^{\sigma}_{l_2, m_2},
\end{equation}
where~\cite{Simon94a} 
\begin{align}
\label{eq: second_quantized_angular_momentum_density_value}
\rho\left(L, M, l_1, m_1, l_2, m_2\right)&=  \int d \boldsymbol{\Omega}~ Y_{L, M}~\left[Y_{l_1, m_1}^Q\right]^*~ Y_{l_2, m_2}^Q\nonumber\\[1ex]
&=(-1)^{Q+l_1+l_2+L+m_1}\nonumber\\[1ex]
&\times\left[\frac{\left(2 l_1+1\right)\left(2 l_2+1\right)\left(2 L+1\right)}{4 \pi}\right]^{\frac{1}{2}}\nonumber\\[1ex]
&\times\left(\begin{array}{ccc}
l_1 & l_2 & L \\
-m_1 & m_2 & M
\end{array}\right)\left(\begin{array}{lll}
l_1 & l_2 & L \\
-Q & Q & 0
\end{array}\right).
\end{align}
Next, we project the density operator to the LLL. This is achieved by restricting the sum in Eqs.~\eqref{eq: second_quantized_density} and \eqref{eq: second_quantized_angular_momentum_density} to the LLL, i.e., by setting $l_1{=}l_2{=}Q$, which leads to the LLL-projected momentum-space density operator
\begin{equation}
\label{eq: second_quantized_projected_angular_momentum_density}
   \bar{\rho}^{~\sigma}_{L,M}=\sum_{\substack{ m_1, m_2}} \bar{\rho}\left(L, M, m_1, m_2\right)~ [\chi^{\sigma}_{Q, m_1}]^{\dagger} \chi^{\sigma}_{Q, m_2},
\end{equation}
where $\bar{\rho}\left(L, M, m_1, m_2\right){=}\rho\left(L, M, l_1{=}Q, m_1, l_2{=}Q, m_2\right)$. The Wigner $3j$ symbol, and consequently $\bar{\rho}\left(L, M, m_1, m_2\right)$, vanishes unless $m_2{+}M{-}m_1{=}0$, simplifying Eq.~\eqref{eq: second_quantized_projected_angular_momentum_density} to
\begin{equation}
\label{eq: simplified_second_quantized_projected_angular_momentum_density}
   \bar{\rho}^{~\sigma}_{L,M}=\sum_{m}\bar{\rho}(L,M,m)~ [\chi^{\sigma}_{ M+m}]^{\dagger} \chi^{\sigma}_{m},
\end{equation}
where $\bar{\rho}(L,M,m){=}\bar{\rho}\left(L, M, m_1{=}m{+}M, m_2{=}m\right)$. We have omitted the LL index from the operators $\chi$ with the understanding that these operators are always projected to the LLL. Within the LLL, the maximum allowed value of $L{=}2Q$ and for $L{>}2Q$ the Wigner $3j$ symbols in $\bar{\rho}(L, M,m)$ are not defined.

The projected density operators generally do not commute with each other. However, very interestingly, we will show in Sec.~\ref{sec: algebra_projected_density_operator} that they form a closed algebra. The algebraic structure of the projected density operators on the planar geometry was discovered and elucidated in detail by GMP in Ref.~\cite{Girvin86}. Next, we define the static structure factor (referred to from here on in as the structure factor), which is the density-density correlator, in an FQH ground state. As we will see later, the knowledge of the structure factor allows us to compute the ground state energy and the dispersion of the GMP mode.

\subsection{ground state structure factor}
\label{ssec: structure_factor}
For a FQH ground state described by the normalized wave function $\Psi_{\nu}$ at filling $\nu$, the structure factor is the angular momentum-space density-density correlation function evaluated in the state $\Psi_{\nu}$~\cite{Kamilla97}, i.e., 
\begin{equation}
    \label{eq: structure_factor}
S^{\sigma}\left(L\right)=\frac{4\pi}{N}\langle\Psi_{\nu}|\left[\rho^{\sigma}_{L,M}\right]^{\dagger}\rho^{\sigma}_{L,M}|\Psi_{\nu}\rangle.
\end{equation}
Due to the rotational invariance of the FQH ground state, the structure factor $S^{\sigma}\left(L\right)$ is independent of $M$. The structure factor satisfies a constraint that at $L{=}0$ it evaluates to the number of particles $N$, i.e., $S^{\sigma}\left(0\right){=}N$, independent of the nature of the state $\Psi_{\nu}$. This is because the $L{=}0$ spherical harmonic is simply a constant, i.e., $Y_{0,0}{=}1/\sqrt{4\pi}$. 

For the intra-LL density-wave excitations, the relevant quantity is the LLL-projected structure factor $\bar{S}^{\sigma}$, which is obtained by replacing $\rho^{\sigma}_{L, M}$ with $\bar{\rho}^{~\sigma}_{L, M}$ in Eq.~\eqref{eq: structure_factor}, i.e., 
\begin{equation}
     \label{eq: projected_structure_factor}
\bar{S}^{\sigma}\left(L\right)=\frac{4\pi}{N}\langle\Psi_{\nu}|\left[\bar{\rho}^{~\sigma}_{L,M}\right]^{\dagger}\bar{\rho}^{~\sigma}_{L,M}|\Psi_{\nu}\rangle.
\end{equation}
Interestingly, if the ground state $\Psi_{\nu}$ resides in the LLL, as is the case for all states considered in this work, then $S^{\sigma}\left(L\right)$ and $\bar{S}^{\sigma}\left(L\right)$ are related by an offset factor $\mathbb{O}(L)$ as (see Appendix~\ref{app: projected_unprojected_Sq})
\begin{align}
\label{eq: relaton_unprojected_projected_structure_factor}
\bar{S}^{\sigma}(L) &= S^{\sigma}(L) - \mathbb{O}(L),\notag\\
\text{where}~~
\mathbb{O}(L)&=1 - (2Q+1)~\left(\begin{array}{ccc}
Q & Q & L \\
-Q & Q & 0
\end{array}\right)^2 . 
\end{align}
The relation in Eq.~\eqref{eq: relaton_unprojected_projected_structure_factor} would be crucial to compute the projected structure factor for large system sizes. Specifically, $\bar{S}^{\sigma}$ as defined in Eq.~\eqref{eq: projected_structure_factor} can only be evaluated for small systems for which the second-quantized Fock-space decomposition of $\Psi_{\nu}$ is available~\cite{Simon94a}. However, $S^{\sigma}$ can be evaluated for comparatively larger system sizes with the first-quantized form of $\Psi_{\nu}$ using Monte Carlo methods~\cite{Kamilla97, Balram17}. Like the unprojected structure factor, the projected structure factor also satisfies the same sum rule, i.e., $\bar{S}^{\sigma}\left(0\right){=}N$, which follows from Eq.~\eqref{eq: relaton_unprojected_projected_structure_factor} since the Wigner $3j$ symbol $\{\{Q, Q,0\},\{-Q, Q,0\}\}{=}({-}1)^{2Q}/\sqrt{2Q{+}1}$. Moreover, at $L{=}1$, $\bar{\rho}^{~\sigma}_{L,M}$ annihilates the state $\Psi_{\nu}$, as a result $\bar{S}^{\sigma}\left(1\right)$ evaluates to zero [see Sec.~\ref{ssec: GMP_state}]. Additionally, since the maximum allowed value of $L$ is $2Q$, by definition $\bar{S}^{\sigma}\left(L{>}2Q\right){=}0$, and the corresponding $S^{\sigma}\left(L{>}2Q\right){=}1$.

\subsection{Interaction Hamiltonian}
\label{ssec: interaction_Hamiltonian}
We restrict the dynamics of particles to the LLL, thereby quenching their kinetic energy. The Hamiltonian comprises only the interparticle interaction term. Let two particles positioned at $\boldsymbol{\Omega}$ and $\boldsymbol{\Omega^{\prime}}$ interact with each other through a rotationally invariant potential $v\left(|\boldsymbol{\Omega} {-}\boldsymbol{\Omega^{'}}|\right)$. Thus, the Hamiltonian of the system is
\begin{equation}
  \label{eq: interaction_Hamiltonian}  
\bar{H}^{\sigma}=\frac{1}{2}\int{d\boldsymbol{\Omega}~d\boldsymbol{\Omega^{'}}}~v\left(|\boldsymbol{\Omega} {-}\boldsymbol{\Omega^{'}}|\right)~\colon\bar{\rho}^{~\sigma}(\boldsymbol{\Omega})~\bar{\rho}^{~\sigma}(\boldsymbol{\Omega^{'}})\colon,
\end{equation}
where the symbol $\colon~\colon$ indicates that the operators inside it are normal-ordered. Here, normal ordering refers to placing all the creation operators in $\bar{\rho}^{~\sigma}(\boldsymbol{\Omega})~\bar{\rho}^{~\sigma}(\boldsymbol{\Omega^{'}})$ to the left of all the annihilation operators. As a result, normal ordering eliminates self-interaction terms from the Hamiltonian. To resolve the interaction potential into its angular momentum components, we expand it in terms of the spherical harmonics as~\cite{Simon94a}
\begin{equation}
\label{eq: interaction_angular_momentum_space}
    v\big(|\boldsymbol{\Omega} {-}\boldsymbol{\Omega^{'}}|\big)=4\pi\sum_{L}~v_{L}\sum_{M{=}-L}^{L}Y_{L,M}(\boldsymbol{\Omega})~Y_{L,M}^{*}(\boldsymbol{\Omega^{'}}).
\end{equation}
The set of harmonics $\{v_{L}\}$ fully parameterizes the interaction potential. Substituting Eq.~\eqref{eq: interaction_angular_momentum_space} into Eq.~\eqref{eq: interaction_Hamiltonian} and rewriting $\colon\bar{\rho}^{~\sigma}(\boldsymbol{\Omega})~\bar{\rho}^{~\sigma}(\boldsymbol{\Omega^{'}})\colon$ in terms of a product of density operators, $\bar{H}^{\sigma}$ can be written as
\begin{equation}
\label{eq: not_normal_ordered_interaction}
    \bar{H}^{\sigma}=\frac{4\pi}{2}\sum_{L}v_{L}\sum_{M{=}-L}^{L}\left[\bar{\rho}^{~\sigma}_{L,M}\right]^{\dagger}\bar{\rho}^{~\sigma}_{L,M}~-~\bar{H}^{(s),\sigma},
\end{equation}
where $\bar{H}^{(s),\sigma}$ is a single-body term included to cancel the self-interaction. The expression for $\bar{H}^{(s),\sigma}$ is given in Appendix~\ref{app: Derivation_not_normal_ordered_interaction_Hamiltonia} where we also present a derivation of Eq.~\eqref{eq: not_normal_ordered_interaction}. 

The energy of a given FQH ground state $\Psi_{\nu}$ for the interaction $\bar{H}^{\sigma}$ can be expressed in terms of its projected structure factor $\bar{S}^{\sigma}$ [see Eq.~\eqref{eq: projected_structure_factor}] as
\begin{align}
 \label{eq: ground_state_energy}
   \langle \bar{H}^{\sigma}\rangle_{\Psi_{\nu}}&=  \frac{N}{2}\sum_{L}(2L+1)v_{L}\bar{S}^{\sigma}\left(L\right)\nonumber\\
    & - \frac{N(2Q+1)}{2}~\sum_{L}(2L+1)v_{L}\left(\begin{array}{ccc}
Q & Q & L \\
-Q & Q & 0
\end{array}\right)^2.
\end{align}
The second line of Eq.~\eqref{eq: ground_state_energy} is ${-}\langle\bar{H}^{(s),\sigma}\rangle_{\Psi_{\nu}}$ [see Appendix~\ref{app: Derivation_not_normal_ordered_interaction_Hamiltonia}]. In Appendix~\ref{app: ground_state_energies}, using the above equation, we compute the energy of various fermionic and bosonic FQH states for the Coulomb and short-range interactions discussed next.

\subsubsection{Values of $v_L$ for different interactions}
\label{sssec: value_of_v_l}
In this article, we consider both model short-range interactions such as the Trugman-Kivelson (TK) ones~\cite{Trugman85}, and the realistic long-range Coulomb interaction.
In the case of the Coulomb interaction $v^{C}(|\boldsymbol{\Omega} {-}\boldsymbol{\Omega^{'}}|){=}e^{2}/(\epsilon\ell\sqrt{Q}|\boldsymbol{\Omega} {-}\boldsymbol{\Omega^{'}}|)$, where $\epsilon$ is the permittivity of the surrounding medium, the corresponding harmonics  $v^{(C)}_{L}$ [superscript $(C)$ is for Coulomb] is~\cite{Simon94a}
\begin{equation}
\label{eq: v_l_Coulomb}
    v^{(C)}_{L}=\frac{1}{\sqrt{Q}(2L+1)}~\frac{e^{2}}{\epsilon \ell}.
\end{equation}
Throughout, we quote energies in units of $e^{2}/(\epsilon \ell)$, so for ease of notation, we drop it from many expressions below.

Next, we consider a general $k$-ranged TK interaction $v^{(k-TK)}(|\boldsymbol{\Omega} {-}\boldsymbol{\Omega^{'}}|){=}(\nabla^{2}_{\boldsymbol{\Omega}})^{k}\delta(\boldsymbol{\Omega} {-}\boldsymbol{\Omega^{'}})$, where the TK interaction's range is specified by a non-negative integer $k$ and the Laplacian operator acts only on the coordinate $\boldsymbol{\Omega}$. Using the completeness of the spherical harmonics, we can express $v^{(k-TK)}$ in the following form:
\begin{align}
\label{eq: contact_interaction_angular_momentum space}
 v^{(k-TK)}(|\boldsymbol{\Omega} {-}\boldsymbol{\Omega^{'}}|)=    \frac{1}{Q}\nabla_{\boldsymbol{\Omega}}^{2k}\sum_{L}\sum_{M{=}-L}^{L}Y_{L,M}(\boldsymbol{\Omega})~Y_{L,M}^{*}(\boldsymbol{\Omega^{'}})\nonumber\\
 =\frac{1}{Q}\sum_{L}\frac{[-L(L+1)]^{k}}{Q^{k}}\sum_{M}Y_{L,M}(\boldsymbol{\Omega})~Y_{L,M}^{*}(\boldsymbol{\Omega^{'}}),
\end{align}
where we have used the fact that the Laplacian operator is related to the square of the orbital angular momentum, i.e., $\nabla^{2}Y_{L, M}(\boldsymbol{\Omega}){=}[-L(L+1)/Q]Y_{L, M}(\boldsymbol{\Omega})$. 

Comparing Eq.~\eqref{eq: contact_interaction_angular_momentum space} with Eq.~\eqref{eq: interaction_angular_momentum_space}, one finds the harmonics of the $k$-ranged TK-interaction, 
\begin{equation}
\label{eq: v_l_contact_interaction}
    v^{(k-TK)}_{L}=\frac{1}{4\pi}\frac{[-L(L+1)]^{k}}{Q^{k+1}}.
\end{equation}
In particular, for the ultra short-range delta function interaction, relevant for bosons, $v^{(0-TK)}_{L}{\equiv}v^{(\delta)}_{L}{=}1/(4\pi Q)$.

\subsection{Haldane pair-pseudopotentials on a sphere}
\label{ssec: pair_pesudopotential}
A spherically symmetric two-body interaction can be parametrized by a set of numbers, called the Haldane pair-pseudopotentials $\{V_{l}\}$~\cite{Haldane83}, where $V_{l}$ is the energy cost for a pair of particles to be in the total angular momentum $l$. In the LLL, $l$ ranges from $0$ to $2Q$. Pseudopotentials could also be specified in terms of relative angular momentum $\mathfrak{m}{=}2Q{-}l$~\cite{Simon07}, allowing one-to-one correspondence between planar pseudopotential $V_{\mathfrak{m}}$ and pair-pseudopotential $V_{2Q{-}l}$ on a sphere. For ease of comparison with the planar geometry, we label the  Haldane pair-pseudopotentials $\{V_{l}\}$ as $\{V_{\mathfrak{m}}\}$, where $\mathfrak{m}$ similarly ranges from $0$ to $2Q$. In the LLL, for an interaction parameterized by harmonics \{$v_{L}\}$, the corresponding pseudopotential $V_{\mathfrak{m}}$ is~\cite{Wooten14}
\begin{align}
\label{eq: pair_pseudopotential}
V_{\mathfrak{m}}= & (-1)^{\mathfrak{m}}(2 Q+1)^2 \nonumber\\
& \times \sum_{L=0}^{2 Q} v_{L}~(2L+1)\left\{\begin{array}{lll}
2Q-\mathfrak{m} & Q & Q \\
L & Q & Q
\end{array}\right\}\left(\begin{array}{ccc}
Q & L & Q \\
-Q & 0 & Q
\end{array}\right)^2,
\end{align}
where the large curly bracket denotes the Wigner $6j$ symbol. The above expression allows us to compute the pseudopotentials of any general interaction. 

For the Coulomb interaction in the LLL, using the harmonics $v^{(C)}_{L}$ from Eq.~\eqref{eq: v_l_Coulomb} in Eq.~\eqref{eq: pair_pseudopotential}, the Haldane pseudopotential $V_{\mathfrak{m}}^{\left(C\right)}$ is~\cite{Fano86}
\begin{equation}
  V_{\mathfrak{m}}^{\left(C\right)}=\frac{2}{\sqrt{Q}}\frac{\binom{2\mathfrak{m}}{\mathfrak{m}}\binom{8Q-2\mathfrak{m}+2}{4Q-\mathfrak{m}+1}}{\binom{4Q+2}{2Q+1}^2}. 
\end{equation}
Note that, unlike in the LLL, there is no known compact expression for the corresponding $V_{\mathfrak{m}}^{\left(C\right)}$ in higher LLs.

The pseudopotentials of the contact interaction $v^{(0-TK)}{=}\delta(\boldsymbol{\Omega} {-}\boldsymbol{\Omega^{'}})$, relevant for bosons, are~\cite{Sharma23}
\begin{equation}
\label{eq: pair_pseudopotential_delta_function}
    V^{(0-TK)}_{\mathfrak{m}{=}0}=\frac{(2Q+1)^2}{4\pi Q(1+4Q)},~ \text{and} ~~V^{(0-TK)}_{\mathfrak{m}}=0~~\forall \mathfrak{m}>0.
\end{equation}
Similarly, for the interaction $v^{(1-TK)}{=}\nabla^{2}_{\boldsymbol{\Omega}}\delta(\boldsymbol{\Omega} {-}\boldsymbol{\Omega^{'}})$, usually relevant for fermions, the LLL pseudopotentials are:
\begin{equation}
\label{eq: pair_pseudopotential_TK_interaction}
V_{\mathfrak{m}}^{(1-TK)} =
\begin{cases}
-\frac{(2Q+1)^2}{4\pi Q(1+4Q)} &  \mathfrak{m} = 0, \\
\frac{(2Q+1)^2}{4\pi Q(4Q-1)}&  \mathfrak{m} =1, \\
~~~~~0 &  \mathfrak{m} >1.
\end{cases}
\end{equation}
In general, for the $k$-ranged TK interaction $V_{\mathfrak{m}}^{(k-TK)}$ vanishes when $\mathfrak{m}{>}k$. Notably, this is different from the pseudopotentials of the $k$-ranged TK interaction $\nabla^{2k}_{\boldsymbol{r}}\delta(\boldsymbol{r}{-}\boldsymbol{r}^{\prime})$ in the planar geometry where there is only one nonzero pseudopotential at relative angular momentum $\mathfrak{m}{=}k$ while others are zero. For a fixed interaction, the spherical and planar pseudopotentials are identical in the thermodynamic limit, i.e., $Q{\to} \infty$, since the sphere becomes a plane in this limit. Nevertheless, the difference in the pseudopotentials show that $k{-}TK$ spherical interaction $\nabla^{2k}_{\boldsymbol{\Omega}}\delta(\boldsymbol{\Omega} {-}\boldsymbol{\Omega^{'}})$, for $k{\geq}1$, is different from the $k{-}TK$ planar interaction $\nabla^{2k}_{\boldsymbol{r}}\delta(\boldsymbol{r}{-}\boldsymbol{r}^{\prime})$ even in the thermodynamic limit. The spherical pseudopotentials of the ultra short-range interaction $\delta(\boldsymbol{\Omega} {-}\boldsymbol{\Omega^{'}})$ ($k{=}0$) in the thermodynamic limit [see Eq.~\eqref{eq: pair_pseudopotential_delta_function}] do match the planar pseudopotentials, apart from an overall constant that can be adjusted to vary the strength of the contact interaction. For fermions (bosons), only odd (even) $\mathfrak{m}$ pseudopotentials are relevant~\cite{Wooten14}. 

The Laughlin state~\cite{Laughlin83} is an exact zero-energy incompressible ground state of the TK interaction. In particular, the $1/2$ bosonic Laughlin state is realized for the model interaction $V_{0}{=}1$ with all other pseudopotentials being zero. Similarly, the fermionic $1/3$ and $1/5$ Laughlin states are stabilized when $V_{1}{=}1$, and $V_{1}{=}1$ and $V_{3}{=}1$, respectively, keeping all the other $V_{\mathfrak{m}}{=}0$. In general, the $\nu{=}1/p$ Laughlin state is realized when $V_{\mathfrak{m}}{=}1$ for $\mathfrak{m}{<}p$, while all other $V_{\mathfrak{m}}{=}0$.

Interestingly, for a given set of pseudopotentials $\{V_{\mathfrak{m}}\}$, an FQH state's energy and the dispersion of its GMP mode can be ascertained from its static structure factor. We remind the readers that the computation of these energies requires the knowledge of the harmonics $\{v_{L}\}$ [see Eq.~\eqref{eq: ground_state_energy}]. This is achieved through the following relation, which uniquely inverts a set of pseudopotentials $\{V_{\mathfrak{m}}\}$ to yield the corresponding $\{v_{L}\}$~\cite{Wooten14} 
\begin{align}
\label{eq: inversion_pair_pseudopotential}
v_L & =\frac{1}{(2 Q+1)^2}\left(\begin{array}{ccc}
Q & L & Q \\
-Q & 0 & Q
\end{array}\right)^{-2}~\sum_{\mathfrak{m}=0}^{2 Q}\bigg[(-1)^{\mathfrak{m}}~ V_{\mathfrak{m}}\nonumber \\
& \times \left(2 \left(2Q-\mathfrak{m}\right)+1\right)\left\{\begin{array}{lll}
Q & Q & ~~~~L \\
Q & Q & 2Q-\mathfrak{m}
\end{array}\right\} \bigg],
\end{align}
Next, we derive the algebra of the projected density operators in the spherical geometry, which would be used to evaluate the dispersion of the GMP mode.

\section{Algebra of lowest Landau level projected density operators}
\label{sec: algebra_projected_density_operator}

In the restricted Hilbert space of the LLL, the projected density operators generally do not commute with one another. In the planar geometry, the commutation algebra of the momentum-space projected density operators was worked out by GMP and is referred to as the GMP algebra~\cite{Girvin86}, which is also closely related to the $\mathcal{W}_{\infty}$ algebra~\cite{Cappelli93, Cappelli21}. Here, we seek the analog of the GMP algebra in the angular momentum space in the spherical geometry. On the plane, the commutator of two projected density operators is a projected density operator. This is not the case in spherical geometry. Nevertheless, encouragingly, we find that the commutation of two projected density operators on the sphere is a \emph{finite} sum of projected density operators. 

Let us begin by considering two arbitrary LLL projected density operators $ \bar{\rho}^{~\sigma}_{L_1, M_1}$ and $\bar{\rho}^{~\sigma}_{L_2, M_2}$~[see Eq.~\eqref{eq: simplified_second_quantized_projected_angular_momentum_density}]. Their commutator is:
\begin{flalign}
\label{eq: GMP_commutator_expanded}
[\bar{\rho}^{~\sigma}_{L_1,M_1},\bar{\rho}^{~\sigma}_{L_2,M_2}] 
& = \sum_{m} \bigg[ \bar{\rho}(L_1, M_1; L_2, M_2; m) \bigg]\left [\chi^{\sigma}_{M+m}\right]^{\dagger} \chi^{\sigma}_{m},
\end{flalign}
\begin{flalign}
\intertext{where~$M{=}M_1{+}M_2$~~and}
\bar{\rho}(L_1, M_1; L_2, M_2; m) 
& = \bar{\rho}(L_1, M_1,M_2+m)~\bar{\rho}(L_2, M_2,m) \nonumber \\
& -\bar{\rho}(L_2, M_2, M_1+m)~\bar{\rho}(L_1, M_1,m).
\label{eq: coefficient_GMP_commutator}
\end{flalign}

The LLL projected density operators form a complete basis, i.e., any single-particle LLL-projected operator can be expressed in terms of them~\cite{Spodyneiko23}. This implies that the commutator in Eq.~\eqref{eq: GMP_commutator_expanded} can be written as a linear sum of projected density operators $\bar{\rho}^{~\sigma}_{L,\tilde{M}}$ over all the allowed values of $L$ and $\tilde{M}$. However, comparing the right-hand side of Eq.~\eqref{eq: GMP_commutator_expanded} with the expression of projected density operator as given in Eq.~\eqref{eq: simplified_second_quantized_projected_angular_momentum_density} requires that $\tilde{M}{=}M{=}M_1{+}M_2$. More precisely, the following equation encodes these discussions:
\begin{equation}
    \label{eq: GMP_algebra}
[\bar{\rho}^{~\sigma}_{L_1,M_1},\bar{\rho}^{~\sigma}_{L_2,M_2}]=\sum_{L}\alpha_{L}^{(L_1,L_2,M_1,M_2)}\bar{\rho}^{~\sigma}_{L,M}.
\end{equation}
The expansion coefficients $\alpha_{L}^{(L_1,L_2,M_1,M_2)}$ are given by
\begin{align}
\label{eq: expression_expansion_coefficients}
  \alpha_{L}^{(L_1,L_2,M_1,M_2)}&=(-1)^{M}\bigg[(-1)^{L_1+L_2+L}-1\bigg]\frac{\mathcal{F}(L_1)\mathcal{F}(L_2)}{\mathcal{F}(L)}\nonumber\\
  &\times(2L+1)\left(\begin{array}{ccc}
L_1 & L_2 & L \\
M_1 & M_2 & -M
\end{array}\right)\left\{\begin{array}{lll}
L_1 & L_2 & L \\
Q & Q & Q
\end{array}\right\},
\end{align}
where the LLL form factor $\mathcal{F}(K)$ is:
\begin{align}
\nonumber
   \mathcal{F}(K)=(2Q+1)\sqrt{\frac{2K+1}{4\pi}} \left(\begin{array}{ccc}
Q & Q & K \\
-Q & Q & 0
\end{array}\right).
\end{align}
In Appendix~\ref{app: derivation_GMP_algebra_coefficients} we provide a complete derivation of Eq.~\eqref{eq: expression_expansion_coefficients}. The expansion coefficients, and thus the algebra of the projected density operators, are identical for bosons and fermions.

One can see from Eq.~\eqref{eq: expression_expansion_coefficients} that for the expansion coefficient to be nonzero, $L$ must satisfy the triangular inequality condition, i.e., $|L_1{-}L_2|\leq L\leq L_1{+}L_2$, which stems from the properties of the Wigner $3j$ symbol. Furthermore, $L_1{+}L_2{+}L$ must be odd, otherwise the factor $\bigg[(-1)^{L_1{+}L_2{+}L}{-}1\bigg]$ in Eq.~\ref{eq: expression_expansion_coefficients} becomes zero. As a result, the summation over $L$ in Eq.~\eqref{eq: GMP_algebra} runs from $|L_1{-}L_2|{+}1$ to $|L_1{+}L_2{-}1|$ in steps of two. We note that within the LLL, the maximum allowed $L$ is $2Q$ and $L$ must also be greater or equal to $M$ for the expansion coefficients to be nonzero.

To this end, the following special cases of the commutation algebra are noteworthy:
\begin{flalign}
\label{eq: commutation_special_case_1}
   &[\bar{\rho}^{~\sigma}_{L_1,0}, ~\bar{\rho}^{~\sigma}_{L_2,0}] = 0, \\
  &[\bar{\rho}^{~\sigma}_{L_1,L_1}, ~\bar{\rho}^{~\sigma}_{L_2,L_2}]= 0, \\
  &[\bar{\rho}^{~\sigma}_{L_1,-L_1}, ~\bar{\rho}^{~\sigma}_{L_2,-L_2}]= 0.
\end{flalign}
The above commutators vanish because, for all of them, the corresponding expansion coefficients $\alpha_{L}^{(L_1, L_2, M_1, M_2)}$ are identically zero. Furthermore, the expansion coefficients of the commutator $[\bar{\rho}^{~\sigma}_{L_1,M_1\leq0},\bar{\rho}^{\sigma}_{L_2,M_2\leq0}]$ are related to that of $[\bar{\rho}^{~\sigma}_{L_1,M_1\geq0},\bar{\rho}^{\sigma}_{L_2,M_2\geq0}]$ by a minus sign, i.e.,
\begin{align}
\label{eq: relation_GMP_algebra_coefficients}
\alpha_{L}^{\left(L_1,L_2,M_1\geq0,M_2\geq0\right)}{=}{-}\alpha_{L}^{\left(L_1,L_2,M_1\leq0,M_2\leq0\right)}.
\end{align}
This can be seen by noting the relation $\left[\bar{\rho}^{~\sigma}_{L,M}\right]^{\dagger}{=}(-1)^{M}\bar{\rho}^{~\sigma}_{L,-M}$ and then taking the Hermitian adjoint of $[\bar{\rho}^{~\sigma}_{L_1,M_1\geq0},\bar{\rho}^{~\sigma}_{L_2,M_2\geq0}]$. Moreover, the anti-symmetric property of the commutator implies that interchanging $(L_1, M_1)$ and $(L_2, M_2)$ in $\alpha_{L}^{\left(L_1, L_2, M_1, M_2\right)}$ introduces a minus sign, i.e.,
\begin{align}
    \label{eq: interchange_relation_GMP_algebra_coefficients}
\alpha_{L}^{\left(L_1,L_2,M_1,M_2\right)}{=}{-}\alpha_{L}^{\left(L_2,L_1,M_2,M_1\right)}.
\end{align}

In Sec.~\ref{ssec: Validation of the GMP algebra}, we validate the correctness of the above algebra of projected density operators. This is done by computing the GMP gap using the algebra, which we will discuss next, and then by comparing this with the gap computed directly from the GMP wave function.

\section{Dispersion of the GMP density-wave collective mode on the sphere}
\label{sec: magnetoroton_dipersion}
Here, we discuss the construction of the GMP state on the sphere and then compute its gap relative to the ground state using the algebra of projected density operators derived in the previous section. 
\subsection{GMP state}
\label{ssec: GMP_state}
The GMP state~\cite{Girvin85, Girvin86} is constructed by applying the momentum-space density operator on the ground state $|\Psi_{\nu}\rangle$, i.e.,
\begin{equation}
    \label{eq: GMP_wave_function}
  |\Psi_{L,M}^{\rm GMP}\rangle = \bar{\rho}_{L,M}^{~\sigma}|\Psi_{\nu}\rangle.
\end{equation}
Since $\bar{\rho}^{~\sigma}_{0,0}$ is proportional to the identity operator [This can be seen by explicitly evaluating the coefficient $\bar{\rho}(L{=}0, M{=}0, m)$ in Eq.~\eqref{eq: simplified_second_quantized_projected_angular_momentum_density}, which is independent of $m$, so it is just a constant.], it follows that $|\Psi_{0,0}^{\rm GMP}\rangle{\propto}|\Psi_{\nu}\rangle$. Similarly, $\bar{\rho}^{\sigma}_{1,0}{\propto}\sum_{i{=1}}^{N}\hat{L}^{z}_{i}$ when written in the second quantized notation in the LL basis~\cite{He94}. Consequently, $|\Psi_{1,0}^{\rm GMP}\rangle$ vanishes, as the ground state is rotationally invariant. Furthermore, $\bar{\rho}_{1,1}$ and $\bar{\rho}_{1,{-}1}$ are proportional to the total angular momentum raising and lowering operators, $\hat{\mathcal{L}}^{+}{=}\sum_{i{=}1}^{N}\hat{L}_{i}^{+}$ and $\hat{\mathcal{L}}^{-}{=}\sum_{i{=}1}^{N}\hat{L}_{i}^{-}$, respectively~\cite{He94}. Therefore, $|\Psi_{1,1}^{\rm GMP}\rangle$ and $|\Psi_{1,{-}1}^{\rm GMP}\rangle$ also vanish implying that the $L{=}1$ GMP state is annihilated. In general, $\bar{\rho}^{~\sigma}_{L{\geq}1,0}$ is proportional to the second quantized expression of the operator $\sum_{i}f_1(\hat{L}_{i}^{z})$. Similarly, $\bar{\rho}^{~\sigma}_{L{\geq}1,M{>}0}$ is proportional to the second-quantized version of the one-body operator $\sum_{i}g(\hat{L}^{+}_i)f_2(\hat{L}_{i}^{z})$. Here $f_1$, $f_2$, and $g$ are polynomial functions. Along with the identity $\left[\bar{\rho}^{~\sigma}_{L, M}\right]^{\dagger}{=}(-1)^{M}\bar{\rho}^{~\sigma}_{L,-M}$, these polynomial expressions would be useful in simplifying expressions involving $\bar{\rho}^{~\sigma}_{L,M}$.

\subsection{GMP gap equation}
\label{ssec: magnetoroton_gap}
To obtain the dispersion of the GMP mode, we compute the energy of the state $\left|\Psi_{L, M}^{\rm GMP}\right\rangle$ [see Eq.~\eqref{eq: GMP_wave_function}] relative to the ground state $\left|\Psi_{\nu}\right\rangle$ for the interaction $\bar{H}^{\sigma}$ [defined in Eq.~\eqref{eq: not_normal_ordered_interaction}]. If $\left|\Psi_{\nu}\right\rangle$ is an exact eigenstate of $\bar{H}^{\sigma}$ with energy $E_{0}$, then the energy gap $\Delta(L)$ of $\left|\Psi_{L,M}^{\rm GMP}\right\rangle$ is:
\begin{equation}
\label{eq: GMP_gap_direct_evaluation}
     \Delta(L)= \frac{\langle \Psi_{\nu}|[\bar{\rho}^{~\sigma}_{L,M}]^{\dagger}~\bar{H}^{\sigma}~\bar{\rho}^{~\sigma}_{L,M}|\Psi_{\nu}\rangle}{\langle\Psi_{\nu}|[\bar{\rho}^{~\sigma}_{L,M}]^{\dagger}\bar{\rho}^{~\sigma}_{L,M}|\Psi_{\nu}\rangle} - E_{0}.
\end{equation}
As expected for a spherically symmetric interaction, the gap $\Delta(L)$ is independent of $M$. Utilizing the rotational invariance of $\bar{H}^{\sigma}$ and $\left|\Psi_{\nu}\right\rangle$, and noting that we assumed $\bar{H}^{\sigma}\left|\Psi_{\nu}\right\rangle{=}E_{0}\left|\Psi_{\nu}\right\rangle$, and using the relation $\left[\bar{\rho}^{~\sigma}_{L,M}\right]^{\dagger}{=}(-1)^{M}\bar{\rho}^{~\sigma}_{L,{-}M}$, one can express $\Delta(L)$ in terms of double commutator as follows:
\begin{align}
\label{eq: gap_equation}
 \Delta(L)&= \frac{4\pi}{N}\frac{\bar{F}^{\sigma}(L)}{\bar{S}^{\sigma}(L)},
 \intertext{where the oscillator strength}
\bar{F}^{\sigma}(L)&=\frac{1}{2}\left\langle \Psi_{\nu}\right|\left[\left[\bar{\rho}^{~\sigma}_{L,M}\right]^{\dagger},\left[\bar{H}^{\sigma},\bar{\rho}^{~\sigma}_{L,M}\right]\right]\left|\Psi_{\nu}\right\rangle\nonumber,
\end{align}
and the denominator in Eq.~\eqref{eq: GMP_gap_direct_evaluation} by definition is the projected structure factor $\bar{S}^{\sigma}$ [see Eq.~\eqref{eq: projected_structure_factor}]. Without loss of generality, we set $M{=}0$. With the aid of the commutation algebra of the projected density operators [see Eq.~\eqref{eq: GMP_algebra}], $\bar{F}^{\sigma}(L)$ can be expressed in terms of $\bar{S}^{\sigma}$ as
\begin{flalign}
\label{eq: numerator_gap_equation}
 \bar{F}^{\sigma}(L)&=\frac{N}{4}\sum_{\Tilde{L}}v_{\Tilde{L}}\sum_{\Tilde{M}>0}\sum_{\lambda}\bigg[4\left(\alpha_{\lambda}^{(\tilde{L},L,\tilde{M},0)}\right)^{2}~\bar{S}^{\sigma}(\lambda)\nonumber\\
    & + 4\alpha_{\lambda}^{(\tilde{L},L,\tilde{M},0)}~\alpha_{\tilde{L}}^{(L,\lambda,0,\tilde{M})}~\bar{S}^{\sigma}(\tilde{L})\bigg],
\end{flalign}
where the angular momentum quantum number $\lambda$ ranges from $\left|L{-}\tilde{L}\right|{+}1$ to $\left|L{+}\tilde{L}{-}1\right|$ in steps of two. We reiterate that $\tilde{L}$ runs from $0$ to $2Q$ in steps of one [see Sec.~\ref{ssec: density_operators}]. The derivation of Eq.~\eqref{eq: numerator_gap_equation} is provided in Appendix~\ref{app: Oscillator strength_sphere}. Eq.~\eqref{eq: gap_equation}, together with Eq.~\eqref{eq: numerator_gap_equation}, constitutes the spherical analog of the planar GMP gap equation~\cite{Girvin86} [see Eq.~\eqref{eq: planar_oscillator_strength}] and is the main result of the current work. 

Although we assumed the ground state is an exact eigenstate of the Hamiltonian in the above discussion, we will use trial wave functions instead for interactions where the exact ground state is difficult to obtain. As long as the trial ansatz is an accurate variational wave function, the gap obtained from the GMP gap equation using the trial wave function's  $\bar{S}^{\sigma,\rm trial}$ will lie close to that obtained using the exact ground state's $\bar{S}^{\sigma,\rm exact}$. In the following, when from the context it is clear that the state under consideration is bosonic or fermionic, for brevity, we drop the symbol $\sigma$ from $S^{\sigma}$, $\bar{S}^{\sigma}$, $\bar{S}^{\sigma,\rm trial}$, and $\bar{S}^{\sigma,\rm exact}$. Next, we take a detour to discuss the composite fermion exciton (CFE), which provides a different description of the lowest-lying neutral excitation for the Jain states. Later on, we will compare the gaps of the neutral excitation obtained from the CFE and GMP states.     

\section{Composite fermion exciton (CFE) gap}
\label{sec: CFE_gap}
In the CF theory~\cite{Jain89}, electrons under a total magnetic flux of $2Q$ (in units of flux quantum) are mapped to CFs carrying $2p$ quantized vortices experiencing an effective flux of $2Q^{*}{=}2Q{-}2p(N{-}1)$. The ground state of electrons at $\nu{=}n/(2np{+}1)$ is the IQH state of CFs filling $n$ CF-LLs or Lambda levels ($\Lambda$Ls). The ground state at $\nu{=}n/(2np{+}1)$ is described by the Jain wave function~\cite{Jain89}
\begin{equation}
    \label{eq: CF_ground_state}
    \Psi^{\rm CF}_{\nu{=}n/(2np{+}1)}=\mathcal{P}_{\rm LLL}\Phi_{1}^{2p}\Phi_{n}.
\end{equation}
Here, $\mathcal{P}_{\rm LLL}$ denotes the projection to the LLL and $\Phi_{n}$ is the Slater determinant IQH wave function of $n$ filled LLs. In particular, the wave function of the lowest filled LL $\Phi_{1}$ is the Laughlin-Jastrow factor, and it raised to the power $2p$ attaches $2p$ vortices to each electron
to turn them into CFs. In our notation, the CF $\Lambda$Ls on the sphere are labeled by $l{=}Q^{*}{+}n$, where $n{=}0,1,2,\cdots$. The CFE is obtained by promoting a CF from the topmost occupied $\Lambda$L with $l{=}Q^{*}{+}n{-}1$ to the lowest unoccupied $\Lambda$L with $l{=}Q^{*}{+}n$, thus creating a CF-hole (CFH) and CF-particle (CFP) pair in the respective $\Lambda$Ls. The CFE state carrying a total angular momentum $L$ and azimuthal quantum number $M$ is given by~\cite{Kamilla96b}
\begin{equation}
\label{eq: CFE_state}
    \Psi^{\rm CFE}_{L,M}= \mathcal{P}_{\rm LLL} \Phi_{1}^{2p}\rho^{\left(n-1\right)\to n}_{L,M}~\Phi_{n}.
\end{equation}
Here $\rho^{\left(n-1\right)\to n}_{L, M}$ creates a pair of CFH and CFP atop the filled CF ground state $\Phi_n$ and can be expressed as~\cite{Kamilla96b}
\begin{equation}
\label{eq: CFE_creation_operator}
  \rho^{\left(n-1\right)\to n}_{L,M}=\sum_{m}\rho^{\rm CF}(L,M,m)~ \chi_{Q^{*}+n, M+m}^{\dagger} ~\chi_{Q^{*}+n-1,m},
\end{equation}
where $\rho^{\rm CF}(L,M,m){=}\rho\left(L,M,l_1, m_1{=}m{+}M,l_2,m_2{=}m\right)$ [see Eq.~\eqref{eq: second_quantized_angular_momentum_density_value}] and $l_1{=}Q^{*}{+}n$, and $l_2{=}Q^{*}{+}n{-}1$. We note that the CFE state at $L{=}1$ vanishes upon projection to the LLL~\cite{Dev92}. In Eq.~\eqref{eq: CFE_creation_operator}, we have omitted the label $\sigma$ from the LL creation/annihilation operator with the understanding that we are considering fermions in this section. However, the above discussion also applies to bosons with the understanding that the number of vortices attached to them to form composite fermions is odd, i.e., $2p{+}1$. 

The CFE gap, which is the energy of the CFE state [see Eq.\eqref{eq: CFE_state}] relative to the CF ground state, is given by
\begin{equation}
\label{eq: CFE_gap}
    \Delta^{\rm CFE}(L)= \frac{\langle\Psi^{\rm CFE}_{L,0}|\bar{H}|\Psi^{\rm CFE}_{L,0}\rangle}{\langle\Psi^{\rm CFE}_{L,0}|\Psi^{\rm CFE}_{L,0}\rangle}-\frac{\langle\Psi_{\nu}|\bar{H}|\Psi_{\nu}\rangle}{\langle\Psi_{\nu}|\Psi_{\nu}\rangle}.
\end{equation}
Owing to the rotational invariance of the interparticle interaction, we have set $M{=}0$ in the CFE state. Using the Metropolis Monte Carlo method, the CFE gap can be evaluated for large systems~\cite{Kamilla96b, Scarola00, Park00a, Balram16d} using the Jain-Kamilla method of projection to the LLL~\cite{Jain97, Jain97b}. 

\section{Results}
\label{sec: results}
\subsection{Validation of the GMP algebra}
\label{ssec: Validation of the GMP algebra}

\begin{figure*}[tbh!]
        \includegraphics[width=0.66\columnwidth]{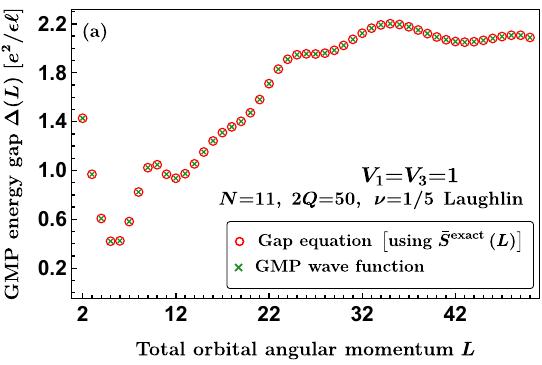}
        \includegraphics[width=0.66\columnwidth]{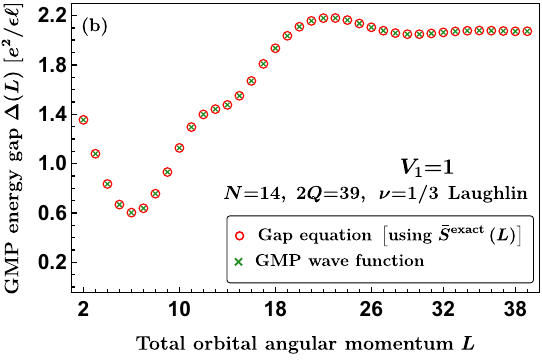}
         \includegraphics[width=0.66\columnwidth]{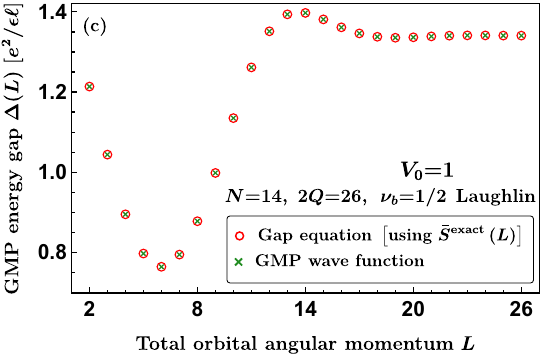}
          \caption{Validation of the GMP algebra on a sphere [see Eq.~\eqref{eq: GMP_algebra}]. Comparison of the short-range model interaction GMP gaps obtained from the GMP wave function [green crosses] and the GMP gap equation [see Eq.~\eqref{eq: gap_equation}] employing the exact projected structure factor [red circles], for the fermionic Laughlin states at fillings 1/5 [panel $(a)$] and 1/3 [panel $(b)$], and the bosonic  Laughlin state at filling  1/2 [panel $(c)$].
          }
          \label{fig: Validation of GMP algebra plots}
        \end{figure*}        
To test the GMP algebra presented in Eq.~\eqref{eq: GMP_algebra}, we first consider model Hamiltonians hosting exact ground states and compute the GMP gap in two distinct ways. First, the GMP gap is calculated by directly evaluating the expectation value in Eq.~\eqref{eq: GMP_gap_direct_evaluation} using the Fock-space representation of both the GMP state $\bar{\rho}^{~\sigma}_{L, M}|\Psi_{\nu}\rangle$ and the Hamiltonian $\bar{H}^{\sigma}$ [see Eq.~\eqref{eq: interaction_Hamiltonian}]. Secondly, we deploy the derived GMP gap equation that uses the exact projected structure factor $\bar{S}^{\sigma,{\rm exact}}\left(L\right)$ of the exact ground state which is also computed in the Fock-space. Although these methods are limited to small systems, they provide a direct way to verify the correctness of the GMP algebra. 

In Figs.~\ref{fig: Validation of GMP algebra plots}(a),~\ref{fig: Validation of GMP algebra plots}(b) and \ref{fig: Validation of GMP algebra plots}(c), we present the gaps computed using the aforementioned two methods for the $1/5$ and $1/3$ fermionic and the $1/2$ bosonic Laughlin states, respectively. The GMP gap computed using two entirely different methods yields identical results, thereby validating the derived GMP algebra of projected density operators for both bosonic and fermionic systems.

In Fig.~\ref{fig: Validation of GMP algebra plots}, one can observe that the GMP mode flattens out as the angular momentum $L$ approaches $2Q$. This saturation of the GMP mode is also consistent with the planar GMP gap which goes to a constant at large momentum~\cite{Girvin86}. The dispersion of the GMP mode at large wavenumber saturates since in this limit the constituent hole and the particle are far away from each other and behave as free particles~\cite{Girvin86}.

\begin{figure*}[tbh!]
\includegraphics[width=0.99\columnwidth]{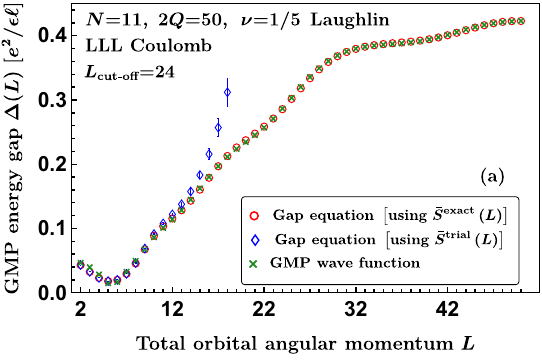}
\includegraphics[width=0.99\columnwidth]{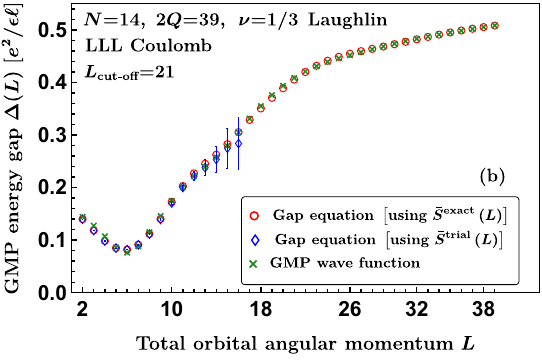}\\
\includegraphics[width=0.99\columnwidth]{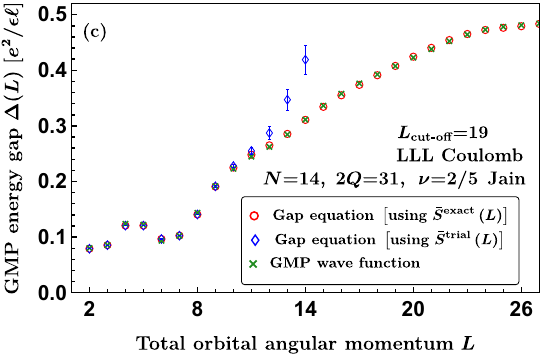}
\includegraphics[width=0.99\columnwidth]{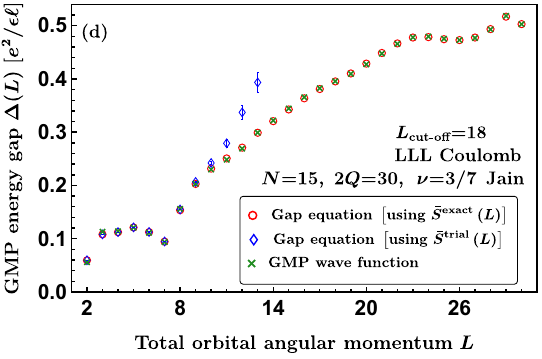}
 \caption{Comparison of the LLL Coulomb GMP gaps obtained in three different ways: $(i)$ using the Fock-space representation of the GMP wave function [gree crosses], $(ii)$ employing the exact projected structure factor $\bar{S}^{\rm exact}$ [red-open circles], and $(iii)$ the projected structure factor $\bar{S}^{\rm trial}$ inferred from the unprojected structure factor [blue-open diamonds] in the GMP gap equation [see Eq.~\eqref{eq: gap_equation}], for the 1/3 and 1/5 Laughlin and the 2/5 and 3/7 Jain states. See the text for the definition of $L_{\rm cut-off}$.}
 \label{fig: Coulomb_GMP_gap_exact_SF}
\end{figure*}

\subsection{Coulomb GMP gaps from near-exact trial states}
\label{ssec: Coulomb_GMP_gap_small_system_size}
In this section, we present the GMP gap for the realistic Coulomb interaction. Here, the GMP state is constructed from a trial wave function, which provides an accurate but not exact representation of the Coulomb ground state. In contrast to the discussion in the previous section for model short-range interactions, the Coulomb gap computed from the GMP gap equation (using the exact projected structure factor), in general, will not exactly match that computed using the Fock-space representation of the GMP state. This discrepancy arises because the trial state is not an exact ground state of the Coulomb interaction. However, the difference between the two gaps will be small for trial states that encode the correlations of the underlying exact ground state accurately.

Figure~\ref{fig: Coulomb_GMP_gap_exact_SF} shows the Coulomb GMP gap for Laughlin and Jain states, which are known to accurately capture the exact LLL Coulomb ground states~\cite{Ambrumenil88, Dev92a, Wu93, Jain07, Balram13, Kusmierz18, Yang19a, Balram20a, Balram20b, Balram21b}. The GMP gap computed from the Fock-space representation of the GMP state (green cross marks) and that obtained from the gap equation using the exact projected structure factor of the trial state, $\bar{S}^{\rm exact}$ (red circles) differ only very slightly from each other all the way up to $L_{\rm max}{=}2Q$. 

To access large systems, we obtain the unprojected structure factor for the trial wave functions using Monte Carlo methods, and discern the projected structure factor $\bar{S}^{\rm trial}$ from that, and then use the GMP gap equation with $\bar{S}^{\rm trial}$. This conversion involves certain approximations that we outline in the next section.

\subsubsection{Approximate projected structure factor from unprojected structure factor}
\label{sssec: approximate_projected_structure_factor}
The unprojected structure factor $S$ is computed from the trial wave function via Monte Carlo integration~\cite{Kamilla97, Balram17} with an accuracy up to certain decimal places. Therefore, the resulting projected structure factor $\bar{S}^{\rm trial}$ obtained from it employing Eq.~\eqref{eq: relaton_unprojected_projected_structure_factor} agrees with the exact projected structure factor of the trial wave function $\bar{S}^{\rm exact}$ only up to certain decimal places. The difference between $\bar{S}^{\rm trial}$ and $\bar{S}^{\rm exact}$ is small up to some angular momentum cut-off $L_{\rm cut-off}$ beyond which the discrepancy between their values is large. The cut-off $L_{\rm cut-off}$ is placed at the $L$ where the number $\left[1{-}\mathbb{O}\left(L\right)\right]$ [see Eq.~\eqref{eq: relaton_unprojected_projected_structure_factor} for the definition of the offset $\mathbb{O}$ that relates $S$ and $\bar{S}^{\rm trial}$] equals the order of the error in the $S$ data. As $L$ increases, both $S$ and $\mathbb{O}$ approach unity ($\mathbb{O}$ is exactly one for the $L{>}2Q$ while $S$, whose exact value is one for $L{>}2Q$, never becomes exactly one due to numerical Monte Carlo statistical errors). To alleviate the unwanted errors in the GMP gap arising from $\bar{S}^{\rm trial}(L)$, we set $S(L){=}1$ and $\mathbb{O}(L){=}1$ for $L{>}L_{\rm cut-off}$. Consequently, the $\bar{S}^{\rm trial}(L){=}0$ for $L{>}L_{\rm cut-off}$.

To illustrate these ideas better, we consider an example of the $1/3$ Laughlin state with $N{=}14$ particles at flux $2Q{=}39$. The statistical Monte Carlo error in our computed $S(L)$ data is of order $10^{-5}$. As $L$ increases from $0$, $(1{-}\mathbb{O})$ decreases and approaches $10^{-5}$ from above first at $L{=}21$ and this is where we place $L_{\rm cut-off}$. Consequently, for this system, we set $\bar{S}^{\rm Laughlin}(L){=}0$ for $L{>}21$.

In Fig.~\ref{fig: Coulomb_GMP_gap_exact_SF}, we show the GMP gap computed using the above approximate $\bar{S}^{\rm trial}$ (see blue-empty diamonds). This gap agrees very well with that obtained from $\bar{S}^{\rm exact}$ (red-empty circles) for small $L$. As $L$ increases, the GMP gap from $\bar{S}^{\rm trial}$ slowly begins to deviate and eventually near the vicinity of $L_{\rm cut-off}$ significantly differs from the gap ascertained using the $\bar{S}^{\rm exact}$. In the following paragraph, we explain this deviation of the GMP gap from its expected value around $L_{\rm cut-off}$.

We begin our discussion with the GMP gap computed using $\bar{S}^{\rm exact}$ since it provides a way to compute the gap exactly. Although the summation over $\tilde{L}$ in Eq.~\eqref{eq: numerator_gap_equation} runs from $0$ to $2Q$, we find that limiting it to $L_{\rm cut-off}$ still gives a very good estimate of the GMP gap, accurate to five decimal places, for all allowed $L$. Thus, for the following discussion, we restrict the $\tilde{L}$ in Eq.~\eqref{eq: numerator_gap_equation} to $0{\leq}\tilde{L}{\leq}L_{\rm cut-off}$. Around $L_{\rm cut-off}$, generally, $\bar{S}^{\rm exact}$ is of order $10^{-6}$ and becomes much smaller as $L{\gg}L_{\rm cut-off}$. To test the importance of the contribution arising from $\bar{S}^{\rm exact}$ for $\tilde{L}{>}L_{\rm cut-off}$ to the GMP gap $\Delta(L)$ as $L$ increases, we set $\bar{S}^{\rm exact}{=}0$ for $\tilde{L}{>}L_{\rm cut-off}$. Comparing the gaps with and without the cut-off, we find that the gaps for $L{\ll}L_{\rm cut-off}$ agree with each other but as $L{\approx}L_{\rm cut-off}$ they differ significantly. This is because as $L{\to}L_{\rm cut-off}$, the total number of $\bar{S}^{\rm exact}$ terms having $\tilde{L}{>}L_{\rm cut-off}$ in the GMP gap equation [see Eq.~\eqref{eq: numerator_gap_equation}] increases (which we have set to zero for investigation) but the coefficients multiplying them (from the GMP algebra) become very large. However, for $L{\ll}L_{\rm cut-off}$, the number of $\bar{S}^{\rm exact}$ terms with $\tilde{L}{>}L_{\rm cut-off}$ in the gap equation $\Delta(L)$ is small and the approximation that $\bar{S}^{\rm exact}{=}0$ for $L{>}L_{\rm cut-off}$ remains valid in this regime [note that here, $\tilde{L}$ runs from $0$ to $L_{\rm cut-off}$]. For this reason, $\Delta(L)$ computed from $\bar{S}^{\rm trial}$ is reliable only in the regime where $L{\ll}L_{\rm cut-off}$. In this regime, the error in the $\Delta(L)$ mostly stems from the Monte Carlo error in $\bar{S}^{\rm trial}$ and thus the error in the estimated $\Delta(L)$ is very small, as shown in Fig.~\ref{fig: Coulomb_GMP_gap_exact_SF}. Another aspect that controls the error in the GMP gap is the form of the real space harmonics $v_{L}$, as evident from the gap equation [see Eq.~\eqref{eq: numerator_gap_equation}]. The decaying nature of $v_{L}{=}1/(2L{+}1)$ for the Coulomb interaction mitigates the errors at large $L$ allowing for a reliable computation of the Coulomb GMP gap. 

Unlike the Coulomb interaction, the harmonics $v_{L}$ of the model short-range interactions $V_{\mathfrak{m}}$ (for $\mathfrak{m}{>}0$) exhibit a polynomial growth. As a result, the GMP gap computed using the approximate $\bar{S}^{\rm trial}$ quickly deviates from its exact value as $L$ and $N$ increase. In Appendix~\ref{app: GMP_gap_short_range_interactions}, we provide results for model short-range interactions by proposing a potential solution to reliably estimate the GMP gap following the method presented in Ref.~\cite{Yutushui24}.

\begin{figure*}[tbh!]
\includegraphics[width=0.99\columnwidth]{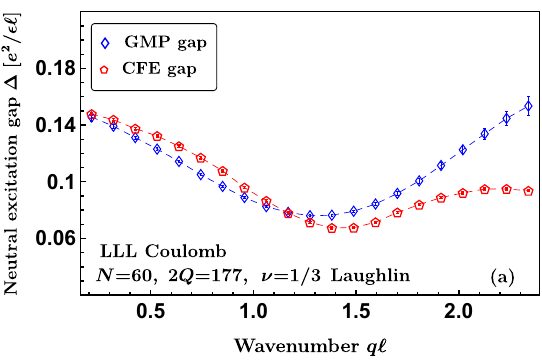}
\includegraphics[width=0.99\columnwidth]{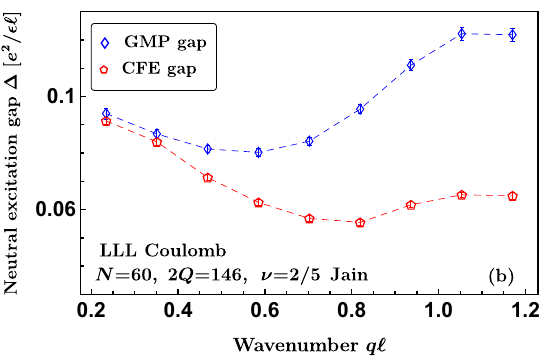} \\
\includegraphics[width=0.99\columnwidth]{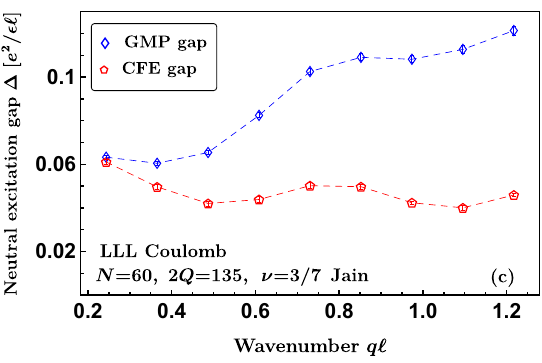}
\includegraphics[width=0.99\columnwidth]{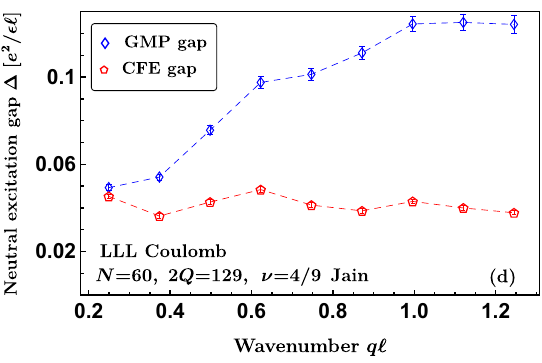}
 \caption{Comparison of the LLL Coulomb GMP and composite fermion exciton (CFE) gaps for states in the primary Jain sequence of $n/(2n{+}1)$. All the computations are done for $N{=}60$ electrons on the sphere.}
 \label{fig: GMP_CFE_gap_large_system_size}
\end{figure*}

\subsection{Comparison of GMP and CF exciton modes}
\label{ssec: CFE_GMP_gap}

\begin{figure}[tbh!]
\includegraphics[width=0.99\columnwidth]{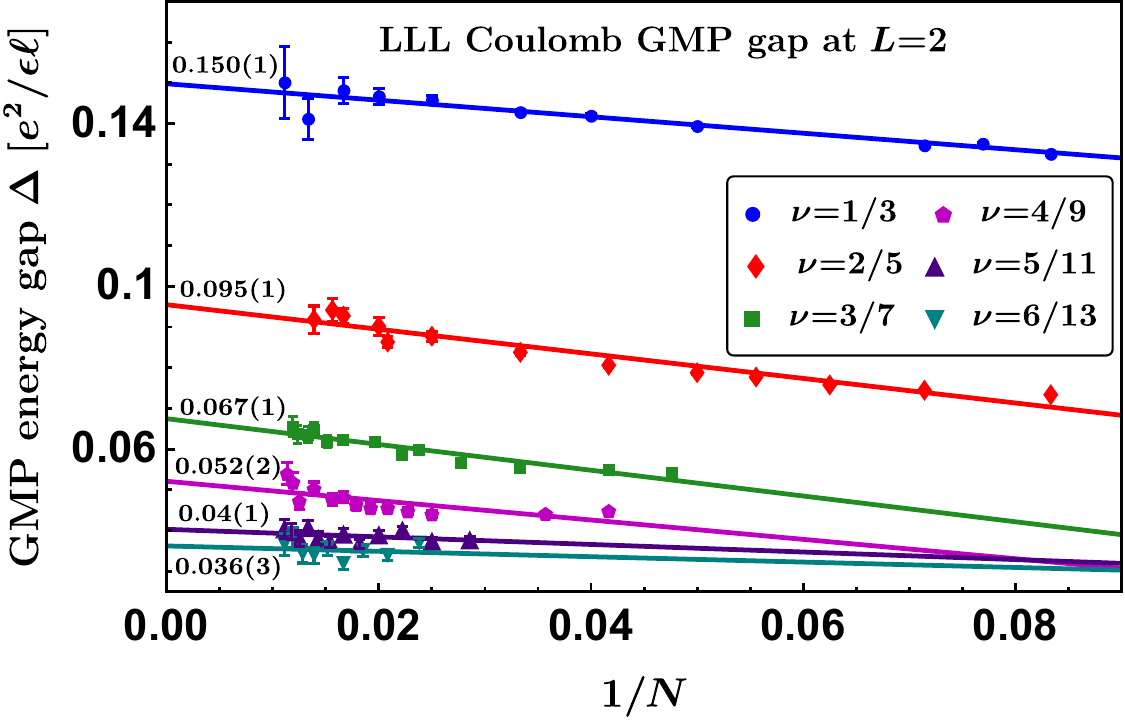}
 \caption{The long-wavelength ($L{=}2$) lowest Landau level Coulomb GMP gaps obtained from the gap equation in the spherical geometry and their extrapolation to the thermodynamic limit as a linear function of $1/N$, where $N$ is the number of electrons. The error bars are obtained from the error in the intercept and are comparable to the size of the symbols.
}
 \label{fig: thermodynamic_limit_GMP_CFE_gap_large_system_size}
\end{figure}

In this section, we compare the CFE and GMP gaps for the primary Jain states. We consider the Coulomb interaction and present our results for a comparatively larger system size than those presented in Sec.~\ref{ssec: Coulomb_GMP_gap_small_system_size}. The CFE gap is computed using Eq.~\eqref{eq: CFE_gap} while the GMP gap is computed using Eq.~\eqref{eq: gap_equation}. We compute the GMP gap using the approximate projected structure factor $\bar{S}^{\rm trial}(L)$ obtained following the procedure outlined in Sec.~\ref{ssec: structure_factor} since the calculation of the exact projected structure factor  $\bar{S}^{\rm exact}(L)$ for the Laughlin and Jain states is challenging as the system size increases [see Sec.~\ref{ssec: structure_factor}]. For the reasons outlined in Sec.~\ref{sssec: approximate_projected_structure_factor}, we compute the GMP gap only for the angular momenta that lie well below the cutoff angular momentum. We note that both the CFE and GMP modes start from $L{=}2$ as both the GMP and CFE states at $L{=}1$ vanish~\cite{Dev92, Simon94a}. To subsequently compare with the planar results, we convert the angular momentum $L$ to linear momentum $q$ using the relation $q\ell{=}L/\sqrt{Q}$ and present the gap as a function of $q\ell$. The gap at $L{=}2$ is referred to as the long-wavelength limit of the gap as it corresponds to the smallest $q\ell$ for a given system size. 

We find that the GMP gap of the $1/3$ Laughlin state in the long-wavelength limit, as shown in Fig.~\ref{fig: GMP_CFE_gap_large_system_size}$(a)$, agrees well with the corresponding CFE gap. This is consistent with the results of Ref.~\cite{Kamilla96b}, where the authors showed that the CFE and GMP excitations of the Laughlin states are identical in the long-wavelength limit. Recently, from studies on small systems, Refs.~\cite{Pu24, Balram24} found that the GMP wave function at $L{=}2$ and $L{=}3$ is exactly equal to the corresponding CFE wave function for Laughlin states. The slight deviation of the GMP gap from the CFE gap at $L{=}2$ and $L{=}3$ in our results is primarily due to the residual error in $\bar{S}^{\rm Laughlin}$. From Fig.~\ref{fig: GMP_CFE_gap_large_system_size}$(a)$, we also observe that the GMP and CFE gaps track each other up to the magnetoroton minimum, beyond which they differ significantly. This illustrates that the GMP mode for the $1/3$ Laughlin state remains a physically relevant neutral mode up to the magnetoroton minimum.

Next, we move to the Jain states at $\nu{=}2/5$,$~3/7,$ and $4/9$, results for which are presented in Figs.~\ref{fig: GMP_CFE_gap_large_system_size}$(b){-}(d)$. We find for these fractions too the long-wavelength limit of the GMP and CFE gaps lie very close. This suggests that similar to the results for the Laughlin state, the long-wavelength limit of the GMP and CFE states for these primary Jain states are also approximately identical, consistent with the recent findings of Ref.~\cite{Balram24}. In contrast to the Laughlin state, the GMP mode of these 2/5, 3/7, and 4/9 Jain states differs significantly from the CFE mode as the momentum increases. Thus, except the long-wavelength limit, the GMP wave function does not provide an accurate description of the lowest-lying neutral excitation in the non-Laughlin primary Jain states. 

Next, we present the thermodynamic limits of the long-wavelength GMP gap by extrapolating the $L{=}2$ GMP gap as a linear function of $1/N$ in Fig.~\ref{fig: thermodynamic_limit_GMP_CFE_gap_large_system_size}. To reduce finite-size effects, we scale the $L{=}2$ GMP gap by a ``density-correction" factor of $\sqrt{(2Q\nu)/N}$~\cite{Morf86} and then extrapolate the gap to $N{\to}\infty$. Moreover, to further mitigate the effects of finite-size, we only consider systems $N{>}10$ for $1/3$ Laughlin and $2/5$ Jain states, and $N{>}20$ for $3/7$ and $4/9$ Jain states [for the thermodynamic extrapolations presented here and in the Appendices]. The estimated thermodynamic $q{\to}0$ GMP gaps are very close to the $L{=}2$ CFE gaps reported in Ref.~\cite{Balram16d} further corroborating that the GMP and CFE descriptions are approximately identical in the long-wavelength limit. Although we do not have a formal proof, it appears that the GMP state $\bar{\rho}_{2,0}|\Psi_{\nu}\rangle$ approximates $\mathcal{P}_{\rm LLL} \Phi_{1}^{2}~\rho_{2,0}~\Phi_{n}$ and the unprojected density operator $\rho_{2,0}$ becomes approximately equal to $\rho^{\left(n-1\right)\to n}_{2,0}$, in the thermodynamic limit. For the Laughlin states, it has been proven that the analogous version of the above argument holds exactly~\cite{Kamilla96b}.

In Appendix~\ref{app: GMP_gap_bosonic_FQH_states}, we provide the Coulomb GMP dispersion for the bosonic FQH states, specifically, $\nu_{b}{=}1/2$ Laughlin state~\cite{Laughlin83} and $\nu_{b}{=}1$ Moore-Read Pfaffian state~\cite{Moore91}, for large systems. 

\section{Planar GMP gap equation and GMP density-mode dispersion}
\label{sec: planar_GMP_gap}

\begin{figure*}[tbh]
\includegraphics[width=0.99\columnwidth]{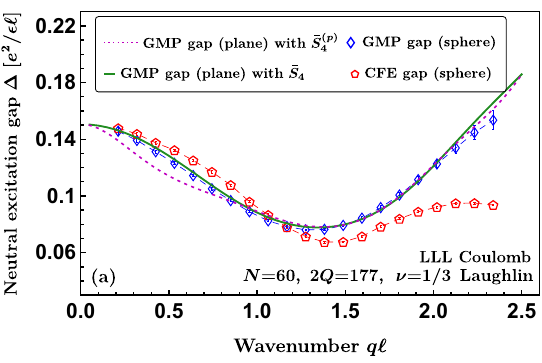}
\includegraphics[width=0.99\columnwidth]{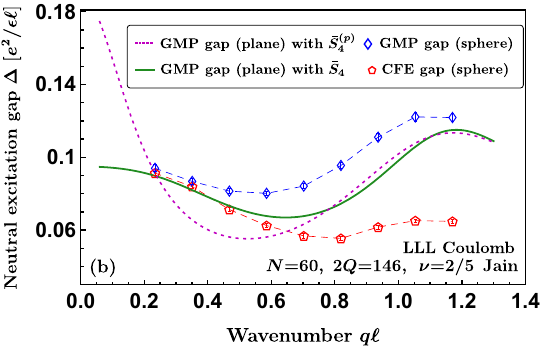}\\
\includegraphics[width=0.99\columnwidth]{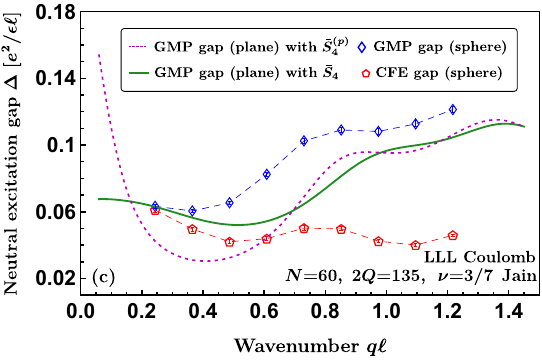}
\includegraphics[width=0.99\columnwidth]{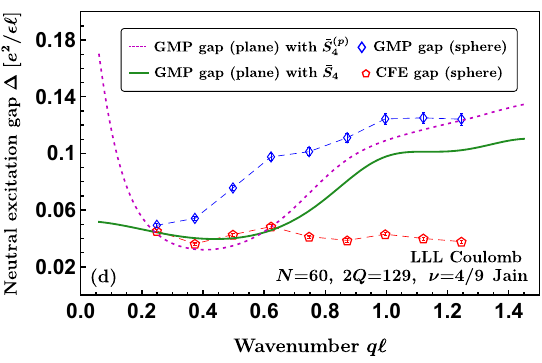}
 \caption{LLL Coulomb GMP gap computed in the planar geometry for primary Jain states using the coefficient $\bar{S}^{p}_{4}$ of Ref.~\cite{Park00a} [purple-dotted line] and the coefficient $\bar{S}_{4}$ considered here [green-solid line]. For comparison, the GMP and CFE gaps computed on the sphere [from Fig.~\ref{fig: GMP_CFE_gap_large_system_size}] are also shown.}
 \label{fig: plane_sphere_GMP_CFE_gap}
\end{figure*}

In Ref.~\cite{Park00a}, the dispersion of the GMP mode for many primary Jain states was computed in the planar geometry. In contrast to our results for the non-Laughlin primary Jain states [see Fig.~\ref{fig: GMP_CFE_gap_large_system_size}], the authors of Ref.~\cite{Park00a} saw a steep growth in the energy of the density mode in the long-wavelength limit (see Fig.~3 in Ref.~\cite{Park00a}) leading them to propose that the GMP mode is inaccurate as $q{\to}0$ for the non-Laughlin primary Jain states. Below, we outline the main procedure employed in their work and suggest suitable modifications to it to resolve the discrepancy between the results in Fig.~\ref{fig: GMP_CFE_gap_large_system_size} and Ref.~\cite{Park00a}. 

On the plane, the gap $\Delta^{(p)}(q)$ of the GMP mode carrying a linear momentum $\boldsymbol{q}$ is again given by an expression of the form $\Delta^{(p)}(q){=}\bar{F}(q)/\bar{S}(q)$. Here, $\bar{S}(q)$ is the projected static structure factor of the underlying FQH ground state $|\Psi_{\nu}\rangle$. In the plane, $\bar{S}(q)$ is related to the corresponding $S(q)$ as~\cite{Girvin86}
\begin{equation}
\label{eq: planar_unprojected_projected_relation}
    \bar{S}(q)= S(q)-(1-e^{-q^{2}/2}).
\end{equation}
$S(q)$ is defined as $S(q){=}N^{-1}\langle \Psi_{\nu}|\rho_q^{\dagger}\rho_q|\Psi_{\nu}\rangle$, where $\rho_q{=}\sum_{j}e^{-i\boldsymbol{q\cdot r_{j}}}$ is the Fourier component of the real space planar density operator. In this section, we have omitted the particle statistics indicator $\sigma$ from the symbols as we specifically consider fermions. We also set magnetic length $\ell{=}1$ in this section. The planar oscillator strength $\bar{F}(q)$ can be expressed in terms of $\bar{S}(q)$ as~\cite{Girvin86}
\begin{align}
\label{eq: planar_oscillator_strength}
\bar{F}(q)=& 2 \int \frac{d^2 \boldsymbol{\mathfrak{q}}}{(2 \pi)^2} v(\mathfrak{q}) \sin ^2\left(\frac{1}{2}|\mathbf{q} \times \boldsymbol{\mathfrak{q}}|\right)\notag\\
&~~~\times\left[e^{\boldsymbol{\mathfrak{q}} \cdot \mathbf{q}} \bar{S}(|\boldsymbol{\mathfrak{q}}+\mathbf{q}|)-e^{-\frac{1}{2} q^2} \bar{S}(\mathfrak{q})\right].
\end{align}
Here, $v(\mathfrak{q}){=}2\pi/\mathfrak{q}$ is the Coulomb interaction in the momentum-space. For comparison with the sphere, Eqs.~\eqref{eq: planar_unprojected_projected_relation} and~\eqref{eq: planar_oscillator_strength} are the planar analogs of Eqs.~\eqref{eq: relaton_unprojected_projected_structure_factor} and~\eqref{eq: numerator_gap_equation}, respectively.

As on the sphere, the GMP gap on the plane requires the knowledge of $\bar{S}(q)$. Following Refs.~\cite{Girvin86, Park00a}, we obtain $\bar{S}(q)$ from the Fourier transform of the pair-correlation function $g(r)$ as follows:
\begin{equation}
\label{eq: structure_factor_pair_correlation_relation}
S(q)=1 + \rho\int d^{2}\boldsymbol{r}~ e^{-i\boldsymbol{q\cdot r}}~[g(r)-1],
\end{equation}
where $\rho{=}\nu/(2\pi)$ is the uniform density. For FQH ground states a nice analytic expression of $g\left(r\right)$ is given by~\cite{Girvin84a}:
\begin{equation}
\label{eq: g(r)_expansion}
    g(r)=1-e^{-r^2/2}+e^{-r^2/4}\sum\limits_{j=1}^{\infty}{^{\prime}}~\frac{2c_{j}}{j!}\left(\frac{r^2}{4}\right)^{j},
\end{equation}
which allows to control the small wavevector i.e., $q{\to}0$ limit of $\bar{S}(q)$, by imposing constraints on the unknown expansion coefficients $c_{j}$. In Eq.~\eqref{eq: g(r)_expansion}, the prime indicates that the summation over $j$ is restricted to odd integers owing to the anti-symmetric nature of the wave function for a pair of electrons. Interestingly, the incompressibility of FQH states demands that in the power series expansion of $\bar{S}(q)$ in  $q{\to}0$ limit, the coefficient of both the constant term $q^0$ and the quadratic term $q^2$ must vanish (note that the rotational invariance of FQH states ensures that the expansion of $\bar{S}(q)$ contains only even powers of $q$). Moreover, recent topological quantum field theories have been used to shed some light on the other higher order terms in $\bar{S}(q)$ expansion~\cite{Can14, Gromov15, Nguyen17, Nguyen17b}. These terms are obtained by embedding the LLL-projected holomorphic FQH state on a curved geometry and computing the linear density response to the change in the underlying curvature. Following the geometric prescription, Ref.~\cite{Can14} obtained the expansion coefficients for Laughlin states up to $q^6$ in terms of their topological quantum numbers such as filling fraction, ``Wen-Zee shift"~\cite{Wen92}, and chiral central charge. Subsequently, Ref.~\cite{Nguyen17} extended the ideas of Ref.~\cite{Can14} to primary Jain states at $\nu{=}n/(2n{+}1)$, where $n$ is a positive integer and conjectured the following series expansion of $\bar{S}(q)$ in the $q{\to}0$ limit:
\begin{eqnarray}
\label{eq: expansion_structure_factor}
    \bar{S}(q)&=&\frac{n+1}{8}(q)^{4}+\frac{n^3+2n^2+2n-2}{48}(q)^6+\cdots \\
    &\equiv& ~~~~~\bar{S}_{4}(q)^{4}+~~~~~~~~~~~~~~~~~~~~~\bar{S}_{6}(q)^{6}+\cdots.\nonumber
\end{eqnarray}
This expansion is reasonably consistent with numerical computations of the structure factor~\cite{Balram17, Kumar24}.

In Ref.~\cite{Park00a}, the authors obtained $S(q)$ data from evaluations in the spherical geometry and used the same data on the plane, which is valid since the systems they considered were fairly large. However, the authors of Ref.~\cite{Park00a} consider a different coefficient for the $q^{4}$ term in the expansion of $\bar{S}(q)$ from the one given in Eq.~\eqref{eq: expansion_structure_factor}, which is $\bar{S}^{p}_{4}{=}(n{+}1)/8n$. This value of $\bar{S}^{p}_{4}$, which only depends on the filling factor $\nu$, was obtained by applying the plasma analogy~\cite{Laughlin83} to a wave function proposed by Lopez and Fradkin~\cite{Lopez92}, who argued that their state correctly captures the small-$q$ properties of any general incompressible state at $\nu{=}n/(2n{+}1)$. We find that the $q{\to}0$ gap of the GMP mode is extremely sensitive to $\bar{S}_{4}$. Therefore, the difference in $\bar{S}_{4}$ and $\bar{S}^{p}_{4}$ is at the heart of the origin of the discrepancy mentioned above in the $q{\to}0$ GMP gaps of the non-Laughlin Jain states ($n{>}1$). Using the expansion for $\bar{S}(q)$ given in Eq.~\eqref{eq: expansion_structure_factor}, we determine the GMP gap and find that it saturates in the long-wavelength limit consistent with numerical results on small systems~\cite{Platzman94, He94, Balram24} as well as the spherical GMP gaps presented in the previous section. Furthermore, we find that in the $q{\to}0$ regime, $\Delta^{(p)}(q)$ is only weakly dependent on the value of $\bar{S}_{6}$. Interestingly, the GMP mode for the Laughlin state $(n{=}1)$ shown in Ref.~\cite{Park00a} does saturate as $q{\to}0$. This stems from the fact that the $q{\to}0$ expansion for $\bar{S}(q)$ considered in Ref.~\cite{Park00a}, correctly describes the Laughlin state $(n{=}1)$, i.e., $\bar{S}^{p}_{4}(n{=}1){=}\bar{S}_{4}(n{=}1)$. The slight differences between our results and those in Ref.~\cite{Park00a} for the GMP gap of the 1/3 Laughlin state [see Fig.~\ref{fig: plane_sphere_GMP_CFE_gap}(a)], arise from the fact that we fit $\bar{S}(q)$ up to $q^{6}$ while in Ref.~\cite{Park00a} the fitting is done only up to $q^{4}$ (more on this below).

The expression of $g(r)$ in Eq.~\eqref{eq: g(r)_expansion} helps to obtain the desired $q{\to}0$ behavior of $\bar{S}(q)$, as given in Eq.~\eqref{eq: expansion_structure_factor}, provided the following four constraints on the $\{c_{j}\}$ are imposed for the $n/(2n{+}1)$ Jain states:
\begin{flalign}
\label{eq: c_m_constraints_Jain_states}
&\sum_{j=1}^{\infty}{'}~\frac{1}{n+1}c_{j}=\frac{-1}{4n},\\ \nonumber
&\sum_{j=1}^{\infty}{'}~\frac{(j+1)}{n+1}c_{j}=\frac{-1}{8n},\\ \nonumber
&\sum_{j=1}^{\infty}{'}~\frac{(j+1)(j+2)}{n+1}c_{j}=\frac{1}{4},\\ \nonumber
&\sum_{j=1}^{\infty}{'}~\frac{(j+1)(j+2)(j+3)}{1-n(n^2-1)(5+2n)}c_{j}=\frac{9}{48n}.
\end{flalign}
These constraints result from first plugging Eq.~\eqref{eq: g(r)_expansion} into Eq.~\eqref{eq: structure_factor_pair_correlation_relation} and then converting $S(q)$ to $\bar{S}(q)$ using Eq.~\eqref{eq: planar_unprojected_projected_relation}. We then perform the two-dimensional spatial integration in Eq.~\eqref{eq: structure_factor_pair_correlation_relation} and expand the resulting $\bar{S}(q)$ in powers of $q$ for $q{\ll}1$. Finally, we compare the coefficients of the first four relevant powers of $q$, i.e., $q^{0}$, $q^{2}$, $q^{4}$, and $q^{6}$ with those given in Eq.~\eqref{eq: expansion_structure_factor} to arrive at Eq.~\eqref{eq: c_m_constraints_Jain_states}.
 
To determine the unknown coefficients $c_{j}$, we perform a least square fitting~\cite{Fulsebakke23} of the numerically computed $g(r)$ data to its analytic expression given in Eq.~\eqref{eq: g(r)_expansion}, subject to the constraints of Eq.~\eqref{eq: c_m_constraints_Jain_states}. The $g(r)$ data is computed numerically on a sphere for $N{=}60$ electrons using the Monte Carlo method. Further, to reduce curvature effects, the distance between electrons is measured along the arc on the sphere. To capture the gradually increasing oscillation in the $g(r)$ data as the filling fraction approaches $1/2$, we incrementally increase the number of coefficients $c_{j}$ in the analytic expression of $g(r)$. Specifically, for the Laughlin and Jain states at $1/3$, $2/5$, $3/7$, and $4/9$ we use $10$, $15$, $20$, and $21$ coefficients, respectively, to fit Eq.~\eqref{eq: g(r)_expansion} [with the constraints in Eq.~\eqref{eq: c_m_constraints_Jain_states} enforced] to the numerical $g(r)$ data. After obtaining $\{c_{j}\}$, we get an analytic expression for $\bar{S}(q)$, which we then use in Eq.~\eqref{eq: planar_oscillator_strength} to compute the oscillator strength $\bar{F}(q)$ through numerical integration. This determines the planar GMP gap $\Delta^{(p)}(q)$. In Appendix~\ref{app: fit_gr_Sq}, we provide the $g(r)$ fitting results, which validates the $g(r)$ expansion given in Eq.~\eqref{eq: g(r)_expansion}. Additionally, we show that the above analytically computed $S(q)$ on the plane closely matches that numerically computed on the sphere for large system sizes. This suggests that $S(q)$ on the sphere, for sufficiently large system sizes, provides a reliable estimation of its thermodynamic value.

In Fig.~\ref{fig: plane_sphere_GMP_CFE_gap}, we present the planar GMP dispersion for many primary Jain states. As shown, the long-wavelength GMP mode obtained using the $\bar{S}(q)$ expansion of Eq.~\eqref{eq: expansion_structure_factor} (solid-green line) approaches a finite value for all the non-Laughlin primary Jain states, compared to those obtained from the $\bar{S}(q)$ expansion in Ref.~\cite{Park00a} (purple-dotted line). As a consistency check, we find that the thermodynamic limit of the extrapolated $L{=}2$ GMP gap computed on the sphere very closely matches the long-wavelength limit of the planar GMP gap. This agreement lends further credence to the long-wavelength limit expansion of the structure factor given in Refs.~\cite{Gromov15, Nguyen17, Nguyen17b} [see Eq.~\eqref{eq: expansion_structure_factor}]. At finite wave numbers, the GMP mode dispersion in the planar and spherical geometries [see Fig.~\ref{fig: plane_sphere_GMP_CFE_gap}] differs, which likely owes its origins to the curvature of the sphere. 

For comparison, in Fig.~\ref{fig: plane_sphere_GMP_CFE_gap} we have also plotted the GMP and CFE dispersions computed entirely on the sphere for $N{=}60$. For non-Laughlin Jain states, although the GMP dispersion on the sphere follows the shape of the planar GMP mode, the spherical and planar GMP gaps for $L{>}2$ do not tally. We expect these gaps to match if one compares the thermodynamic limits of the extrapolated spherical and planar GMP gap as done for the $L{=}2$ case. We do not explore this further here since we are mainly interested in the $q{\to}0$ gap as the GMP mode, which generically provides an accurate description of the true lowest-lying neutral excitation and is thus physically relevant, only in this long-wavelength limit~\cite{Platzman94, He94, Balram24}. 

\section{Discussion}
\label{sec: discussions}
In this work, we described the neutral excitations of the primary Jain states on the Haldane sphere. Following Girvin-MacDonald-Platzman (GMP), we constructed the neutral excitation using the single-mode approximation (SMA) on the spherical geometry. Similar to the planar geometry, we determined an analogous algebra of the LLL-projected density operators on the sphere, which is crucial for evaluating the GMP gap. The computation of the GMP gap further requires the static structure factor of the ground state as input, which we numerically obtained from Laughlin and Jain trial wave functions on the sphere. Previously, the GMP gap of the primary Jain states was computed in the planar geometry using the static structure factor derived from the pair-correlation function, for large systems on the sphere. In contrast, our computations are entirely carried out on the sphere.

This enabled us to compare the GMP mode with the composite fermion exciton (CFE) mode constructed on the sphere for the same system. While, compared to the GMP mode, the CFEs provide a more accurate representation of the lowest-lying neutral excitation at all wavevectors in the primary Jain states, the two descriptions become approximately equivalent in the long-wavelength limit. Specifically, we find that the GMP and CFE energies lie close to each other as the wavenumber approaches zero. Our results for large systems (in the thermodynamic extrapolation of $q{\rightarrow} 0$ GMP gap for primary Jain states, we have used systems of up to $N_{\rm max}{\sim}72~\text{to}~90$ electrons) are consistent with previous results for the Laughlin states~\cite{Girvin85, Park00a, Scarola00} and small system-size (up to $N_{\rm max}{\sim}14~\text{to}~18$ electrons) results of Ref.~\cite{Balram24}. Since the long-wavelength limit of the GMP mode corresponds to the fluctuations of the underlying geometric degrees of freedom in the FQH state, our findings suggest that the geometrical description of the primary Jain states can be understood using the CF theory. In contrast to primary Jain states, where our results suggest that the SMA remains valid in the small-wavenumber limit, Refs.~\cite{Balram21d, Nguyen22, Balram24} showed that for secondary Jain states, SMA fails to provide a valid description even in the long-wavelength limit. Generically, in the secondary Jain states, there are two distinct gravitons instead of the single GMP graviton- CFE graviton and parton graviton~\cite{Balram21d, Balram24}.

Furthermore, we resolve a long-standing issue of the sharp rise of the planar GMP mode as the wavevector approaches zero for the non-Laughlin primary Jain states as reported in Refs.~\cite{Park00a, Scarola00} which results in a significant mismatch with the corresponding CFE gap. The resolution is rooted in using an accurate long-wavelength expansion of the structure factor for the Jain states. 

In our GMP gap computation on a sphere, the only source of error stems from Monte Carlo statistical uncertainty in the static structure factor data of trial wave functions. By improving its accuracy, we can reliably compute the GMP gap for a wide range of angular momenta. For the Coulomb interaction, owing to its decaying harmonics $v_{L}$ with $L$, we can compute very accurately the GMP gap in the long-wavelength limit  (i.e., $L{=}2$), as well as the GMP gap at other higher $L$ lying well below a suitably defined $L_{\rm cut{-}off}$. Our approach to computing the GMP gap equally applies to other rotationally invariant interactions, including those parameterized by a set of Haldane pseudopotentials. The LLL GMP gap computation presented here can also be generalized to higher isolated LLs. However, suppose the harmonics $v_{L}$ corresponding to a LL-projected interaction grow rapidly with $L$. In that case, the interaction needs to be appropriately modified as discussed in Ref.~\cite{Yutushui24} and Appendix~\ref{app: GMP_gap_short_range_interactions}, to reduce the amplification of the statistical error in the structure factor data. To this end, it is worth exploring interactions such as the realistic Coulomb in the second LL of GaAs [see Appendix~\ref{app: GMP_gap_fermionic_MR_Pf_SLL}], LLs of graphene, and its multilayer incarnations, as well as, generalizations to model short-range three- and higher-body interactions.

The method outlined here for computing the GMP gap can also be applied to other fermionic and bosonic single-component FQH states, including, the Moore-Read Pfaffian~\cite{Moore91} (see Appendices~\ref{app: GMP_gap_fermionic_MR_Pf_SLL} and \ref{app: GMP_gap_bosonic_FQH_states}), as well as a broad class of parton states~\cite{Jain89b, Wu17, Balram18, Balram18a, Balram19, Faugno19, Balram21c, Dora22, Bose23, Balram24a, Bose25}, which have recently been shown to capture many experimentally observed FQH states. It would be useful to generalize the current method to multicomponent FQH states, where the different components could represent spin, valley, orbital, layer, LL, etc. degrees of freedom. For FQH states with multiple LL degrees of freedom, which can occur in multi-layer graphene, our work naturally motivates the search for an analog of the ``GMP-like" algebra presented here for the LLL but extended to encompass multiple LLs. 

The structure factor and its constraints play a central role in the recent bootstrap approach to the quantum Hall effect~\cite{Gao24}. Presumably, the structure factor that we and others~\cite{Kamilla97, Balram17, Kumar24, Herviou24} have computed from accurate trial wave functions for large systems could aid in further narrowing the constraints, thereby tightening the bounds obtained from the bootstrap approach. We leave an exploration of these and other directions to future work.

\begin{acknowledgments}
We acknowledge useful discussions with Nicolas Regnault. The work was made possible by financial support from the Science and Engineering Research Board (SERB) of the Department of Science and Technology (DST) via the Mathematical Research Impact Centric Support (MATRICS) Grant No. MTR/2023/000002. Computational portions of this research work were conducted using the Nandadevi and Kamet supercomputers maintained and supported by the Institute of Mathematical Science's High-Performance Computing Center. Some numerical calculations were performed using the DiagHam package~\cite{diagham}, for which we are grateful to its authors.
\end{acknowledgments}

\appendix
\begin{widetext}

\section{Relation between the projected and unprojected static structure factors}
\label{app: projected_unprojected_Sq}
In this appendix, we derive the relationship between the projected and unprojected static structure factors stated in Eq.~\eqref{eq: relaton_unprojected_projected_structure_factor}. This is achieved by observing that the expectation value of a normal-ordered operator in a LLL projected state remains the same regardless of whether the operator is itself projected onto the LLL or not. The normal-ordered projected and unprojected density operators are related to the corresponding bare density operators as
\begin{align}
\label{eq: relation_projected_normal_ordered_bare_density_operator}
    \colon\bar{\rho}^{~\sigma}(\boldsymbol{\Omega})~\bar{\rho}^{~\sigma}(\boldsymbol{\Omega^{'}})\colon&=\bar{\rho}^{~\sigma}(\boldsymbol{\Omega})~\bar{\rho}^{~\sigma}(\boldsymbol{\Omega^{'}}) - \big[\bar{\Psi}^{\sigma}(\boldsymbol{\Omega}),\left[\bar{\Psi}^{\sigma}(\boldsymbol{\Omega^{\prime}})\right]^{\dagger}\big]_{\pm}~\left[\bar{\Psi}^{\sigma}(\boldsymbol{\Omega})\right]^{\dagger}\bar{\Psi}^{\sigma}(\boldsymbol{\Omega^{\prime}}), ~~\text{and}\\
    \label{eq: relation_unprojected_normal_ordered_bare_density_operator}
    \colon\rho^{\sigma}(\boldsymbol{\Omega})~\rho^{\sigma}(\boldsymbol{\Omega^{'}})\colon&=\rho^{\sigma}(\boldsymbol{\Omega})~\rho^{\sigma}(\boldsymbol{\Omega^{'}}) - \big[\Psi^{\sigma}(\boldsymbol{\Omega}),\left[\Psi^{\sigma}(\boldsymbol{\Omega^{\prime}})\right]^{\dagger}\big]_{\pm}~\left[\Psi^{\sigma}(\boldsymbol{\Omega})\right]^{\dagger}\Psi^{\sigma}(\boldsymbol{\Omega^{\prime}}).
\end{align}
Here, $\colon\bar{\rho}^{~\sigma}(\boldsymbol{\Omega})~\bar{\rho}^{~\sigma}(\boldsymbol{\Omega^{'}})\colon{=}\left[\bar{\Psi}^{\sigma}(\boldsymbol{\Omega})\right]^{\dagger}\left[\bar{\Psi}^{\sigma}(\boldsymbol{\Omega^{\prime}})\right]^{\dagger}\bar{\Psi}^{\sigma}(\boldsymbol{\Omega^{\prime}})\bar{\Psi}^{\sigma}(\boldsymbol{\Omega})$ is the normal-ordered product of projected density operators, while $\colon\rho^{\sigma}(\boldsymbol{\Omega})~\rho^{\sigma}(\boldsymbol{\Omega^{'}})\colon{=}\left[\Psi^{\sigma}(\boldsymbol{\Omega})\right]^{\dagger}\left[\Psi^{\sigma}(\boldsymbol{\Omega^{\prime}})\right]^{\dagger}\Psi^{\sigma}(\boldsymbol{\Omega^{\prime}})\Psi^{\sigma}(\boldsymbol{\Omega})$ is the normal-ordered product of unprojected density operators. In the above equations, the sign $+(-)$ in $[\cdot\cdot]_{\pm}$ stands for anticommutator (commutator) when $\sigma{=}f$ ($\sigma{=}b$) represents fermions (bosons). The real space creation ($[\Psi^{\sigma}]^{\dagger}$) and annihilation ($\Psi^{\sigma}$) operators [see Eq.~\eqref{eq: creation_operator_landau_level_basis}] satisfy the usual commutation or anti-commutation algebra,
\begin{align}
\label{eq: unprojected_anticommutation_commutation_relation}
 \left[\Psi^{\sigma}(\boldsymbol{\Omega}),\left[\Psi^{\sigma}(\boldsymbol{\Omega^{\prime}})\right]^{\dagger}\right]_{\pm}&=\delta\left(\boldsymbol{\Omega}-\boldsymbol{\Omega^{\prime}}\right).
\end{align}
Upon projection to the LLL, the algebra of the corresponding projected real space creation ($[\bar{\Psi}^{\sigma}]^{\dagger}$) and annihilation ($\bar{\Psi}^{\sigma}$) operators is modified. The LLL projected real space creation operator $[\bar{\Psi}^{\sigma}]^{\dagger}$ is obtained by restricting the summation in Eq.~\eqref{eq: creation_operator_landau_level_basis} to the LLL, i.e., $l{=}Q$, and is given by
\begin{align}
\label{eq: projected_creation_operator_in_landau_level_basis}
 \left[\bar{\Psi}^{\sigma}(\boldsymbol{\Omega})\right]^{\dagger}=\sum_{m}\left[Y^{Q}_{Q,m}(\boldsymbol{\Omega})\right]^{*}~\left[\chi^{\sigma}_{m}\right]^{\dagger}.   
\end{align}
Here, $m$ is the azimuthal quantum number corresponding to the LL index $l$ and for the LLL, it ranges from $-Q$ to $Q$. To be consistent with the notation used in the main text, we have omitted the LLL index from $\left[\chi^{\sigma}_{m}\right]^{\dagger}$. We use Eq.~\eqref{eq: projected_creation_operator_in_landau_level_basis} to evaluate the commutator/anticommutator of real space projected density operators to arrive at
\begin{align}
\label{eq: projected_anticommutation_commutation_relation}
 \left[\bar{\Psi}^{\sigma}(\boldsymbol{\Omega}),\left[\bar{\Psi}^{\sigma}(\boldsymbol{\Omega^{\prime}})\right]^{\dagger}\right]_{\pm}&=\sum_{m} \left[Y^{Q}_{Q,m}(\boldsymbol{\Omega^{\prime}})\right]^{*}~Y^{Q}_{Q,m}(\boldsymbol{\Omega}),
\end{align}
where we have used the usual algebra of the LL creation and annihilation operators, i.e., $\big[\chi^{\sigma}_{m},[\chi^{\sigma}_{m^{\prime}}]^{\dagger}\big]_{\pm}{=}\delta_{m,m^{\prime}}$.

For later convenience, we express the second term on the right-hand side of both Eqs.~\eqref{eq: relation_projected_normal_ordered_bare_density_operator} and~\eqref{eq: relation_unprojected_normal_ordered_bare_density_operator} in terms of the LL creation and annihilation operators as
\begin{align}
\label{eq: offset_unprojected_normal_ordered_operator}
 \left[\Psi^{\sigma}(\boldsymbol{\Omega}),\left[\Psi^{\sigma}(\boldsymbol{\Omega^{\prime}})\right]^{\dagger}\right]_{\pm}~\left[\Psi^{\sigma}(\boldsymbol{\Omega})\right]^{\dagger}\Psi^{\sigma}(\boldsymbol{\Omega^{\prime}})&=\delta(\boldsymbol{\Omega}-\boldsymbol{\Omega^{\prime}})~\sum_{\substack{l_1, m_1,\\l_2, m_2}}\left[Y^{Q}_{l_1,m_1}(\boldsymbol{\Omega})\right]^{*}~Y^{Q}_{l_2,m_2}(\boldsymbol{\Omega^{\prime}})~\left[\chi^{\sigma}_{l_1,m_1}\right]^{\dagger}\chi^{\sigma}_{l_2,m_2},\\
 \vspace{0.7cm}
\label{eq: offset_projected_normal_ordered_operator}
 \left[\bar{\Psi}^{\sigma}(\boldsymbol{\Omega}),\left[\bar{\Psi}^{\sigma}(\boldsymbol{\Omega^{\prime}})\right]^{\dagger}\right]_{\pm}~\left[\bar{\Psi}^{\sigma}(\boldsymbol{\Omega})\right]^{\dagger}\bar{\Psi}^{\sigma}(\boldsymbol{\Omega^{\prime}})&=\sum_{m_1,m_2,m_3} \left[Y^{Q}_{Q,m_1}(\boldsymbol{\Omega^{\prime}})\right]^{*}~Y^{Q}_{Q,m_3}(\boldsymbol{\Omega^{\prime}})~\left[Y^{Q}_{Q,m_2}(\boldsymbol{\Omega})\right]^{*}~Y^{Q}_{Q,m_1}(\boldsymbol{\Omega})~\left[\chi^{\sigma}_{m_2}\right]^{\dagger}\chi^{\sigma}_{m_3},
\end{align}
where to obtain Eq.~\eqref{eq: offset_unprojected_normal_ordered_operator}, we have used Eqs.~\eqref{eq: creation_operator_landau_level_basis} and~\eqref{eq: unprojected_anticommutation_commutation_relation}. Similarly, in Eq.~\eqref{eq: offset_projected_normal_ordered_operator}, we have used Eqs.~\eqref{eq: projected_creation_operator_in_landau_level_basis} and~\eqref{eq: projected_anticommutation_commutation_relation}. In the above equations, $m_1$, $m_2$, and $m_3$ represent the azimuthal quantum number dummy indices, which range from $-Q$ to $Q$. In the angular momentum-space, the above equations can be written as 
\begin{align}
\label{eq: offset_LL_basis_1}
\mathcal{O}^{\sigma}_{L,M}&=\int{d\boldsymbol{\Omega}~d\boldsymbol{\Omega^{'}}}~Y_{L,M}(\boldsymbol{\Omega})~Y^{*}_{L,M}(\boldsymbol{\Omega^{\prime}})~\left[\Psi^{\sigma}(\boldsymbol{\Omega}),\left[\Psi^{\sigma}(\boldsymbol{\Omega^{\prime}})\right]^{\dagger}\right]_{\pm}~\left[\Psi^{\sigma}(\boldsymbol{\Omega})\right]^{\dagger}\Psi^{\sigma}(\boldsymbol{\Omega^{\prime}})\notag\\
&=\sum_{\substack{l_1, m_1,\\l_2, m_2}}\left[\int{d\boldsymbol{\Omega}}~A^{l_2,m_2}_{l_1,m_1}(\boldsymbol{\Omega})~\left|Y_{L,M}\left(\boldsymbol{\Omega}\right)\right|^{2}\right]~\left[\chi^{\sigma}_{l_1,m_1}\right]^{\dagger}\chi^{\sigma}_{l_2,m_2},
\end{align}
where $A^{l_2,m_2}_{l_1,m_1}(\boldsymbol{\Omega}){=}\left[Y^{Q}_{l_1,m_1}(\boldsymbol{\Omega})\right]^{*}~Y^{Q}_{l_2,m_2}(\boldsymbol{\Omega})$. Similarly, 
\begin{align}
    \label{eq: offset_LL_basis_2}\bar{\mathcal{O}}^{\sigma}_{L,M}&=\int{d\boldsymbol{\Omega}~d\boldsymbol{\Omega^{'}}}~Y_{L,M}(\boldsymbol{\Omega})~Y^{*}_{L,M}(\boldsymbol{\Omega^{\prime}})~\left[\bar{\Psi}^{\sigma}(\boldsymbol{\Omega}),\left[\bar{\Psi}^{\sigma}(\boldsymbol{\Omega^{\prime}})\right]^{\dagger}\right]_{\pm}~\left[\bar{\Psi}^{\sigma}(\boldsymbol{\Omega})\right]^{\dagger}\bar{\Psi}^{\sigma}(\boldsymbol{\Omega^{\prime}})\notag\\
    &=\sum_{m_1,m_2,m_3}B_{L,M}^{m_1,m_2}~C_{L,M}^{m_1,m_3}~\left[\chi^{\sigma}_{m_2}\right]^{\dagger}\chi^{\sigma}_{m_3},
\end{align}
where
\begin{align}
\label{eq: coefficients_one_body_HAmiltonian_1}
    B_{L,M}^{m_1,m_2}&=\int d\boldsymbol{\Omega}~Y_{L,M}(\boldsymbol{\Omega})~Y^{Q}_{Q,m_1}(\boldsymbol{\Omega})~\left[Y^{Q}_{Q,m_2}(\boldsymbol{\Omega})\right]^{*},~~\text{and}\\
    \label{eq: coefficients_one_body_HAmiltonian_2}
    C_{L,M}^{m_1,m_3}&=\int d\boldsymbol{\Omega^{\prime}}~~ \big[Y_{L,M}(\boldsymbol{\Omega^{\prime}})\big]^{*}~\left[Y^{Q}_{Q,m_1}(\boldsymbol{\Omega^{\prime}})\right]^{*}~Y^{Q}_{Q,m_3}(\boldsymbol{\Omega^{\prime}}).
\end{align}
We now compute the expectation value of the normal-ordered operators given in Eqs.~\eqref{eq: relation_projected_normal_ordered_bare_density_operator} and~\eqref{eq: relation_unprojected_normal_ordered_bare_density_operator} in a LLL projected FQH ground state $|\Psi_{\nu}\rangle$. As stated in the beginning of this section, the two expectation values are equal to each other, i.e., $\langle\Psi_{\nu}|\colon\bar{\rho}^{~\sigma}(\boldsymbol{\Omega})~\bar{\rho}^{~\sigma}(\boldsymbol{\Omega^{'}})\colon|\Psi_{\nu}\rangle{=}\langle\Psi_{\nu}|\colon\rho^{\sigma}(\boldsymbol{\Omega})~\rho^{\sigma}(\boldsymbol{\Omega^{'}})\colon|\Psi_{\nu}\rangle$ and yield the following relation:
\begin{align}
\label{eq: relation_projected_unprojected_structure_factor_0}
\left\langle\Psi_{\nu}\big|\left[\bar{\rho}^{~\sigma}_{L,M}\right]^{\dagger}\bar{\rho}^{~\sigma}_{L,M}\big|\Psi_{\nu}\right\rangle -\left\langle\Psi_{\nu}\big| \bar{\mathcal{O}}^{\sigma}_{L,M}\big|\Psi_{\nu}\right\rangle=\left\langle\Psi_{\nu}\big|\left[\rho^{~\sigma}_{L,M}\right]^{\dagger}\rho^{~\sigma}_{L,M}\big|\Psi_{\nu}\right\rangle -\left\langle\Psi_{\nu}\big| \mathcal{O}^{\sigma}_{L,M}\big|\Psi_{\nu}\right\rangle.
\end{align}
Here, $\left[\bar{\rho}^{~\sigma}_{L,M}\right]^{\dagger}{=}
\int d\boldsymbol{\Omega}~\left[Y_{L,M}(\boldsymbol{\Omega})\right]^{*}~\bar{\rho}^{~\sigma}(\boldsymbol{\Omega}),~$ and a similar expression holds for $\left[\rho^{~\sigma}_{L,M}\right]^{\dagger}$ with the projected density operator $\bar{\rho}^{~\sigma}(\boldsymbol{\Omega})$ replaced by the corresponding unprojected density operator $\rho^{\sigma}(\boldsymbol{\Omega})$. The expectation value of the operators $\mathcal{O}^{\sigma}_{L,M}$ and $\mathcal{\bar{O}}^{\sigma}_{L,M}$ are
\begin{align}
    \label{eq: expectation_offset__unprojected}
    \left\langle\Psi_{\nu}\big| \mathcal{O}^{\sigma}_{L,M}\big|\Psi_{\nu}\right\rangle &= \left\langle \mathcal{O}^{\sigma}_{L,M} \right\rangle_{\Psi_{\nu}}= \frac{N}{4\pi},~~\text{and}\\
\label{eq: expectation_offset__projected}
    \left\langle\Psi_{\nu}\big| \mathcal{\bar{O}}^{\sigma}_{L,M}\big|\Psi_{\nu}\right\rangle&= \left\langle \mathcal{\bar{O}}^{\sigma}_{L,M} \right\rangle_{\Psi_{\nu}}=\frac{N (2Q+1)}{4\pi}~\left(\begin{array}{ccc}
Q & Q & L \\
-Q & Q & 0
\end{array}\right)^2.
\end{align}
In the next two subsections, we present a derivation of the above equations. Using the definition of structure factor [see Eqs.~\eqref{eq: structure_factor} and~\eqref{eq: projected_structure_factor}] and the above Eqs.~\eqref{eq: expectation_offset__unprojected} and~\eqref{eq: expectation_offset__projected} into Eq.~\eqref{eq: relation_projected_unprojected_structure_factor_0}, we obtain the desired relation between the projected and unprojected structure factor [presented in Eq.~\eqref{eq: relaton_unprojected_projected_structure_factor} of the main text], which is 
\begin{align}
S^{\sigma}(L) &= \bar{S}^{\sigma}(L)+1 - (2Q+1)~\left(\begin{array}{ccc}
Q & Q & L \\
-Q & Q & 0
\end{array}\right)^2 . 
\end{align}

\subsection{Evaluation of $~\left\langle \mathcal{O}^{\sigma}_{L,M} \right\rangle_{\Psi_{\nu}}$}
\label{ssec: expectation_unprojected}
The expectation value of $\mathcal{O}^{\sigma}_{L, M}$ [see Eq.~\eqref{eq: offset_LL_basis_1}] in the LLL projected FQH ground state $|\Psi_{\nu}\rangle$ is determined from the expectation value of the constituent LL operators $\left[\chi^{\sigma}_{l_1,m_1}\right]^{\dagger}\chi^{\sigma}_{l_2,m_2}$, which is given by $\left\langle\left[\chi^{\sigma}_{l_1,m_1}\right]^{\dagger}\chi^{\sigma}_{l_2,m_2}\right\rangle_{\Psi_{\nu}}{=}\bar{\nu}\delta_{l_1, Q}~\delta_{l_2, Q}~\delta_{m_1,m_2}$. This follows from the fact that the state $|\Psi_{\nu}\rangle$ resides entirely in the LLL, and the action of the LL operators having indices different from the LLL index $l{=}Q$ annihilate the state. Furthermore, for $m_{1}{\neq} m_{2}$, the action of the operator $\left[\chi^{\sigma}_{Q,m_1}\right]^{\dagger}\chi^{\sigma}_{Q,m_2}$ on the uniform state $|\Psi_{\nu}\rangle$ yields a nonuniform state, resulting in a zero overlap with the corresponding uniform bra state $\langle\Psi_{\nu}|$. When $m_{1}{=}m_{2}{=}m$, the above operator becomes the number operator $\left[\chi^{\sigma}_{Q,m}\right]^{\dagger}\chi^{\sigma}_{Q,m}$ and yields the average occupancy $\bar{\nu}$ in the single-particle LL state $|Q,m\rangle$. Since the many-body state $|\Psi_{\nu}\rangle$ is uniform, $\bar{\nu}$ is the same for all the single-particle states and equals $N/(2Q{+}1)$ as there are $N$ particles distributed in a total of $2Q{+}1$ single-particle states in the LLL. In the thermodynamic limit, $\bar{\nu}$ converges to the filling $\nu$ of $|\Psi_{\nu}\rangle$. With this understanding, $\left\langle \mathcal{O}^{\sigma}_{L,M} \right\rangle_{\Psi_{\nu}}$ [see Eq.~\eqref{eq: offset_LL_basis_1}] simplifies to
\begin{align}
\label{eq: expectation_value_offset_1}
     \left\langle \mathcal{O}^{\sigma}_{L,M} \right\rangle_{\Psi_{\nu}} &=\frac{N}{2Q+1}\sum_{\substack{m}}\left[\int{d\boldsymbol{\Omega}}~A^{Q,m}_{Q,m}(\boldsymbol{\Omega})~\left|Y_{L,M}\left(\boldsymbol{\Omega}\right)\right|^{2}\right].
\end{align}
To further simplify the above equation, we note the identity $\sum_{\substack{m}}A^{Q,m}_{Q,m}(\boldsymbol{\Omega}){=}(2Q{+}1)/(4\pi)$. This can be seen from the explicit expression of the monopole spherical harmonics~\cite{Jain_2007},
\begin{equation}
\label{eq: monopole_spherical_harmonics_spinor}
    Y^{Q}_{Q,m}(\theta,\phi)=\phi_{k}(u,v)=\sqrt{\frac{2Q+1}{4\pi}\binom{2Q}{k}}(-1)^k v^k u^{2Q-k},
\end{equation}
where the spinor coordinates $(u,v){=}\left(\cos{(\theta/2)}~e^{i\phi/2},\sin{(\theta/2)}~e^{-i\phi/2}\right)$~\cite{Haldane83}. Here, the index $k{=}Q{-}m$ and runs from $0$ to $2Q$. Using Eq.~\eqref{eq: monopole_spherical_harmonics_spinor} into the expression of $A^{Q,m}_{Q,m}$ [see just below Eq.~\eqref{eq: offset_LL_basis_1}] and using the binomial theorem, we obtain
\begin{align}
 \label{eq: identity_spherical_harmonica}
     \sum_{\substack{m}}A^{Q,m}_{Q,m}(\boldsymbol{\Omega})&=\frac{2Q+1}{4\pi}\sum_{k{=}0}^{2Q} \binom{2Q}{k}\left(|v|^2\right)^{k}\left(|u|^2\right)^{2Q-k}=\frac{2Q+1}{4\pi}\left(|u|^2+|v|^2\right)^{2Q}=\frac{2Q+1}{4\pi},
\end{align}
reproducing the above identity. Finally, we substitute Eq.~\eqref{eq: identity_spherical_harmonica} into Eq.~\eqref{eq: expectation_value_offset_1} and use the fact that the spherical harmonic $Y_{L, M}\left(\boldsymbol{\Omega}\right)$ is normalized to unity. This yields $\left\langle \mathcal{O}^{\sigma}_{L,M} \right\rangle_{\Psi_{\nu}}$ as given in Eq.~\eqref{eq: expectation_offset__unprojected}. 

\subsection{Evaluation of $~\left\langle \mathcal{\bar{O}}^{\sigma}_{L,M} \right\rangle_{\Psi_{\nu}}$}
\label{ssec: expectation_projected}
To evaluate $\left\langle \mathcal{\bar{O}}^{\sigma}_{L,M} \right\rangle_{\Psi_{\nu}}$ [see Eq.~\eqref{eq: offset_LL_basis_2}], we use $\left\langle\left[\chi^{\sigma}_{m_2}\right]^{\dagger}\chi^{\sigma}_{m_3}\right\rangle_{\Psi_{\nu}}{=}\bar{\nu}\delta_{m_2,m_3}$ [see Appendix~\ref{ssec: expectation_unprojected}] to obtain
\begin{align}
\label{eq: expectation_value_offset_2}
\left\langle \mathcal{\bar{O}}^{\sigma}_{L,M} \right\rangle_{\Psi_{\nu}}=\frac{N}{2Q+1}\sum_{m_1,m_2}B_{L,M}^{m_1,m_2}~C_{L,M}^{m_1,m_2}.
\end{align}
The integrals in $B_{L,M}^{m_1,m_2}$ [see Eq.~\eqref{eq: coefficients_one_body_HAmiltonian_1}] and $C_{L,M}^{m_1,m_2}$ [see Eq.~\eqref{eq: coefficients_one_body_HAmiltonian_2}] are evaluated using Eq.~\eqref{eq: second_quantized_angular_momentum_density_value} from the main text by noting that $\left[Y^{Q}_{Q,m}(\boldsymbol{\Omega})\right]^{*}{=}(-1)^{Q+m}~Y^{-Q}_{Q,-m}(\boldsymbol{\Omega})$ and that $Y_{L,M}(\boldsymbol{\Omega}){=}Y^{Q{=}0}_{L,M}(\boldsymbol{\Omega})$. We simplify the resulting expression using the following properties of the Wigner $3j$ symbol. First, extracting a negative sign from all the azimuthal quantum numbers results in multiplying the  Wigner $3j$ symbol by a factor of $(-1)$ raised to the power of the sum of the angular momenta. Second, interchanging an odd number of columns similarly multiplies the Wigner $3j$ symbol by a factor of $(-1)$ raised to the power of the sum of the angular momenta. Consequently, Eq.~\eqref{eq: expectation_value_offset_2} simplifies to
\begin{align}
    \label{eq: simplified_expectation_value_offset_2}
  \left\langle \mathcal{\bar{O}}^{\sigma}_{L,M} \right\rangle_{\Psi_{\nu}}=\frac{N(2Q+1)(2L+1)}{4\pi}\left(\begin{array}{lll}
Q & Q & L \\
-Q & Q & 0
\end{array}\right)^{2}\sum_{m_1,m_2}\left(\begin{array}{ccc}
Q & Q & L \\
m_1 & -m_2 & M
\end{array}\right)^2.
\end{align}
In obtaining the above equation, we have substituted $(-1)^{2Q{+}M{+}m_{1}{+}m_2}{=}1$. This follows from the requirement that $m_2{=}m_1{+}M$ for the Wigner $3j$ symbol to be nonzero, and since $Q{+}m_1$ is always an integer, the resulting exponent $2(Q{+}m_{1}){+}2M$ is even. [Note that $M$ takes only integral values (see Sec.~\ref{ssec: density_operators}).] Interestingly, the summation in Eq.~\eqref{eq: simplified_expectation_value_offset_2} evaluates to $1/(2L{+}1)$, which is also one of the special properties of the Wigner $3j$ symbol. As a result, we obtain the  simplified expression for $\left\langle \mathcal{\bar{O}}^{\sigma}_{L,M} \right\rangle_{\Psi_{\nu}}$ as presented in Eq.~\eqref{eq: expectation_offset__projected}.

\section{Derivation of the single-body term that cancels the
self-interaction in Hamiltonian [see Eq.~\eqref{eq: not_normal_ordered_interaction}] and ground state energy}
\label{app: Derivation_not_normal_ordered_interaction_Hamiltonia}
In this appendix, we rewrite Eq.~\eqref{eq: interaction_Hamiltonian}, which represents the normal-ordered interaction Hamiltonian, in terms of the corresponding bare density operators to derive Eq.~\eqref{eq: not_normal_ordered_interaction}. Subsequently, we express the energy of the resulting Hamiltonian in terms of the projected structure factor of a ground state $|\Psi_{\nu}\rangle$. 
For convenience, we reproduce Eq.~\eqref{eq: interaction_Hamiltonian} below:
\begin{equation}
  \label{eq: reiterate_interaction_Hamiltonian}  
\bar{H}^{\sigma}=\frac{1}{2}\int{d\boldsymbol{\Omega}~d\boldsymbol{\Omega^{'}}}~v\left(|\boldsymbol{\Omega} {-}\boldsymbol{\Omega^{'}}|\right)~\colon\bar{\rho}^{~\sigma}(\boldsymbol{\Omega})~\bar{\rho}^{~\sigma}(\boldsymbol{\Omega^{'}})\colon.
\end{equation} 
Here, we outline the steps required to obtain Eq.~\eqref{eq: not_normal_ordered_interaction} from Eq.~\eqref{eq: reiterate_interaction_Hamiltonian}. First, we use Eq.~\eqref{eq: relation_projected_normal_ordered_bare_density_operator} to express the normal-ordered operator $\colon\bar{\rho}^{~\sigma}(\boldsymbol{\Omega})~\bar{\rho}^{~\sigma}(\boldsymbol{\Omega^{'}})\colon$ in terms of the bare projected density operators. We then use Eq.~\eqref{eq: offset_projected_normal_ordered_operator} followed by the use of the angular-momentum-space representation of the interaction potential $v\left(|\boldsymbol{\Omega} {-}\boldsymbol{\Omega^{'}}|\right)$ as given in Eq.~\eqref{eq: interaction_angular_momentum_space}, to arrive at Eq.~\eqref{eq: not_normal_ordered_interaction},
\begin{equation}
\label{eq: reiterate_not_normal_ordered_interaction}
    \bar{H}^{\sigma}=\frac{4\pi}{2}\sum_{L}v_{L}\sum_{M{=}-L}^{L}\left[\bar{\rho}^{~\sigma}_{L,M}\right]^{\dagger}\bar{\rho}^{~\sigma}_{L,M}~-~\bar{H}^{(s),\sigma}.
\end{equation}
The single-body Hamiltonian $\bar{H}^{(s),\sigma}$ can be expressed in terms of the operator $\bar{\mathcal{O}}^{\sigma}_{L,M}$ [see Eq.~\eqref{eq: offset_LL_basis_2}] as
\begin{equation}
\label{eq: one_body_Hamiltonian}
    \bar{H}^{(s),\sigma}=\frac{4\pi}{2}\sum_{L}v_{L}\sum_{M{=}-L}^{L}~\bar{\mathcal{O}}^{\sigma}_{L,M}.
\end{equation}
Finally, we compute the expectation value of the Hamiltonian $\bar{H}^{\sigma}$ for a FQH ground sate $|\Psi_{\nu}\rangle$ as
\begin{align}
\label{eq: reiterate_interaction_Hamiltonian_angular_momentum_space}
 \left\langle \bar{H}^{\sigma}\right\rangle_{\Psi_{\nu}} =\frac{4\pi}{2}\sum_{L}v_{L}\sum_{M{=}-L}^{L}\left\langle \Psi_{\nu}\big|\left[\bar{\rho}^{~\sigma}_{L,M}\right]^{\dagger}\bar{\rho}^{~\sigma}_{L,M}\big|\Psi_{\nu}\right\rangle - \frac{4\pi}{2}\sum_{L}v_{L}\sum_{M{=}-L}^{L}~\left\langle \Psi_{\nu}\big|\bar{\mathcal{O}}^{\sigma}_{L,M}\big|\Psi_{\nu}\right\rangle.
\end{align}
By applying the definition of $\bar{S}^{\sigma}(L)$ as given in Eq.~\eqref{eq: projected_structure_factor} and using Eq.~\eqref{eq: expectation_offset__projected}, we obtain the expression of the ground state energy as given in Eq.~\eqref{eq: ground_state_energy}. We note that both $\bar{S}^{\sigma}(L)$ and $\left\langle \mathcal{\bar{O}}^{\sigma}_{L,M}\right\rangle_{\Psi_{\nu}}$ are independent of the azimuthal quantum number $M$, as a result, the sum over $M$ just gives a factor of $2L{+}1$. In the following Appendix~\ref{app: ground_state_energies}, using Eq.~\eqref{eq: reiterate_interaction_Hamiltonian_angular_momentum_space} or equivalently Eq.~\eqref{eq: ground_state_energy} from the main text, we compute the ground state energy of various FQH states.

\section{Trial wave function energies for Coulomb and short-range interactions}
\label{app: ground_state_energies}
\begin{figure}[tbh]
        \includegraphics[width=0.49\columnwidth]{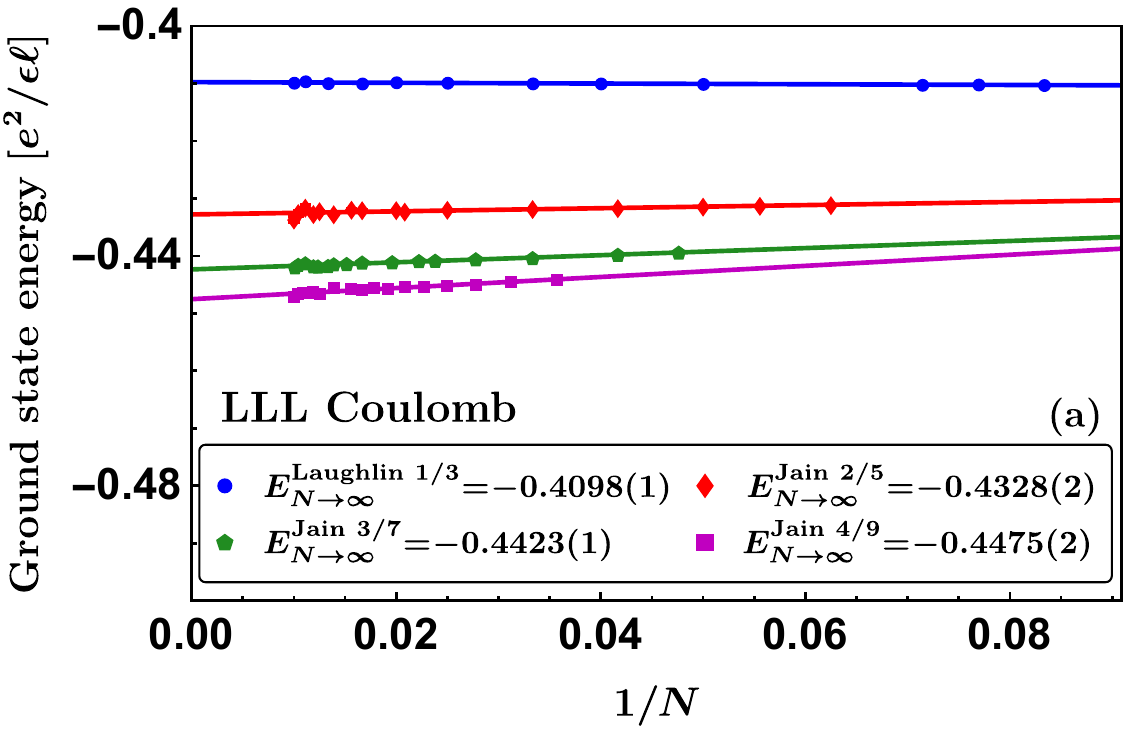}
        \includegraphics[width=0.49\columnwidth]{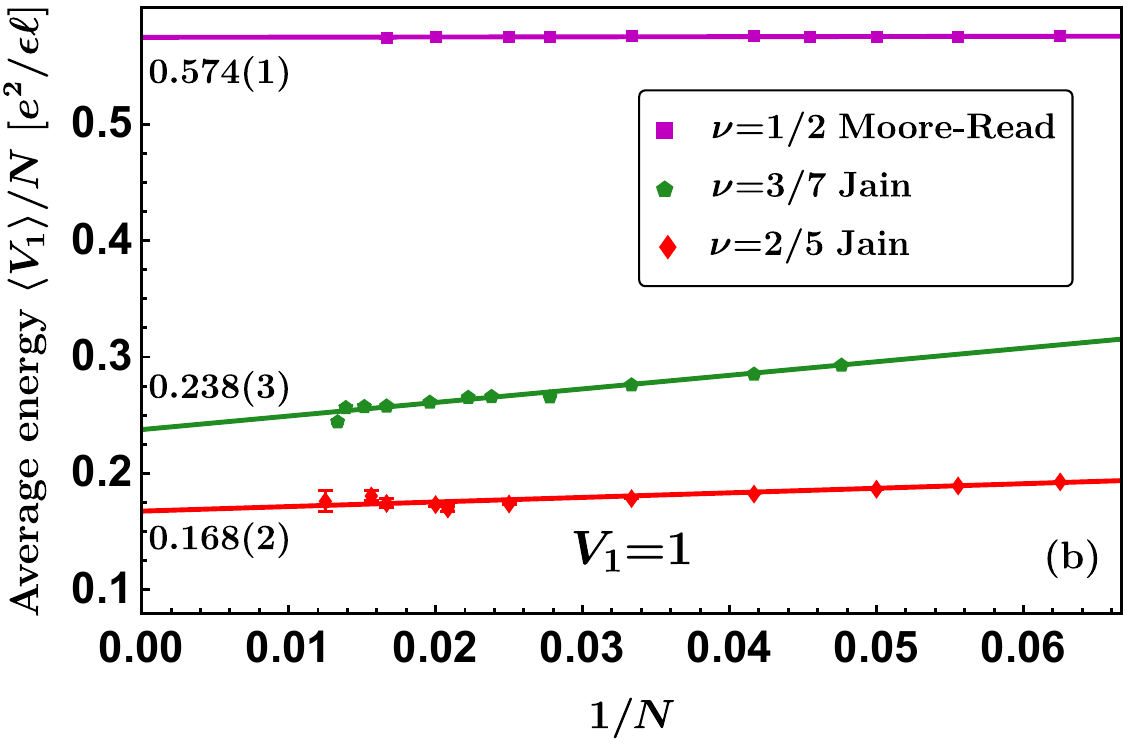}\\
         \includegraphics[width=0.49\columnwidth]{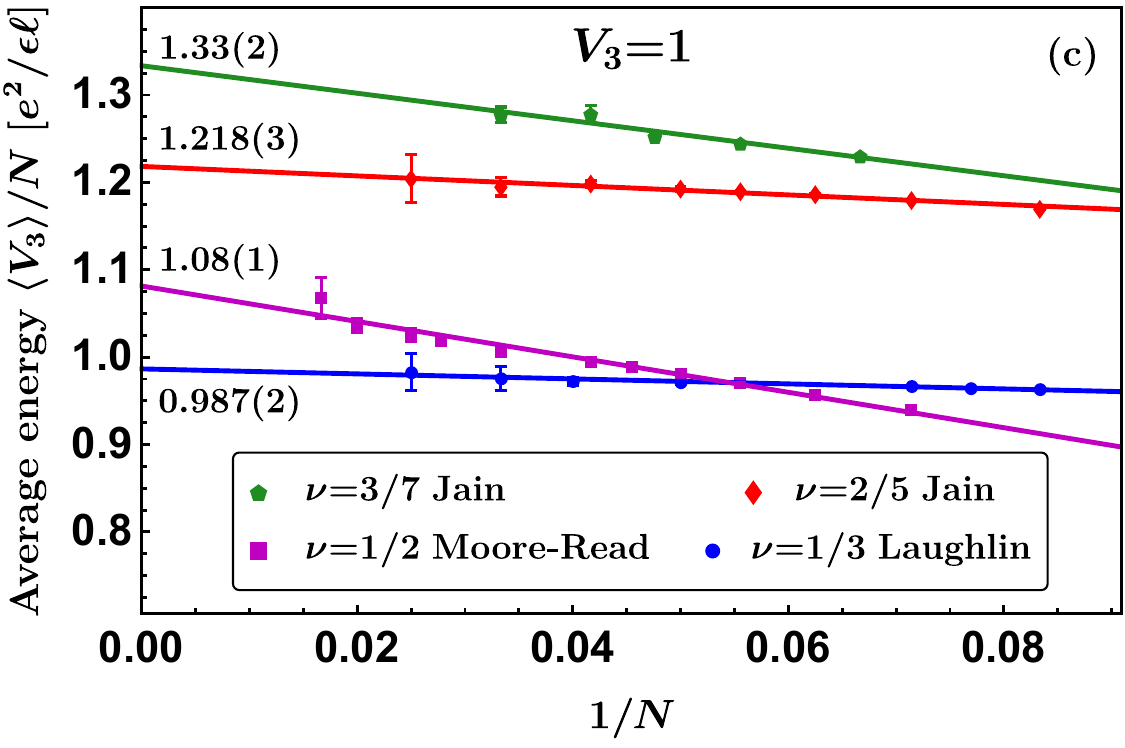}
         \includegraphics[width=0.49\columnwidth]{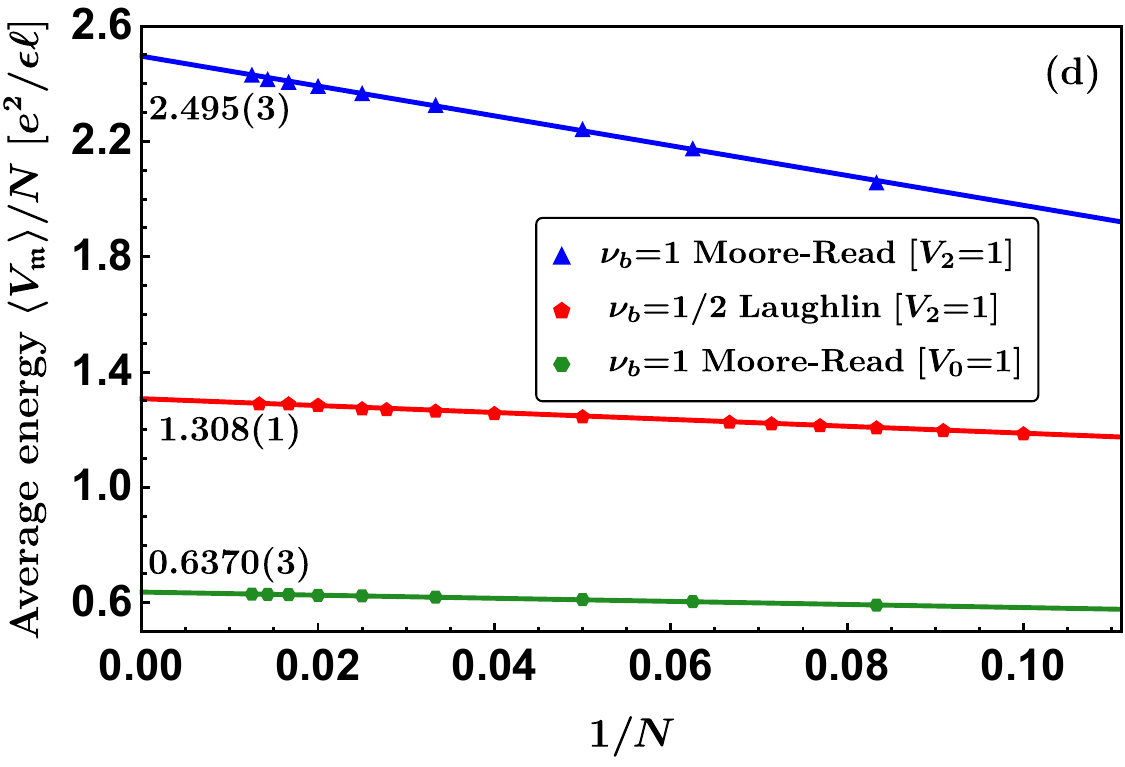}
          \caption{(a) Thermodynamic extrapolation of the density-corrected background subtracted per-particle LLL Coulomb energies for the primary Jain states. The per-particle density-corrected bare (no background subtraction is done) $V_{1}$ (b) and $V_{3}$ (c) energies or the average pair amplitude in relative angular momentum $\mathfrak{m}{=}1$ (b) and $\mathfrak{m}{=}3$ (c) channels for the 1/3 Laughlin and 2/5 and 3/7 Jain states, and $\nu{=}1/2$ Moore-Read state. Panel (d) shows the $\mathfrak{m}{=}0,2$ average pair-amplitudes for the bosonic $\nu_{b}{=}1/2$ Laughlin (the $\mathfrak{m}{=}0$ pair-amplitude is trivially zero here so has not been shown) and $\nu_{b}{=}1$ Moore-Read states.}
          \label{fig: Coulomb_model_interaction_energies}
\end{figure}   

The energy of a FQH state for a given density-density interaction potential is solely determined from its projected structure factor $\bar{S}^{\sigma}\left(L\right)$, as evident from  Eq.~\eqref{eq: ground_state_energy}. Instead of using the trial state's exact $\bar{S}^{\sigma,{\rm exact}}\left(L\right)$, which is computationally challenging to obtain, we use the approximate $\bar{S}^{\sigma, {\rm trial}}\left(L\right)$ obtained from the unprojected $S^{\sigma}\left(L\right)$, as discussed in Sec.~\ref{sssec: approximate_projected_structure_factor}. For the Coulomb interaction, we plug in the expression of the harmonics $v^{(C)}_{L}$ into Eq.~\eqref{eq: ground_state_energy}, to determine the Coulomb energy of a FQH state. Here, we present the Coulomb energy for primary Jain states. To facilitate the comparison with results available in the literature, a constant term $N^{2}/(2\sqrt{Q})$~\cite{Jain07}, representing the contribution from the uniformly distributed positive background charge on the sphere, is subtracted from $\langle \bar{H}^{\sigma}\rangle_{\Psi_{\nu}}$. Furthermore, to mitigate finite-size effects, the resulting per-particle energy is density corrected, i.e., a factor of $\sqrt{(2Q\nu)/N}$ is multiplied to it~\cite{Morf86}, and then extrapolated to the thermodynamic limit as a linear function of $1/N$, i.e., the quantity $\sqrt{(2Q\nu)/N}\left[\langle \bar{H}^{\sigma}\rangle_{\Psi_{\nu}}/N{-}N/(2\sqrt{Q})\right]$ is extrapolated to $1/N{\to}0$ by a linear fit in $1/N$.

Interestingly, our formalism can be readily used to compute the energy for an arbitrary interaction parameterized by a set of Haldane pseudopotentials $\{V_{\mathfrak{m}}\}$. This contrasts with the direct Monte Carlo integration-based approach to evaluate the energy of trial wave functions, which relies on the smooth behavior of the real-space interaction potential. In general, to compute the energy, we invert the set of $\{V_{\mathfrak{m}}\}$ using Eq.~\eqref{eq: inversion_pair_pseudopotential} to obtain the corresponding set of harmonics $\{v_{L}\}$, which are then substituted into Eq.~\eqref{eq: ground_state_energy}. Unlike for the Coulomb interaction, the energy for a single isolated pseudopotential $V_{\mathfrak{m}}$ $(\mathfrak{m}{>}1)$ is highly susceptible to the statistical error in $\bar{S}^{\sigma}\left(L\right)$ data, which increases with both the system size and $\mathfrak{m}$, posing an obstacle to extrapolating its energy to the thermodynamic limit. This is because the corresponding harmonics $v_{L}$ for $V_{\mathfrak{m}}$ grows polynomially as $[L(L{+}1)]^{\mathfrak{m}}$ [contrast this with Coulomb where $v_{L}$ decays with $L$] resulting in $\langle \bar{H}^{\sigma}\rangle_{\Psi_{\nu}}$ deviating significantly from its actual value. Encouragingly, the recent Ref.~\cite{Yutushui24} offers a neat solution to control the rapid growth of $\{v_{L}\}$ as $L$ increases. Reference~\cite{Yutushui24} utilizes the fact that adding a set of pseudopotentials $\boldsymbol{V^{\prime}}{=}\{V^{\prime}_{\mathfrak{m}'}\}$, where $\mathfrak{m}'$ is even (odd), to the given set of $\boldsymbol{V}{=}\{V_{\mathfrak{m}}\}$, where $\mathfrak{m}$ is odd (even) does not change the energy for a fermionic (bosonic) system. The new set of pseudopotentials $\boldsymbol{\tilde{V}}{\equiv}\boldsymbol{V^{\prime}}{\cup}\boldsymbol{V}$ results in a different set of harmonics $\{\tilde{v}_{L}\left(\boldsymbol{V^{\prime}}\right)\}$ [obtained using Eq.~\eqref{eq: inversion_pair_pseudopotential}], whose behavior with $L$ can be tuned by adjusting $\boldsymbol{V^{\prime}}$. To obtain an optimal value of the set of parameters $\boldsymbol{V^{\prime}}$ that restrict the growth of  $\{\tilde{v}_{L}\left(\boldsymbol{V^{\prime}}\right)\}$ as $L$ increases, Ref.~\cite{Yutushui24} minimizes the following ``cost function" with respect to $\boldsymbol{V^{\prime}}$,
\begin{align}
\label{eq: cost_function}
    \mathbb{F}\left(\boldsymbol{V^{\prime}}\right)&= \sum_{L=0}^{2Q} [\tilde{v}_{L}\left(\boldsymbol{V^{\prime}}\right)]^{2} (2L+1)^{\alpha},
\end{align}
where the exponent $\alpha{\geq} 2$. Plugging the resulting optimal values of $\boldsymbol{V^{\prime}}$ into  $\tilde{v}_{L}\left(\boldsymbol{V^{\prime}}\right)$, one obtains the desired modified harmonics $\{\tilde{v}_{L}\}$. 

As applications of these ideas, we mention that this technique can be used to compute the average pair amplitudes in the relative angular momentum $\mathfrak{m}'$ channel (as these are equivalent to computing the energy of the $V_{\mathfrak{m}}{=}\delta_{\mathfrak{m},\mathfrak{m}'}$ interaction) for trial wave functions. These pair amplitudes are useful in determining the phase diagram where the short-range part of the Coulomb interaction is altered due to residual interactions, such as lattice-scale interactions in graphene~\cite{An24, An24a} or determining the stability of a phase when certain pseudopotentials are varied~\cite{Yutushui24}. Next, we present energies of many well-known bosonic and fermionic states.

\subsection{Fermionic Laughlin, Jain, and Moore-Read states}
\label{app: energies_fermionic_states}
This section will present results on the Coulomb ground state energies and average pair amplitudes $\left\langle V_{\mathfrak{m}}\right\rangle$ of the fermionic FQH states. The thermodynamic extrapolation of the per-particle density-corrected and background subtracted Coulomb energy is presented in Fig.~\ref{fig: Coulomb_model_interaction_energies}$(a)$, for various FQH states, including, $1/3$ Laughlin, $2/5$, $3/7$, and $4/9$ Jain states. We find an excellent agreement between our results and those presented in Refs.~\cite{Ciftja03, Balram20b, Jain97b, Balram17, Dora23}, where $\langle \bar{H}^{\sigma}\rangle_{\Psi_{\nu}}$ was computed directly using the respective Laughlin and Jain wave functions [see Eq.~\eqref{eq: CF_ground_state}] via either computation of the expectation of the states in Fock-space or by Monte Carlo integration. To this end, we point out that, even if one uses the entire set of $\bar{S}^{\sigma}\left(L\right)$ data in Eq.~\eqref{eq: ground_state_energy}, without assuming it to zero above $L_{\rm cut{-}off}$ [see Sec.~\ref{sssec: approximate_projected_structure_factor}], the Coulomb energy evaluates very accurately, owing to the decaying nature of the harmonics $v^{(C)}_{L}$ as a function of $L$.

We set $\alpha{=}2$ in Eq.~\eqref{eq: cost_function} and follow the aforementioned method of Ref.~\cite{Yutushui24} to compute the energy of the $V_{\mathfrak{m}}{=}\delta_{\mathfrak{m},1}$ interaction (referred to as the $V_{1}$ interaction) for the $2/5$ and $3/7$ Jain states and $\nu{=}1/2$ Moore-Read state. The extrapolated thermodynamic bare (no background subtraction is done) energy is shown in Fig.~\ref{fig: Coulomb_model_interaction_energies}$(b)$ which shows that we can reliably compute the energy for fairly large systems using the approximate $\bar{S}^{\sigma, {\rm trial}}\left(L\right)$ data. Surprisingly, even setting $\alpha{=}0$ in Eq.~\eqref{eq: cost_function}, which treats all $L$ equally in $\mathbb{F}\left(\boldsymbol{V^{\prime}}\right)$, we find a good estimate of the energy. This is because, even with $\alpha{=}0$, the resulting $\{\tilde{v}_{L}\}$ do not exhibit polynomial growth with $L$; instead, they decay and remain small at large $L$.

Similarly, the energy of the $V_{\mathfrak{m}}{=}\delta_{\mathfrak{m},3}$ interaction (referred to as the $V_{3}$ interaction) in the thermodynamic limit is presented in Fig.~\ref{fig: Coulomb_model_interaction_energies}$(c)$ for the $\nu{=}1/3$ Laughlin and $\nu{=}2/5$ and $3/7$ Jain states, and $\nu{=}1/2$ Moore-Read state, where again we set $\alpha{=}2$. Unlike for the $V_1$ interaction, we find that for the $V_{3}$ interaction, the optimized harmonics $\{\tilde{v}_{L}\}$ for $L$ comparatively less than $2Q$ increase rapidly, restricting us to smaller system sizes than those for which we could compute the $V_{1}$ energies. Generally, for an isolated pseudopotential $V_{\mathfrak{m}}$, the accessible system size decreases with increasing $\mathfrak{m}$. This issue can potentially be resolved by observing that the optimization in Eq.~\eqref{eq: cost_function}, for a given system size, only ensures that the harmonics $\{\tilde{v}_{L}\}$ decreases with $L$. Thus, constructing an appropriate ``cost function" that accounts for the rapid growth of $\{\tilde{v}_{L}\}$ with system size as $\mathfrak{m}$ increases can resolve this issue. We leave an exploration of this matter in detail to future work.

\subsection{Bosonic Laughlin and Moore-Read states}
\label{app: energies_bosonic_states}
Here we present the average bare energies (without background subtraction) of the $V_{\mathfrak{m}}$-only interaction, where $\mathfrak{m}$ is even, for bosonic FQH states. In Fig.~\ref{fig: Coulomb_model_interaction_energies}$(d)$, we show the thermodynamic extrapolation of the energy of $V_{\mathfrak{m}}{=}\delta_{\mathfrak{m},0}$ (referred to as the $V_{0}$) interaction for the $\nu_{b}{=}1$ Moore-Read state [green solid hexagons]. The energy of the $V_0$ interaction is obtained directly from its harmonics $v_{L}$ since they remain constant as a function of $L$ [see Eq.~\eqref{eq: v_l_contact_interaction}].

In Fig.~\ref{fig: Coulomb_model_interaction_energies}$(d)$, the average energies of the $V_{\mathfrak{m}}{=}\delta_{\mathfrak{m},2}$ (referred to as the $V_{2}$) interaction for the $\nu_{b}{=}1/2$ Laughlin and $\nu_{b}{=}1$ Moore-Read states, are also shown. Since the harmonics $v_L$ of the $V_2$ interaction grow as $[L(L{+}1)]^2$, we determine an optimized set of harmonics $\tilde{v}_{L}$ [setting $\alpha{=}2$ in Eq.~\eqref{eq: cost_function}] and use them to compute the corresponding energy. As mentioned above, these computations give us access to the average pair amplitudes in the relative angular momentum $2$, and relative angular momentum $0,2$ channels in the bosonic 1/2 Laughlin and $\nu_{b}{=}1$ Moore-Read states, respectively. The Coulomb interaction energy of the bosonic states [not presented here] can be readily computed by following the procedure outlined for fermionic states in Appendix~\ref{app: energies_fermionic_states}.

\section{Derivation of expansion coefficients in GMP algebra}
\label{app: derivation_GMP_algebra_coefficients}
In this appendix, we present a derivation of expansion coefficients $\alpha_{L}^{(L_1, L_2, M_1, M_2)}$. We begin by expressing Eq.~\eqref{eq: GMP_algebra} in a form similar to Eq.~\eqref{eq: GMP_commutator_expanded} by substituting Eq.~\eqref{eq: simplified_second_quantized_projected_angular_momentum_density} into it. Consequently, one obtains
\begin{align}
 \label{eq: GMP_commutator_expanded2}
[\bar{\rho}^{~\sigma}_{L_1,M_1},\bar{\rho}^{~\sigma}_{L_2,M_2}]&=\sum_{m}\bigg[\sum_{L} \alpha_{L}^{(L_1,L_2,M_1,M_2)}\bar{\rho}(L,M,m)\bigg]~ \left [\chi^{\sigma}_{M+m}\right]^{\dagger} \chi^{\sigma}_{m}.
\end{align} 
Next, we compare Eq.~\eqref{eq: GMP_commutator_expanded} with Eq.~\eqref{eq: GMP_commutator_expanded2}, which results
\begin{align}
\label{eq: linear_eq}
   \sum_{L} \alpha_{L}^{(L_1,L_2,M_1,M_2)}~ \bar{\rho}(L,M,m) &= \bar{\rho}(L_1, L_2, M_1, M_2, m).
\end{align}
To evaluate the expansion coefficients $\alpha_{L}^{(L_1,L_2,M_1,M_2)}$, it is useful to note the explicit expression of $\bar{\rho}(L,M,m)$ [see also Sec.~\ref{ssec: density_operators}],
\begin{align}
\label{eq: explicit_expression_coefficient_density_operator}
  \bar{\rho}(L,M,m)= (-1)^{m}\zeta(L,M)\mathcal{F}(L)\left(\begin{array}{ccc}
Q & Q & L \\
-m-M & m & M
\end{array}\right),
\end{align}
where $\zeta(L,M){=}(-1)^{L+M-Q}$ and $\mathcal{F}(L)$ is the LLL form factor as defined in Eq.~\eqref{eq: expression_expansion_coefficients}, which is
\begin{align}
   \mathcal{F}(L)=(2Q+1)\sqrt{\frac{2L+1}{4\pi}} \left(\begin{array}{ccc}
Q & Q & L \\
-Q & Q & 0
\end{array}\right),
\end{align}
Using Eq.~\eqref{eq: explicit_expression_coefficient_density_operator}, we rewrite Eq.~\eqref{eq: linear_eq} as
\begin{align}
\sum_{L} (-1)^m\zeta(L,M)\mathcal{F}(L)\alpha_{L}^{(L_1,L_2,M_1,M_2)}\left(\begin{array}{ccc}
Q & Q & L \\
-m-M & m & M
\end{array}\right)
&=\bar{\rho}(L_1, M_1; L_2, M_2; m).
\end{align}
Next, we further rewrite the above equation as
\begin{align}
\label{eq: GMP_algebra_derivation_1}
\sum_{L} \zeta(L,M)\mathcal{F}(L)\alpha_{L}^{(L_1,L_2,M_1,M_2)}\sum_{m^{\prime}}(-1)^{m^{\prime}-M}\left(\begin{array}{ccc}
Q & Q & L \\
-m^{\prime} & m & M
\end{array}\right)
&=\bar{\rho}(L_1, M_1; L_2, M_2; m),
\end{align}
where we have replaced $m{+}M$ in the Wigner $3j$ symbol by $m^{\prime}$ and summed over it across $-Q$ to $Q$, with the understanding that the Wigner $3j$ symbol vanishes unless $m^{\prime}{=}m{+}M$. To extract the expansion coefficients, we multiply both sides of Eq.~\eqref{eq: GMP_algebra_derivation_1} by $(-1)^{-m}\left(\begin{array}{ccc}
Q & Q & L^{\prime} \\
-m^{\prime} & m & M^{\prime}
\end{array}\right)$ and sum over the entire range of $m$ (i.e., $|m|{\leq} Q$), and then use the following orthogonal property of the Wigner $3j$ symbols:
\begin{align}
    \sum_{m,m^{\prime}}\left(\begin{array}{ccc}
Q & Q & L \\
-m^{\prime} & m & M
\end{array}\right)\left(\begin{array}{ccc}
Q & Q & L^{\prime} \\
-m^{\prime} & m & M^{\prime}
\end{array}\right)&=\frac{\delta_{L,L^{\prime}}}{2L+1}\delta_{M,M^{\prime}}.
\end{align}
Consequently, one obtains
\begin{align}
\label{eq: GMP_algebra_derivation_3}
\alpha_{L}^{(L_1,L_2,M_1,M_2)} &=\frac{2L+1}{\zeta(L,M)~\mathcal{F}(L)}\sum_{m}(-1)^{-m}\bigg[\bar{\rho}(L_1, M_1; L_2, M_2; m)
  \left(\begin{array}{ccc}
Q & Q & L \\
-m-M & m & M
\end{array}\right) \bigg].
\end{align}
Interestingly, the above expression can be further simplified by noting the explicit expression of $\bar{\rho}(L_1, M_1; L_2, M_2; m)$ [see Eqs.~\eqref{eq: coefficient_GMP_commutator} and~\eqref{eq: expression_expansion_coefficients}],
\begin{align}
    \bar{\rho}(L_1, M_1; L_2, M_2; m)&=\mathcal{F}(L_1)\mathcal{F}(L_2)\zeta(L_1,M_1)\zeta(L_2,M_2)(-1)^{2m}\bigg[(-1)^{M_1}\left(\begin{array}{ccc}
Q & Q & L_{1} \\
-M-m & M_2+m & M_{1}
\end{array}\right)\left(\begin{array}{ccc}
Q & Q & L_{2} \\
-M_2-m & m & M_{2}
\end{array}\right)\nonumber\\
&-(-1)^{M_2}\left(\begin{array}{ccc}
Q & Q & L_{1} \\
-M_1-m & m & M_{1}
\end{array}\right)\left(\begin{array}{ccc}
Q & Q & L_{2} \\
-M-m & M_1+m & M_{2}
\end{array}\right)\bigg].
\end{align}
Next, we employ the same trick in obtaining Eq.~\eqref{eq: GMP_algebra_derivation_1}. In particular, we introduce additional summations in the above equation at no cost, i.e.,
\begin{align}
    \bar{\rho}(L_1, M_1; L_2, M_2; m)&=\mathcal{F}(L_1)\mathcal{F}(L_2)\zeta(L_1,M_1)\zeta(L_2,M_2)(-1)^{2m}\bigg[(-1)^{M_1}\sum_{m_1,m_2}\left(\begin{array}{ccc}
Q & Q & L_{1} \\
-m_1 & m_2 & M_{1}
\end{array}\right)\left(\begin{array}{ccc}
Q & Q & L_{2} \\
-m_2 & m & M_{2}
\end{array}\right)\nonumber\\
&-(-1)^{M_2}\sum_{m_1,m_3}\left(\begin{array}{ccc}
Q & Q & L_{1} \\
-m_3 & m & M_{1}
\end{array}\right)\left(\begin{array}{ccc}
Q & Q & L_{2} \\
-m_1 & m_3 & M_{2}
\end{array}\right)\bigg].
\end{align}
This is because the Wigner $3j$ symbols vanish unless $m_{1}{=}M{+}m$, $m_{2}{=}M_2{+}m$, and $m_3{=}M_1{+}m$.
To this end, we substitute the above equation into Eq.~\eqref{eq: GMP_algebra_derivation_3} and obtain
\begin{align}
\label{eq: GMP_algebra_derivation_4}
\alpha_{L}^{(L_1,L_2,M_1,M_2)}&=\mathcal{C}\Bigg[(-1)^{M_1}\sum_{m,m_1,m_2}(-1)^{-m}\left(\begin{array}{ccc}
Q & Q & L_{2} \\
m & -m_2 & M_{2}
\end{array}\right)\left(\begin{array}{ccc}
Q & Q & L_{1} \\
m_1 & -m_2 & -M_{1}
\end{array}\right)\left(\begin{array}{ccc}
Q & Q & L \\
m_1 & -m & -M
\end{array}\right)\nonumber\\
&-(-1)^{M_2}\sum_{m_,m_1,m_3}(-1)^{-m}\left(\begin{array}{ccc}
Q & Q & L_{1} \\
m & -m_3 & M_{1}
\end{array}\right)\left(\begin{array}{ccc}
Q & Q & L_{2} \\
m_1 & -m_3 & -M_{2}
\end{array}\right)\left(\begin{array}{ccc}
Q & Q & L \\
m_1 & -m & -M
\end{array}\right)\Bigg],
\end{align}
where 
\begin{align}
\nonumber
    \mathcal{C}{=}(-1)^{L_1+L_2+L}(2L+1)\frac{\mathcal{F}(L_1)\mathcal{F}(L_2)\zeta(L_1,M_1)\zeta(L_2,M_2)}{\zeta(L,M)~\mathcal{F}(L)}.
\end{align}
In obtaining Eq.~\eqref{eq: GMP_algebra_derivation_4}, we have used $(-1)^m{=}(-1)^{-m{+}2Q}$ and rearranged columns in Wigner $3j$ symbols and also extracted a minus sign from each of the azimuthal quantum numbers in some Wigner $3j$ symbols. This allows to use the following identity in Eq.~\eqref{eq: GMP_algebra_derivation_4}~\cite{Wigner3jsymbol}
\begin{align}
&\sum_{m_1,m_2,m_3}(-1)^{-m_1}\left(\begin{array}{ccc}
l_1 & l_2 & j_{1} \\
m_1 & m_2 & -n_{1}
\end{array}\right)\left(\begin{array}{ccc}
l_3 & l_2 & j_{2} \\
m_3 & m_2 & -n_{2}
\end{array}\right)\left(\begin{array}{ccc}
l_3 & l_1 & j_3 \\
m_3 & -m_1 & -n_3
\end{array}\right)\nonumber\\
&=(-1)^{3l_2-l_1-l_3+2j_1-n_1-n_3}\left(\begin{array}{ccc}
j_1 & j_3 & j_{2} \\
n_1 & n_3 & -n_{2}
\end{array}\right)
\left\{\begin{array}{lll}
l_1 & l_2 & j_1 \\
j_2 & j_3 & l_2
\end{array}\right\}.
\end{align}
Subsequently, we obtain the desired simplified expression of $\alpha_{L}^{(L_1,L_2,M_1,M_2)}$,
\begin{align} \alpha_{L}^{(L_1,L_2,M_1,M_2)}&=(-1)^{M}(2L+1)\bigg[(-1)^{L_1+L_2+L}-1\bigg]\frac{\mathcal{F}(L_1)\mathcal{F}(L_2)}{\mathcal{F}(L)}
 \left(\begin{array}{ccc}
L_1 & L_2 & L \\
M_1 & M_2 & -M
\end{array}\right)\left\{\begin{array}{lll}
L_1 & L_2 & L \\
Q & Q & Q
\end{array}\right\}.
\end{align}
\section{Oscillator strength $\bar{F}^{\sigma}(L)$ on sphere}
\label{app: Oscillator strength_sphere}
In this appendix, we present a derivation of Eq.~\eqref{eq: numerator_gap_equation} for the oscillator strength $\bar{F}^{\sigma}(L)$ defined as
\begin{align}
\label{eq: reiterate_gap_equation}
\bar{F}^{\sigma}(L)&=\frac{1}{2}\left\langle \Psi_{\nu}\right|\left[\left[\bar{\rho}^{~\sigma}_{L,M}\right]^{\dagger},\left[\bar{H}^{\sigma},\bar{\rho}^{~\sigma}_{L,M}\right]\right]\left|\Psi_{\nu}\right\rangle,~\text{where}~\bar{\rho}^{~\sigma}_{L,M}=\sum_{m}\bar{\rho}(L,M,m)~ [\chi^{\sigma}_{ M+m}]^{\dagger} \chi^{\sigma}_{m}~\text{[see Eq.~\eqref{eq: simplified_second_quantized_projected_angular_momentum_density}]}.
\end{align}
Let us write the interaction Hamiltonian $\bar{H}^{\sigma}$ [see Eq.~\eqref{eq: reiterate_interaction_Hamiltonian_angular_momentum_space}] as $\bar{H}^{\sigma}{=}\bar{H}^{(t),\sigma} {+} \bar{H}^{(s),\sigma}$, where
\begin{align}
\label{eq: interaction_Hamiltonian_short_form}
  \bar{H}^{(t),\sigma}&= \frac{4\pi}{2}\sum_{L}v_{L}\sum_{M{=}-L}^{L}~\bar{\mathcal{K}}^{\sigma}_{L,M},~~\text{and}~~\bar{\mathcal{K}}^{\sigma}_{L,M}=\left[\bar{\rho}^{~\sigma}_{L,M}\right]^{\dagger}\bar{\rho}^{~\sigma}_{L,M},
\end{align}
with the single-body Hamiltonian $\bar{H}^{(s),\sigma}$ is given by Eq.~\eqref{eq: one_body_Hamiltonian}. For operator $\bar{\mathcal{O}}^{\sigma}_{L,M}$ [see Eq.~\eqref{eq: offset_LL_basis_2}] in $\bar{H}^{(s),\sigma}$, one can straightforwardly check that $\left\langle \Psi_{\nu}\right|\left[\left[\bar{\rho}^{~\sigma}_{L,M}\right]^{\dagger},\left[\bar{\mathcal{O}}^{\sigma}_{L,M},\bar{\rho}^{~\sigma}_{L,M}\right]\right]\left|\Psi_{\nu}\right\rangle{=}0$. In deriving this, one uses
\begin{align}
    \left[\left[\chi^{\sigma}_{ M+m}\right]^{\dagger} \chi^{\sigma}_{M+m},\left[\chi^{\sigma}_{\tilde{M}+\tilde{m}}\right]^{\dagger} \chi^{\sigma}_{\tilde{m}}\right]&=\left(\delta_{M+m,\tilde{M}+\tilde{m}}\right)
    \left[\chi^{\sigma}_{M+m}\right]^{\dagger}\chi^{\sigma}_{\tilde{m}} - \left(\delta_{M+m,\tilde{m}}\right)
    \left[\chi^{\sigma}_{\tilde{M}+\tilde{m}}\right]^{\dagger}\chi^{\sigma}_{M+m}
\end{align}
along with the fact that the state $\left|\Psi_{\nu}\right\rangle$ is uniform, i.e., $\left\langle\left[\chi^{\sigma}_{m}\right]^{\dagger}\chi^{\sigma}_{m}\right\rangle_{\Psi_{\nu}}{=}\bar{\nu}$, for all $m$ in the LLL [see Appendix~\ref{ssec: expectation_unprojected}]. Thus, $\left\langle \Psi_{\nu}\right|\left[\left[\bar{\rho}^{~\sigma}_{L,M}\right]^{\dagger},\left[\bar{H}^{(s),\sigma},\bar{\rho}^{~\sigma}_{L,M}\right]\right]\left|\Psi_{\nu}\right\rangle{=}0$ and only $\bar{H}^{(t),\sigma}$ [see Eq.~\eqref{eq: interaction_Hamiltonian_short_form}] contributes to $\bar{F}^{\sigma}(L)$. Consequently, 
\begin{align}
   \bar{F}^{\sigma}(L)&= \frac{4\pi}{4}\sum_{\tilde{L}}v_{\tilde{L}}\sum_{\tilde{M}{=}-\tilde{L}}^{\tilde{L}}\left\langle \Psi_{\nu}\right|\left[\left[\bar{\rho}^{~\sigma}_{L,M}\right]^{\dagger},\left[\bar{\mathcal{K}}^{\sigma}_{\tilde{L},\tilde{M}},\bar{\rho}^{~\sigma}_{L,M}\right]\right]\left|\Psi_{\nu}\right\rangle.
\end{align}
In the following, we first compute the commutator $\left[\bar{\mathcal{K}}^{\sigma}_{\tilde{L},\tilde{M}},\bar{\rho}^{~\sigma}_{L,M}\right]$ and use it to compute $\left[\left[\bar{\rho}^{~\sigma}_{L,M}\right]^{\dagger},\left[\bar{\mathcal{K}}^{\sigma}_{\tilde{L},\tilde{M}},\bar{\rho}^{~\sigma}_{L,M}\right]\right]$. Leveraging the algebra of the density operators $\bar{\rho}^{~\sigma}_{L,M}$, as presented in Eq.~\eqref{eq: GMP_algebra}, we obtain
\begin{align}
\label{eq: commutator_step_1}
\left[\bar{\mathcal{K}}^{\sigma}_{\tilde{L},\tilde{M}},\bar{\rho}^{~\sigma}_{L,M}\right]&=(-1)^{\tilde{M}}\sum_{\lambda} \left[\alpha_{\lambda}^{(\tilde{L},L,\tilde{M},M)}\bar{\rho}^{~\sigma}_{\tilde{L},-\tilde{M}}~\bar{\rho}^{~\sigma}_{\lambda,\tilde{M}+M}  + \alpha_{\lambda}^{(\tilde{L},L,-\tilde{M},M)}\bar{\rho}^{~\sigma}_{\tilde{L},\tilde{M}}~\bar{\rho}^{~\sigma}_{\lambda,M-\tilde{M}} \right].
\end{align}
In the above equation, we have also used the relation $\left[\bar{\rho}^{~\sigma}_{\tilde{L},\tilde{M}}\right]^{\dagger}{=}(-1)^{\tilde{M}}\bar{\rho}^{~\sigma}_{\tilde{L},{-}\tilde{M}}$. The summation over angular momentum $\lambda$ in Eq.~\eqref{eq: commutator_step_1} ranges from $\left|L{-}\tilde{L}\right|{+}1$ to $\left|L{+}\tilde{L}{-}1\right|$ in steps of two, as discussed in Sec.~\ref{sec: algebra_projected_density_operator}. In general, for the symbol $\alpha_{L_1}^{(L_2,L_3,M_2,M_3)}$, $L_1$ runs from $\left|L_2{-}L_3\right|{+}1$ to $\left|L_2{+}L_3{-}1\right|$ in steps of two. Similarly, we obtain
\begin{align}
\label{eq: commutator_step_2}
\left[\left[\bar{\rho}^{~\sigma}_{L,M}\right]^{\dagger},\left[\bar{\mathcal{K}}^{\sigma}_{\tilde{L},\tilde{M}},\bar{\rho}^{~\sigma}_{L,M}\right]\right]&=(-1)^{M}\sum_{\lambda}\alpha_{\lambda}^{(\tilde{L},L,\tilde{M},M)}\sum_{\mu}\alpha_{\mu}^{(L,\lambda,-M,\tilde{M}+M)}~\left[\bar{\rho}^{~\sigma}_{\tilde{L},\tilde{M}}\right]^{\dagger}~\bar{\rho}^{~\sigma}_{\mu,\tilde{M}}\notag\\
&+(-1)^{M}\sum_{\lambda}\alpha_{\lambda}^{(\tilde{L},L,-\tilde{M},M)}\sum_{\mu}\alpha_{\mu}^{(L,\lambda,-M,M-\tilde{M})}~\left[\bar{\rho}^{~\sigma}_{\tilde{L},-\tilde{M}}\right]^{\dagger}~\bar{\rho}^{~\sigma}_{\mu,-\tilde{M}}\notag\\
&+\sum_{\lambda}\alpha_{\lambda}^{(\tilde{L},L,\tilde{M},M)}\sum_{\mu}\alpha_{\mu}^{(L,\tilde{L},-M,-\tilde{M})}~\left[\bar{\rho}^{~\sigma}_{\mu,\tilde{M}+M}\right]^{\dagger}~\bar{\rho}^{~\sigma}_{\lambda,\tilde{M}+M}\notag\\
&+\sum_{\lambda}\alpha_{\lambda}^{(\tilde{L},L,-\tilde{M},M)}\sum_{\mu}\alpha_{\mu}^{(L,\tilde{L},-M,\tilde{M})}~\left[\bar{\rho}^{~\sigma}_{\mu,M-\tilde{M}}\right]^{\dagger}~\bar{\rho}^{~\sigma}_{\lambda,M-\tilde{M}}.
\end{align}
Next, to compute $\bar{F}^{\sigma}(L)$, we evaluate the expectation value of the above equation for the state $\left|\Psi_{\nu}\right\rangle$. The resulting expression can be simplified by noting that $\left\langle\Psi_{\nu}\right|\left[\bar{\rho}^{~\sigma}_{L_1, M}\right]^{\dagger}~\bar{\rho}^{~\sigma}_{L_2, M}\left|\Psi_{\nu}\right\rangle{=}(N/4\pi)\bar{S}^{\sigma}(L_1)~\delta_{L_1, L_2}$. This is because for $L_1{\neq} L_2$, $\bar{\rho}^{~\sigma}_{L_i, M}\left|\Psi_{\nu}\right\rangle$ ($i{=}1,2$), generates two orthogonal states resulting in a zero overlap between them. However, for $L_1{=}L_2$, their overlap is proportional to $\bar{S}^{\sigma}(L_1)$ [see Sec.~\ref{ssec: structure_factor}]. Furthermore, we use $\alpha_{L_1}^{(L_2,L_3,M_2,M_3)}{=}\alpha_{L_1}^{(L_3,L_2,-M_3,-M_2)}$, which follows from the combined action of Eqs.~\eqref{eq: relation_GMP_algebra_coefficients} and~\eqref{eq: interchange_relation_GMP_algebra_coefficients}, to get the following expression for the oscillator strength:
\begin{align}
\label{eq: commutator_step_3}
    \bar{F}^{\sigma}(L)&= \frac{N}{4}\sum_{\tilde{L}}v_{\tilde{L}}\sum_{\tilde{M}{=}-\tilde{L}}^{\tilde{L}}\sum_{\lambda}\bm{\Bigg[}(-1)^{M}\left[\alpha_{\lambda}^{(\tilde{L},L,\tilde{M},M)}~\alpha_{\tilde{L}}^{(L,\lambda,-M,\tilde{M}+M)}+\alpha_{\lambda}^{(\tilde{L},L,-\tilde{M},M)}~\alpha_{\tilde{L}}^{(L,\lambda,-M,M-\tilde{M})}\right]\bar{S}^{\sigma}(\tilde{L})\notag\\
   &~~~~~~~~~~~~~~~~~~~~~~~~~~~~~~~~~+\left[\left(\alpha_{\lambda}^{(\tilde{L},L,\tilde{M},M)}\right)^{2}+\left(\alpha_{\lambda}^{(\tilde{L},L,-\tilde{M},M)}\right)^{2} \right]\bar{S}^{\sigma}(\lambda)\bm{\Bigg]}.
\end{align}
The above equation for $\bar{F}^{\sigma}(L)$ is then used to evaluate the gap of the GMP state $|\Psi_{L, M}^{\rm GMP}\rangle$. As explained in Sec.~\ref{ssec: magnetoroton_gap}, $\bar{F}^{\sigma}(L)$ is independent of $M$. Setting $M{=}0$, simplifies the expression of the oscillator strength further, leading to
\begin{align}
\label{eq: commutator_step_4}
    \bar{F}^{\sigma}(L)&= \frac{N}{4}\sum_{\tilde{L}}v_{\tilde{L}}\sum_{\tilde{M}{=}-\tilde{L}}^{\tilde{L}}\sum_{\lambda}\bm{\Bigg[}2\alpha_{\lambda}^{(\tilde{L},L,\tilde{M},0)}~\alpha_{\tilde{L}}^{(L,\lambda,0,\tilde{M})}~\bar{S}^{\sigma}(\tilde{L})~+~2\left(\alpha_{\lambda}^{(\tilde{L},L,\tilde{M},0)}\right)^{2}~\bar{S}^{\sigma}(\lambda)\bm{\Bigg]},
\end{align}
 where we have used $\alpha_{\lambda}^{(\tilde{L},L,\tilde{M},0)}~\alpha_{\tilde{L}}^{(L,\lambda,0,\tilde{M})}{=}\alpha_{\lambda}^{(\tilde{L},L,-\tilde{M},0)}~\alpha_{\tilde{L}}^{(L,\lambda,0,-\tilde{M})}$ and $\left(\alpha_{\lambda}^{(\tilde{L},L,\tilde{M},0)}\right)^{2}{=}\left(\alpha_{\lambda}^{(\tilde{L},L,-\tilde{M},0)}\right)^{2}$, following Eq.~\eqref{eq: relation_GMP_algebra_coefficients}.
Interestingly, again employing Eq.~\eqref{eq: relation_GMP_algebra_coefficients} and noting that $\alpha_{\lambda}^{(L_1, L_2,0,0)}{=}0$, for any $L_1$ and $L_2$ [see Eq.~\eqref{eq: commutation_special_case_1}], the sum over $\tilde{M}$ in Eq.~\eqref{eq: commutator_step_4} can be restricted to positive integers only, as follows:

\begin{align}
\label{eq: commutator_step_5}
    \bar{F}^{\sigma}(L)&= \frac{N}{4}\sum_{\tilde{L}}v_{\tilde{L}}\sum_{\tilde{M}{=}1}^{\tilde{L}}\sum_{\lambda}\bm{\Bigg[}4\alpha_{\lambda}^{(\tilde{L},L,\tilde{M},0)}~\alpha_{\tilde{L}}^{(L,\lambda,0,\tilde{M})}~\bar{S}^{\sigma}(\tilde{L})~+~4\left(\alpha_{\lambda}^{(\tilde{L},L,\tilde{M},0)}\right)^{2}~\bar{S}^{\sigma}(\lambda)\bm{\Bigg]}.
\end{align}
This completes our derivation of Eq.~\eqref{eq: numerator_gap_equation}.
\section{GMP gap for short-range interactions}
\label{app: GMP_gap_short_range_interactions}

\begin{figure}[tbh]
         \includegraphics[width=0.49\columnwidth]{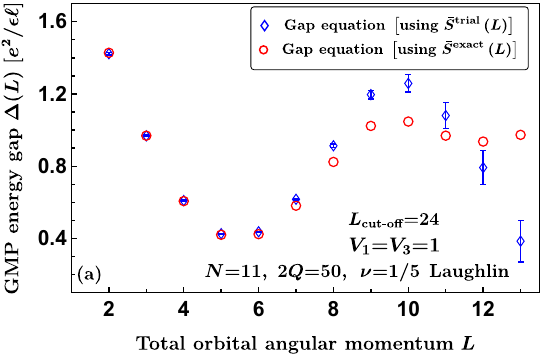}
        \includegraphics[width=0.49\columnwidth]{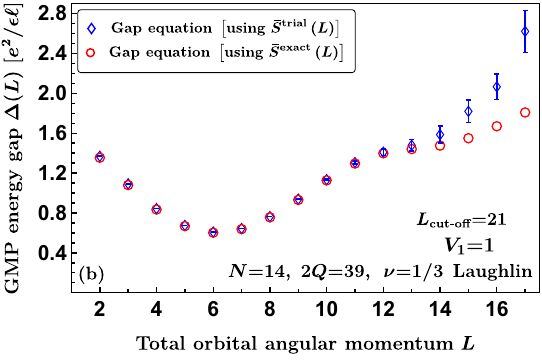}
       \\
         \includegraphics[width=0.49\columnwidth]{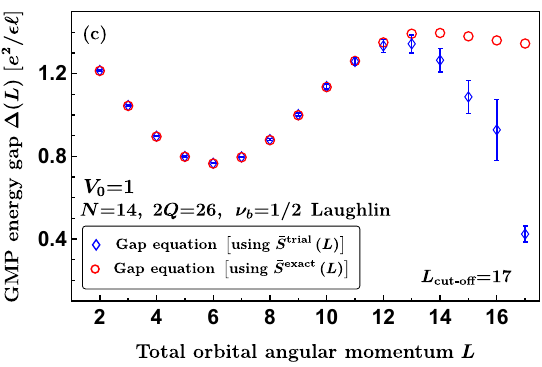}
           \includegraphics[width=0.49\columnwidth]{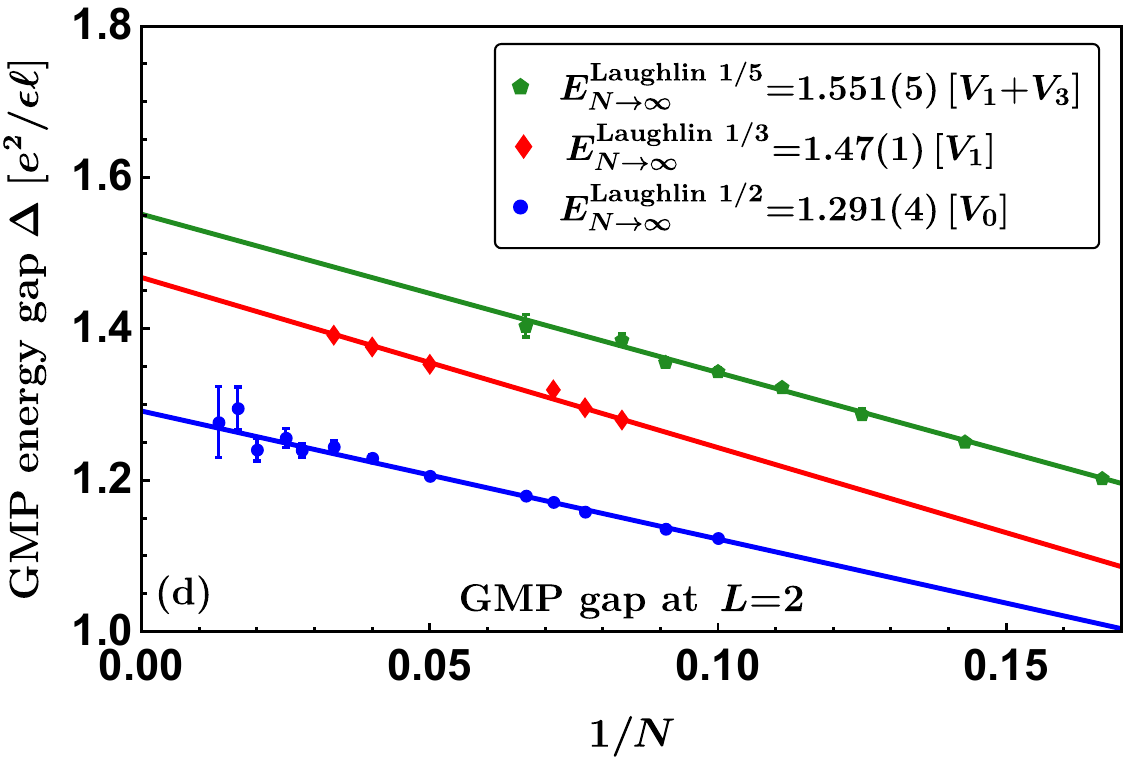}
          \caption{Panels $(a){-}(c)$: Comparison of the GMP gaps for model short-range interactions obtained from the gap equation [see Eq.~\eqref{eq: gap_equation}] using the approximate projected structure factor $\bar{S}^{\rm trial}$ [blue-open diamonds] and exact projected structure factor $\bar{S}^{\rm exact}$ [red-open circles], for the $1/5$ and $1/3$ fermionic and $1/2$ bosonic Laughlin states. See the main text for the definition of $L_{\rm cut{-}off}$. $(d)$ Thermodynamic extrapolation of the corresponding density-corrected $L{=}2$ GMP gap is obtained through a linear fit of the gaps to $1/N$, where $N$ is the number of particles.}
          \label{fig: GMP_gap_model_interaction}
\end{figure} 

This appendix presents the GMP gap for model short-range interactions, parameterized by only a few nonzero Haldane pseudopotentials. These  include the $V_{0}$ (by this we mean the $V_{\mathfrak{m}}{=}\delta_{\mathfrak{m},0}$ interaction), $V_{1}$ (the $V_{\mathfrak{m}}{=}\delta_{\mathfrak{m},1}$ interaction), and $V_1{+}V_{3}$ (the $V_{\mathfrak{m}}{=}\delta_{\mathfrak{m},1}{+}\delta_{\mathfrak{m},3}$ interaction), for the $\nu_{b}{=}1/2$ bosonic Laughlin and $\nu{=}1/3$, and $1/5$ fermionic Laughlin states, respectively. Similar to the computation of the average energy, calculation of the GMP gap also requires the harmonics $v_{L}$ corresponding to these interactions as is evident from Eq.~\eqref{eq: numerator_gap_equation}. The harmonics $v_{L}$ of the $V_0$-only interaction are independent of $L$ [see Eq.~\eqref{eq: pair_pseudopotential_delta_function}]. On the other hand, the harmonics $v_{L}$ representing $V_1$ and $V_1{+}V_{3}$ interactions exhibit polynomial growth as $L(L{+}1)$ and $[L(L{+}1)]^3$, respectively. To mitigate this polynomial growth, which amplifies the statistical error in $\bar{S}^{\rm trial}(L)$ data leading to inaccuracies in the GMP gap, we follow the method outlined in Ref.~\cite{Yutushui24}, as discussed in Appendix~\ref{app: ground_state_energies}, to obtain an optimized set of harmonics $\{\tilde{v}_{L}\}$ [optimized interactions are obtained by setting the exponent $\alpha{=}2$ in Eq.~\eqref{eq: cost_function}]. 

Next, using the harmonics $v_{L}{=}(1{+}4Q)/(2Q{+}1)^2$ for the $V_0$ interaction, and optimized set of harmonics $\{\tilde{v}_{L}\}$ for the $V_1$ and $V_1{+}V_3$ interactions, along with the $\bar{S}^{\rm trial}(L)$ data [see Sec.~\ref{sssec: approximate_projected_structure_factor}] in Eq.~\eqref{eq: gap_equation}, we compute the corresponding GMP gaps. The results are presented in Figs.~\ref{fig: GMP_gap_model_interaction}$(a{-}c)$ for small systems to facilitate its comparison with the GMP gap obtained from $\bar{S}^{\rm exact}(L)$. The two gaps are in good agreement with each other for small $L$ but for large $L$ they deviate from each other as the error from the $\bar{S}^{\rm trial}(L)$ data builds up there. Notably, the GMP gap for the longer-range $V_1{+}V_3$ interaction obtained using $\bar{S}^{\rm trial}(L)$ deviates more rapidly from the exact GMP gap than those for the ultra short-range $V_{0}$ and short-range $V_{1}$ interactions. 

\begin{figure*}[tbh]
         
          \includegraphics[width=0.32\columnwidth]{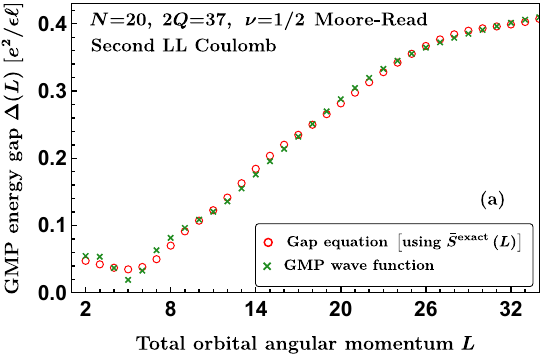}
         \includegraphics[width=0.32\columnwidth]{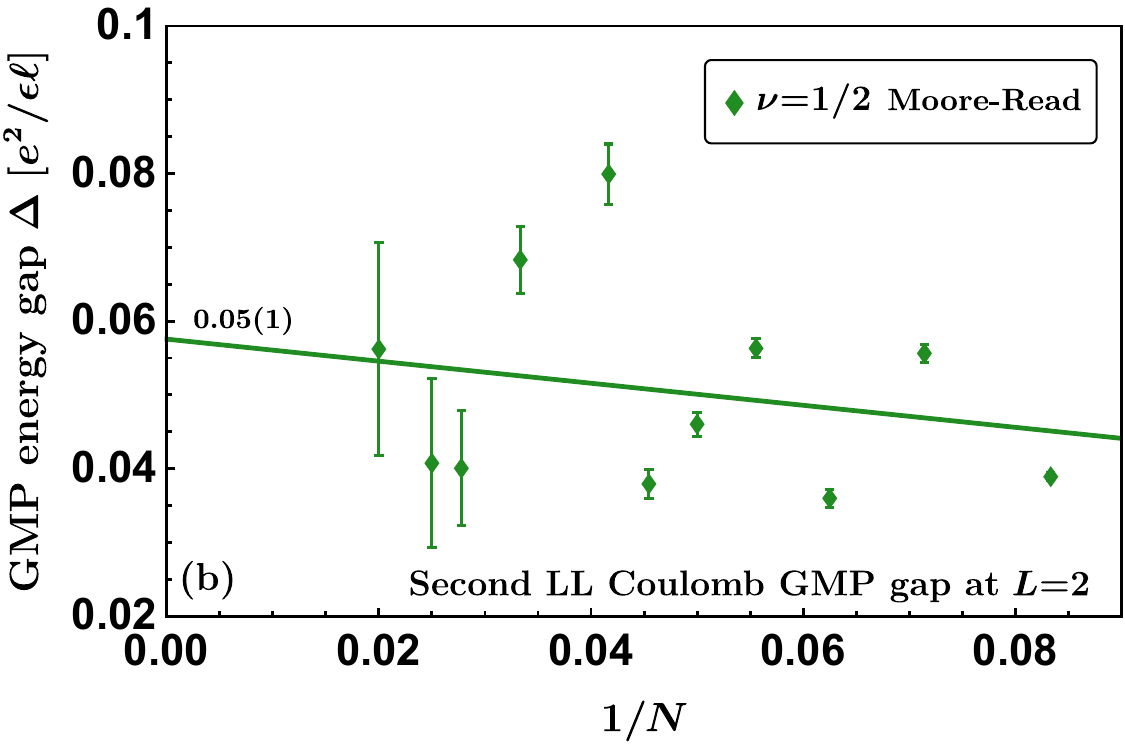}
        \includegraphics[width=0.32\columnwidth]{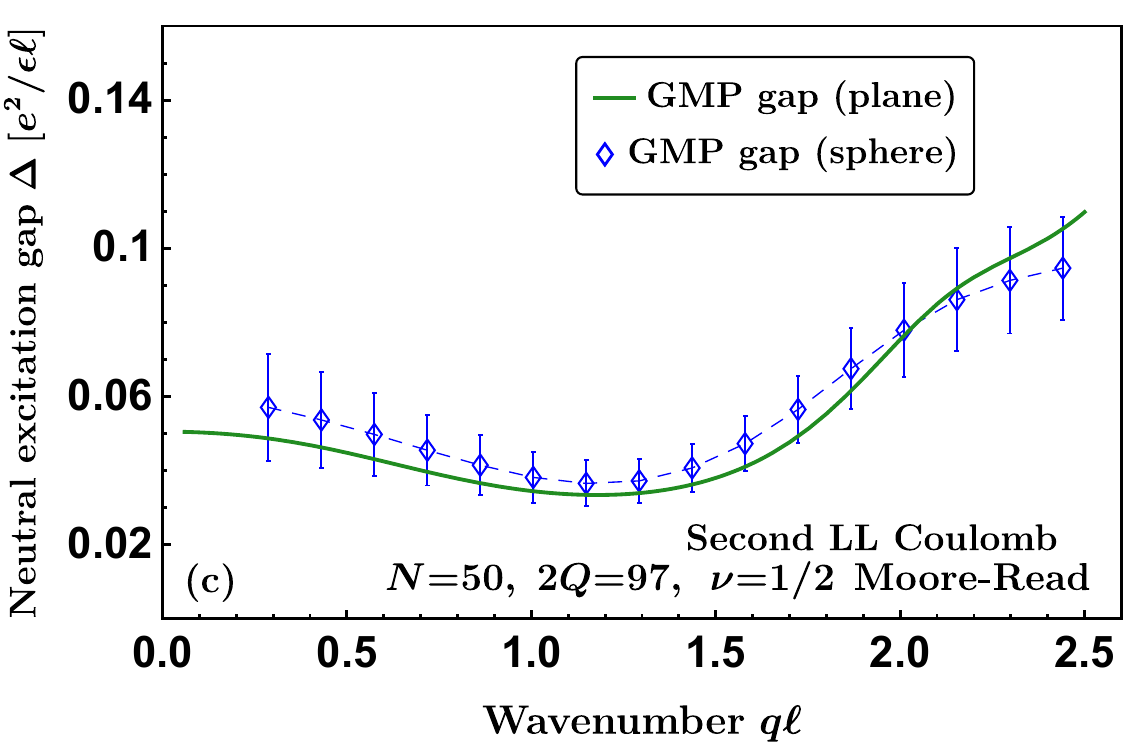}
       \caption{$(a)$ The second Landau level Coulomb GMP gap for the $\nu{=}1/2$ fermionic Moore-Read Pfaffian state obtained using the GMP gap equation [red circles] and GMP wave function [green crosses]. $(b)$ Thermodynamic limit of the density-corrected second Landau level Coulomb $L{=}2$ GMP gap for the $\nu{=}1/2$ fermionic Moore-Read Pfaffian state. $(c)$ Second Landau level Coulomb planar GMP gap of $\nu{=}1/2$ Moore-Read Pfaffian state.}
          \label{fig: more_fermionic_Coulomb_GMP_gap}
\end{figure*}


\begin{figure*}[tbh]
        \includegraphics[width=0.49\columnwidth]{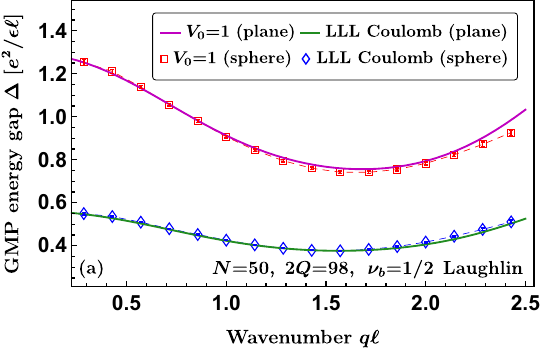}
         \includegraphics[width=0.49\columnwidth]{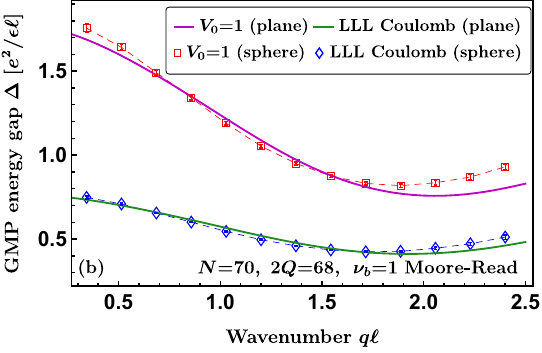}\\
          \includegraphics[width=0.49\columnwidth]{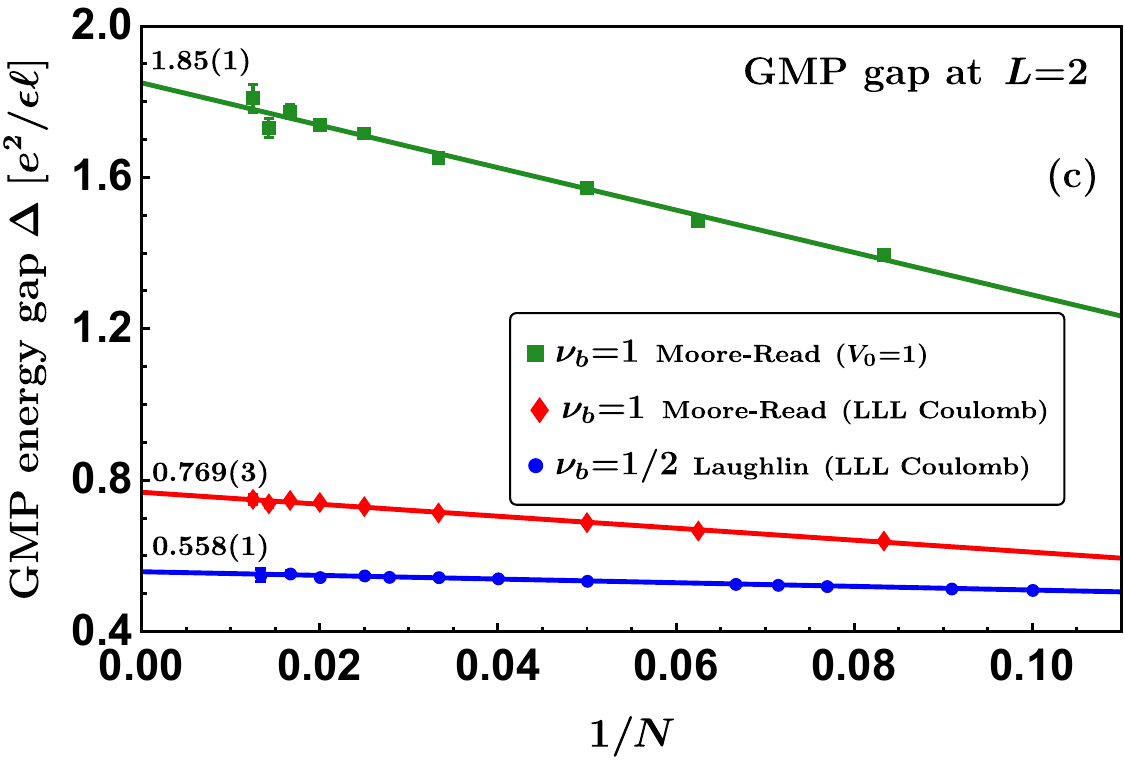}
          \includegraphics[width=0.49\columnwidth]{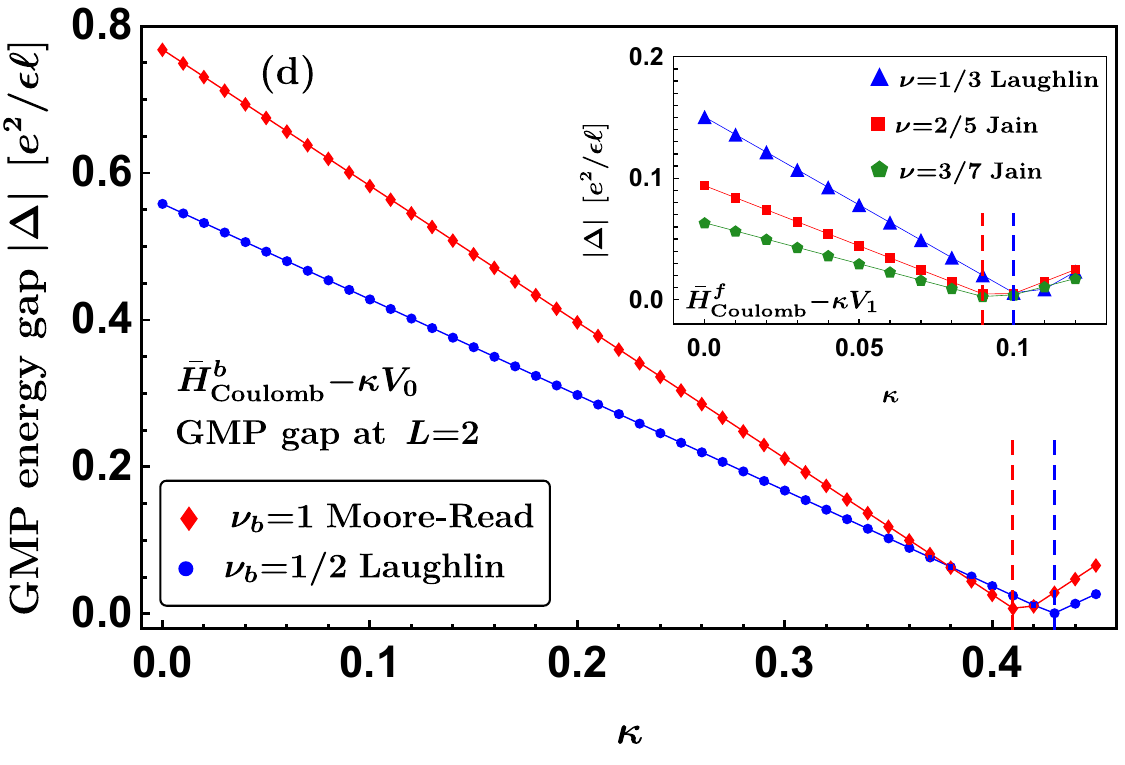}
          \caption{Panels $(a{-}b)$: The LLL Coulomb and $V_{0}$ GMP gaps for the bosonic $(a)$ $\nu_b{=}1/2$ Laughlin and $(b)$ $\nu_b{=}1$ Moore-Read states for a large system on the sphere, along with their planar GMP gaps. $(c)$ Thermodynamic extrapolation of the $L{=}2$ LLL Coulomb and $V_0$ density-corrected GMP gaps for the $\nu_b{=}1$ Moore-Read state, along with the $L{=}2$ LLL Coulomb density-corrected GMP gap for the $\nu_b{=}1/2$ Laughlin state. [The thermodynamic extrapolation of $V_0$ GMP gap at $L{=}2$ for the $\nu_b{=}1/2$ Laughlin state is shown in Fig.~\ref{fig: GMP_gap_model_interaction}$(d)$.] $(d)$ Thermodynamic limit density-corrected $L{=}2$ GMP gap of $\nu_b{=}1$ Moore-Read [red-filled diamonds] and $\nu_b{=}1/2$ Laughlin states [blue-filled circles] as a function of $\kappa$ for the interaction $\bar{H}^{b}_{\rm Coulomb}{-}\kappa V_0$. Red and blue vertical dashed lines indicate the critical values $\kappa^{\left(\rm MR\right)}_{c}{=}0.41$ for $\nu_b{=}1$ Moore-Read and $\kappa^{\left(\rm Laughlin~1/2\right)}_{c}{=}0.43$ for $\nu_b{=}1/2$ Laughlin states, respectively, at which the gap vanishes. The inset in panel $(d)$ shows the analogous data for fermionic states for the interaction $\bar{H}^{f}_{\rm Coulomb}{-}\kappa V_1$, wherein the red and blue vertical-dashed lines indicate critical points $\kappa^{\left(\rm Jain~2/5\right)}_{c}{=}\kappa^{\left(\rm Jain~3/7\right)}_{c}{=}0.09$ and  $\kappa^{\left(\rm Laughlin~1/3\right)}_{c}{=}0.1$, respectively.}
          \label{fig: bosonic_GMP_gap}
\end{figure*}

In Fig.~\ref{fig: GMP_gap_model_interaction}$(d)$, we show the density-corrected~\cite{Morf86} thermodynamic extrapolation of the $L{=}2$ GMP gap for the model pseudopotential interactions. Owing to the constant nature of the harmonics $v_{L}$ for the $V_{0}$ interaction, we can compute the $L{=}2$ GMP gap for fairly large systems. We can access large systems for the $V_{1}$ Hamiltonian too. Here, the GMP gap in the long-wavelength limit that we find for the 1/3 Laughlin state is consistent with that ascertained from calculating the dynamical structure factor, which is the density-density correlation, of the 1/3 Laughlin state on an infinite cylinder using matrix product states~\cite{Liu24}. For the $V_1{+}V_{3}$ interaction, owing to the reasons mentioned in Appendix~\ref{app: ground_state_energies}, the accessible system sizes are limited.

\section{Second LL Coulomb GMP gap of Moore-Read Pfaffian state at $\nu{=}1/2$}
\label{app: GMP_gap_fermionic_MR_Pf_SLL}
One of the leading candidates to explain the experimentally observed $5/2$ FQH state~\cite{Willett87} is the Moore-Read Pfaffian state~\cite{Moore91}. In this appendix, we compute the GMP gap of $\nu{=}1/2$ Moore-Read Pfaffian state for the second LL (SLL) Coulomb interaction. The SLL Coulomb GMP gap on the sphere is obtained by replacing $v^{(C)}_{L}$ [the superscript $C$ is for the Coulomb interaction, see also Eq.~\eqref{eq: v_l_Coulomb}] in Eq.~\eqref{eq: numerator_gap_equation} by~\cite{Yutushui24}
\begin{equation}
v^{(C)}_{L}\rightarrow  v^{C}_{L}\frac{\sqrt{Q}}{\sqrt{Q-1}} \left(\begin{array}{ccc}
Q & Q & L \\
-Q+1 & Q-1 & 0
\end{array}\right)^2\times \left(\begin{array}{ccc}
Q & Q & L \\
-Q & Q & 0
\end{array}\right)^{-2}=v^{C}_{L}\frac{\sqrt{Q}}{\sqrt{Q-1}} \left(\frac{L(L+1)-2Q}{2Q}\right)^2.
\end{equation}
In Fig.~\ref{fig: more_fermionic_Coulomb_GMP_gap}(a), we present a comparison between the SLL Coulomb GMP gap on the sphere obtained using the GMP gap equation and that obtained using the GMP wave function [see Sec.~\ref{ssec: Validation of the GMP algebra}]. The agreement between the two is fairly decent though not as good as that seen for the 1/3 Laughlin state in the lowest LL. This is because while we use the GMP gap equation [Eq.~\eqref{eq: gap_equation}], which assumes the ground state is an eigenstate of the Hamiltonian, the $\nu{=}1/2$ Moore-Read state is not as good a representative of the exact ground state for the SLL Coulomb interaction as the Laughlin state is for the LLL Coulomb interaction~\cite{Kusmierz18, Balram20b}. In Fig.~\ref{fig: more_fermionic_Coulomb_GMP_gap}(b), we show the thermodynamic extrapolated $L{=}2$ SLL GMP gap for the Moore-Read state, which is in good agreement with previous results obtained directly from the GMP wave function~\cite{Pu23}. 

In Fig.~\ref{fig: more_fermionic_Coulomb_GMP_gap}(c), we show the planar GMP gap of the $\nu{=}1/2$ Moore-Read Pfaffian state. The planar GMP gap is computed following the same procedure as discussed in Sec.~\ref{sec: planar_GMP_gap} and using the structure factor expansion of the Moore-Read Pfaffian state at filling $\nu$, as provided in Ref.~\cite{Dwivedi19}, which is
\begin{equation}
\label{eq: pfaffian_S_q_expansion}
S_{\text {Pfaffian }}(q{\rightarrow}0)=\frac{q^2}{2}+\frac{1-\nu}{8 \nu} q^4+\frac{(1-2 \nu)(2-\nu)}{64 \nu^2} q^6+\text{order}~q^8~\text{terms}.
\end{equation}
The disk and spherical GMP gaps of the 1/2 Moore-Read state for the SLL Coulomb interaction are consistent.

\section{GMP gap for bosonic FQH states}
\label{app: GMP_gap_bosonic_FQH_states}
To further demonstrate the applicability of our GMP gap equation, we consider bosonic FQH states. Here, we present the GMP gaps for large systems for the $\nu_{b}{=}1/2$ Laughlin and the $\nu_{b}{=}1$ Moore-Read (MR) states. In the LLL, both these states are stabilized by the hard-core $V_{0}$-only and the LLL Coulomb interactions~\cite{Cooper01, Regnault03, Regnault04, Sharma23}. For these interactions, in Figs.~\ref{fig: bosonic_GMP_gap}$(a{-}b)$, we show the dispersion of the GMP mode for the bosonic Laughlin and MR states, respectively, computed on both the planar and sphere geometry. Analogous to the GMP gap computation for the primary Jain states in Sec.~\ref{ssec: CFE_GMP_gap}, the sphere GMP gaps here also are computed with the GMP gap equation [see Eq.~\eqref{eq: gap_equation}] using the projected structure factor obtained from the unprojected structure factor. The planar GMP gaps are computed following Sec.~\ref{sec: planar_GMP_gap} in conjunction with Appendix~\ref{app: fit_gr_Sq}. As evident from Figs.~\ref{fig: bosonic_GMP_gap}$(a{-}b)$, the sphere GMP gap and planar GMP gap agree with each other up to the magnetoroton minimum, similar to the $1/3$ Laughlin in Fig.~\ref{fig: GMP_CFE_gap_large_system_size}($a$). Figure~\ref{fig: bosonic_GMP_gap}$(c)$ shows the extrapolation to the thermodynamic limit of the corresponding $L{=}2$ density-corrected GMP gaps~\cite{Morf86}, providing an estimate of the long-wavelength limit of the GMP mode. These extrapolated long-wavelength GMP gaps are consistent with the corresponding planar GMP gaps presented in  Figs.~\ref{fig: bosonic_GMP_gap}$(a{-}b)$.

As an application of our framework, we revisit the recent work of Ref.~\cite{Pu24}, which demonstrates that by tuning the $V_{0}$ pseudopotential of the LLL Coulomb interaction, the FQH state of bosons at $\nu_{b}{=}1/2$ becomes unstable to a nematic phase, where the long-wavelength limit of the neutral gap vanishes but the charge gap remains finite. To corroborate their result, we compute the $L{=}2$ GMP gap of the $\nu_{b}{=}1/2$ Laughlin state for the interaction Hamiltonian $\bar{H}^{b}_{\rm Coulomb}{-}\kappa V_0$. Here, $\bar{H}^{b}_{\rm Coulomb}$ is the Coulomb interaction Hamiltonian for bosons, obtained by substituting $v_{L}{=}v_{L}^{(c)}$ [given in Eq.~\eqref{eq: v_l_Coulomb}] into Eq.~\eqref{eq: not_normal_ordered_interaction} for $\sigma{=}b$ and $\kappa$ is a tuning parameter. As shown in Fig.~\ref{fig: bosonic_GMP_gap}$(d)$, we find that the density corrected $L{=}2$ GMP gap in the thermodynamic limit becomes vanishingly small around the critical point $\kappa^{\left(\rm Laughlin~1/2\right)}_{c}{=}0.43$, consistent with the results of Ref.~\cite{Pu24}, which were obtained by direct Monte Carlo integration of the GMP wave function. Analogously, the $L{=}2$ GMP mode of the $\nu_{b}{=}1$ MR state softens as one approaches the critical point $\kappa^{\left(\rm MR\right)}_{c}{=}0.41$ [see Fig.~\ref{fig: bosonic_GMP_gap}$(d)$] signaling a transition to a nematic phase.

A similar phenomenon also arises in fermionic FQH states when the $V_1$ pseudopotential is reduced from its Coulomb value. For the fermionic Coulomb interaction Hamiltonian $\bar{H}^{f}_{\rm Coulomb}$ (obtained similarly as $\bar{H}^{b}_{\rm Coulomb}$ but with $\sigma{=}f$), lowering its $V_1$ pseudopotential through the interaction $\bar{H}^{f}_{\rm Coulomb}{-}\kappa V_1$, we find that the $L{=}2$ GMP gap of the $1/3$ Laughlin state vanishes around the critical point $\kappa^{\left(\rm Laughlin~1/3\right)}_{c}{=}0.1$ [see Fig.~\ref{fig: bosonic_GMP_gap}$(d)$] consistent with the findings of Ref.~\cite{Pu24}. Similarly, for the $2/5$ and $3/7$ Jain states, the long-wavelength limit of the GMP gap becomes vanishingly small around the critical point $\kappa^{\left(\rm Jain~2/5\right)}_{c}{=}\kappa^{\left(\rm Jain~3/7\right)}_{c}{=}0.09$ [see Fig.~\ref{fig: bosonic_GMP_gap}$(d)$].

\section{Fitting of numerical pair-correlation and comparison of structure factor data to their analytic expansions}
\label{app: fit_gr_Sq}
In this appendix, we demonstrate that the pair-correlation function $g(r)$ and the unprojected structure factor $S(q)$ of many bosonic and fermionic states, computed on the sphere for large system sizes, provide reliable approximations to their thermodynamic limits. For fermionic states, we fit the numerically computed $g(r)$ data points on the sphere to its analytic expansion in planar geometry given in Eq.~\eqref{eq: g(r)_expansion}. Note that since the Laughlin state $\Psi^{\rm Laughlin}_{1/p}$ at $\nu{=}1/p$ vanishes as $r^{p}$ ($p$ is an odd integer) as two particles with relative distance $r$ approach each other, the corresponding $g(r)$, which is proportional to $|\Psi^{\rm Laughlin}_{1/p}|^2$, vanishes as $r^{2p}$. To impose the constraint that $g(r)$ vanishes as $r^{2p}$, we set the expansion coefficients $c_{j}{=}{-}1$ for $j{<}p$ for the Laughlin states~\cite{Girvin84a} [see Eq.~\eqref{eq: g(r)_expansion}] in addition to the set of constraints given in Eq.~\eqref{eq: c_m_constraints_Jain_states} to capture the correct $q{\rightarrow}0$ behavior of $S(q)$. [Note that constraints in Eq.~\eqref{eq: c_m_constraints_Jain_states} valid only for $\nu{=}n/(2n{+}1)$, which does not include $1/5$ fermionic Laughlin state. Below in Eq.~\eqref{eq: Laughlin_S_q_expansion} we provide $S(q{\rightarrow}0)$ expansion for the general Laughlin state from which the analogous constraints on $c_j$ can be derived for the $1/5$ Laughlin state.] On the other hand, as the $g(r)$ of Jain states vanishes only as $r^{2}$, we do not need any additional constraints on $c_{j}$ on top of that given in Eq.~\eqref{eq: c_m_constraints_Jain_states}. The fitting results, presented in Fig.~\ref{fig: comparision_fitted_computed_gr_Sq}, demonstrate an excellent agreement between the numerical $g(r)$ data and fitted $g(r)$ indicating that the expansion we use does an accurate job of parametrizing the $g(r)$.

Next, we discuss the $g(r)$ fitting for bosonic states such as the $\nu_{b}{=}1/2$ Laughlin and $\nu_{b}{=}1$ Moore-Read Pfaffian states. It may appear that for bosonic states, the summation over odd $j$ in the $g(r)$ expansion [see Eq.~\eqref{eq: g(r)_expansion}] could be replaced by a summation over even $j$, as the bosonic states involve only even relative angular momenta. Surprisingly, including only even $j$ in $g(r)$ expansion does not yield an accurate fit to the numerical $g(r)$ data. This discrepancy likely arises because the factor $1{-}e^{{-}r^2/2}$ in $g(r)$, which ensures $g(r{\rightarrow}\infty){=}1$, corresponds to the pair-correlation function of a fermionic $\nu{=}1$ state, while the rest of the terms in the summation, for even $j$, are specific to bosonic states. The presence of both fermionic and bosonic terms in $g(r)$ restricts the expansion to even $j$, which is incompatible with the bosonic states' numerical $g(r)$ data. To further illustrate this issue, consider the case of $\nu_{b}{=}1/p$ bosonic Laughlin state (where $p$ is an even integer${\geq}2$). Similar to the fermionic Laughlin state, $g(r)$ of bosonic Laughlin state also vanishes as $r^{2p}$, consequently $g(0){=}0$. However, when only even $j$ terms are included in the $g(r)$ expansion, there is no solution of $c_{j}$ consistent with the constraints that $g(0){=}0$ and its leading term is $r^{2p}$. This is evident from the Taylor-expansion of $g(r)$ as given in Eq.~\eqref{eq: g(r)_expansion} with only even $j$ terms in it, around $r{=}0$,
\begin{align}
    g(r\rightarrow0)&=2c_0 - \frac{1}{2}(c_0-1)r^2 +\frac{1}{8}\left(\frac{c_0+c_2}{2}-1\right)r^4 + \mathcal{O}\left(r^6\right).
\end{align}
For the $\nu_{b}{=}1/2$ Laughlin state, the leading order term in $g(r)$ is proportional to $r^4$, which simultaneously requires $c_{0}{=}0$ and $c_{0}{=}1$, which is impossible to satisfy. To circumvent these issues, we instead include all the even and odd $j$ terms in the $g(r)$ expansion, i.e.,
\begin{equation}
\label{eq: g(r)_expansion_bosons}
    g(r)=1-e^{-r^2/2}+e^{-r^2/4}\sum\limits_{j=0}^{\infty}~\frac{2c_{j}}{j!}\left(\frac{r^2}{4}\right)^{j}.
\end{equation}
Furthermore, to ensure the correct $q{\rightarrow}0$ behavior of $S(q)$, we impose a set of constraints on $c_{j}$ as discussed in Sec.~\ref{sec: planar_GMP_gap}. These constraints, for the Laughlin state, can be read off from its $S(q{\rightarrow}0)$ expansion, given by~\cite{Kalinay00, Can14}
\begin{equation}
\label{eq: Laughlin_S_q_expansion}
S_{\rm {Laughlin }}(q{\rightarrow}0)=\frac{q^2}{2}+\frac{1-2 v}{8 v} q^4+\frac{(1-3 v)(3-4 v)}{96 v^2} q^6+O\left(q^8\right),
\end{equation}
and similarly, for the $\nu_{b}{=}1$ Moore-Read state, from Eq.~\eqref{eq: pfaffian_S_q_expansion}. With these modifications, we find the $g(r)$ expansion fits very accurately with the numerical $g(r)$ data of bosonic states as depicted in Fig.~\ref{fig: comparision_fitted_computed_gr_Sq}.

From the analytic expansion of the $g(r)$, we obtain an analytic expression of $S(q)$ by Fourier transforming it on the plane through Eq.~\eqref{eq: structure_factor_pair_correlation_relation}. The $S(q)$ obtained this way along with the $S(q)$ computed in the spherical geometry are also depicted in Fig.~\ref{fig: comparision_fitted_computed_gr_Sq}. The structure factors obtained in these two ways are not in perfect agreement since we compare a finite-size system's $S(q)$ computed in the spherical geometry with the thermodynamic one obtained by Fourier transforming the $g(r)$ in the planar geometry. Nevertheless, we see that the agreement between the fitted $S(q)$ and computed $S(q)$ is very good suggesting that the systems we are using on the sphere are accurate representatives of the thermodynamic limit. Note that in the numerical computation of the structure factor in the spherical geometry, we have set $S(L{=}0){=}0$ [i.e., $S(L){\rightarrow}S(L){-}N\delta_{L,0}$] to match previous conventions and results~\cite{Kamilla97, Balram17}. This is different from the sum-rule $S(L{=}0){=}N$ that we discussed in Sec.~\ref{ssec: structure_factor}. Furthermore, from this fitting procedure, we can extract the expansion coefficients of the next two leading, i.e., $q^{8}$ and $q^{10}$, terms in the structure factor. These values are tabulated in Table~\ref{tab: q8_expansion_Sq} for the Laughlin and Moore-Read states. In the future, it would be interesting to generalize the methods of Refs.~\cite{Kalinay00, Can14, Dwivedi19} to see if these expansion coefficients can also be analytically derived and how they relate to the topological quantum numbers of the underlying FQH state.

\begin{figure*}[tbh]
        \includegraphics[width=0.24\columnwidth]{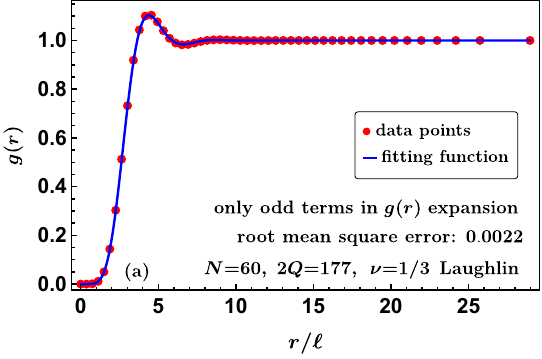}
        \includegraphics[width=0.24\columnwidth]{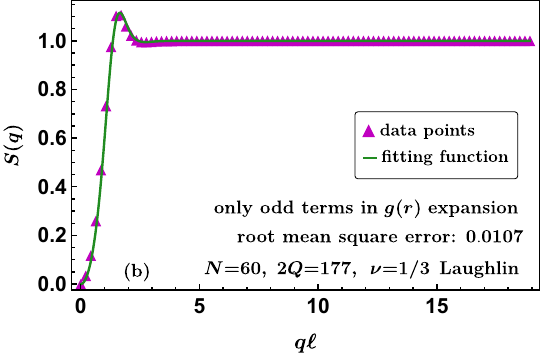}
        \includegraphics[width=0.24\columnwidth]{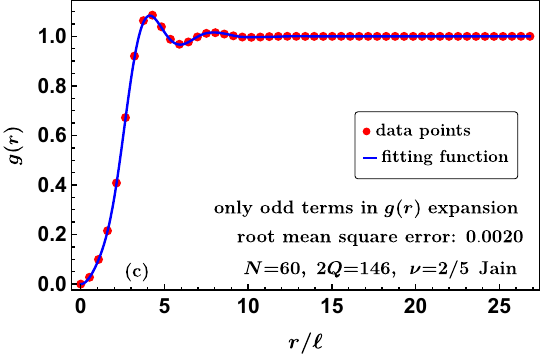}
        \includegraphics[width=0.24\columnwidth]{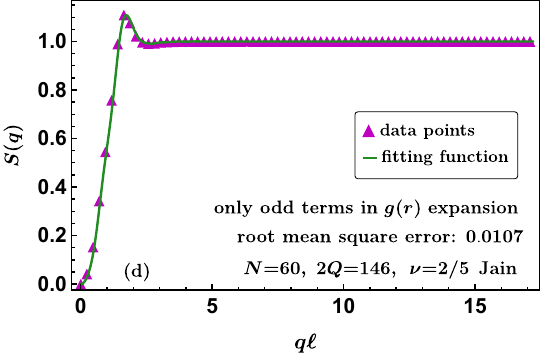} \\
        \includegraphics[width=0.24\columnwidth]{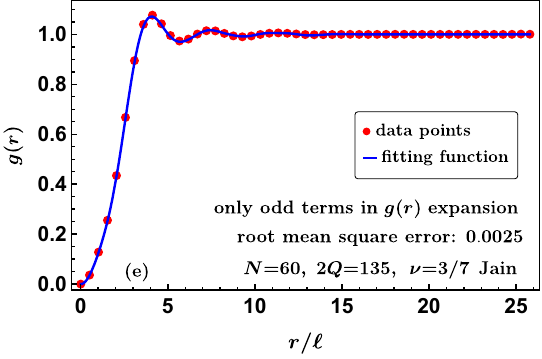}
        \includegraphics[width=0.24\columnwidth]{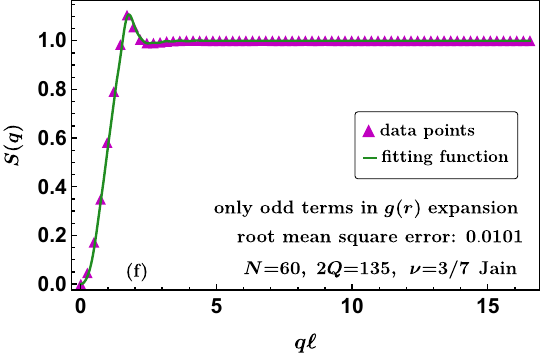}
        \includegraphics[width=0.24\columnwidth]{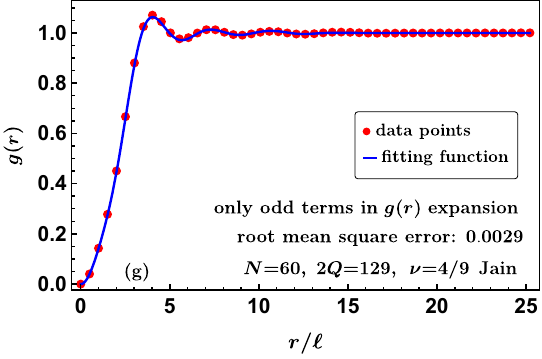}
        \includegraphics[width=0.24\columnwidth]{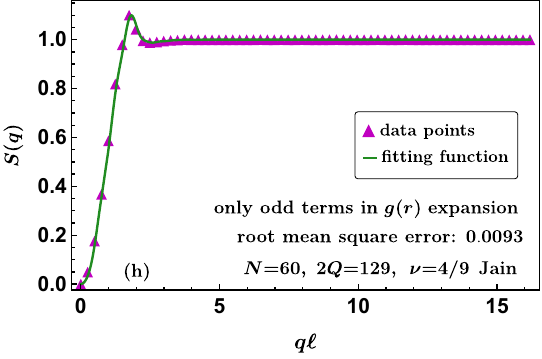} \\
        \includegraphics[width=0.24\columnwidth]{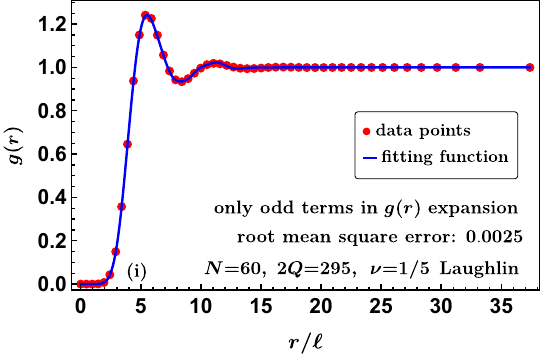}
        \includegraphics[width=0.24\columnwidth]{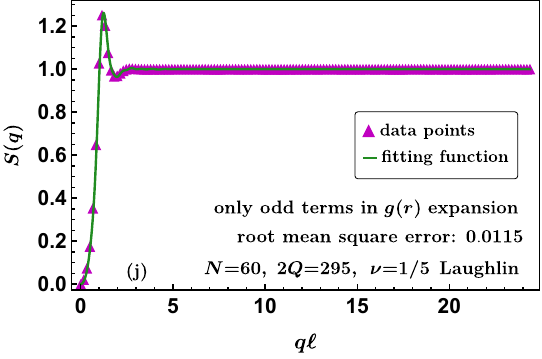}
        \includegraphics[width=0.24\columnwidth]{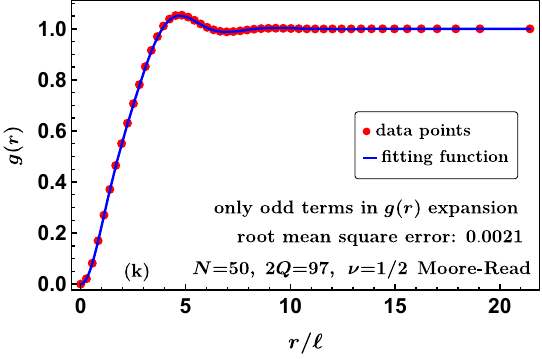}
        \includegraphics[width=0.24\columnwidth]{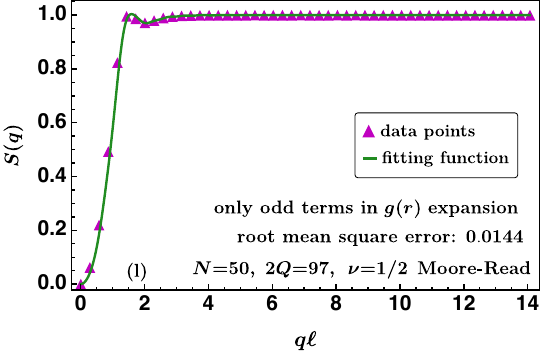}\\
        \includegraphics[width=0.24\columnwidth]{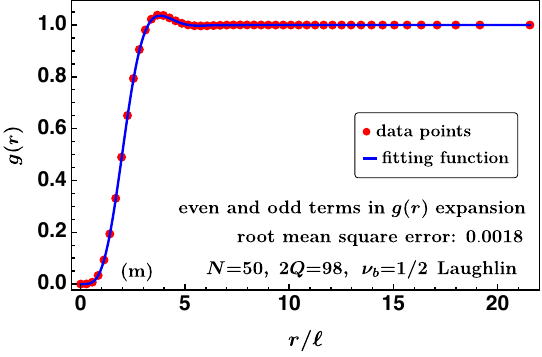}
        \includegraphics[width=0.24\columnwidth]{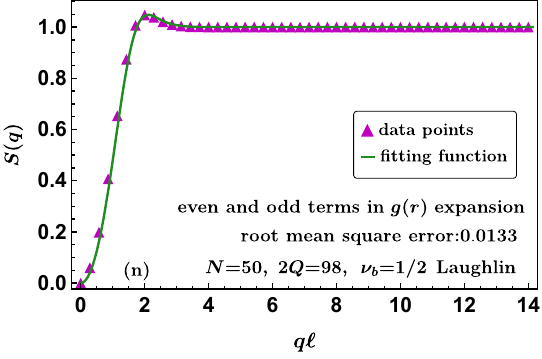}
        \includegraphics[width=0.24\columnwidth]{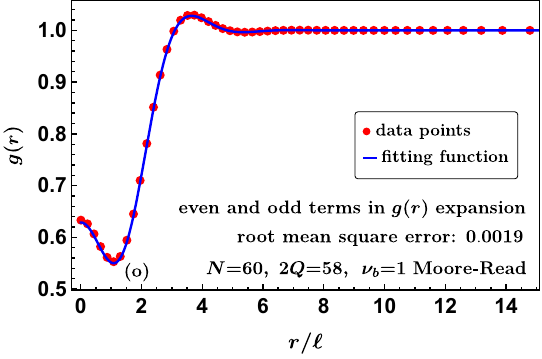}
        \includegraphics[width=0.24\columnwidth]{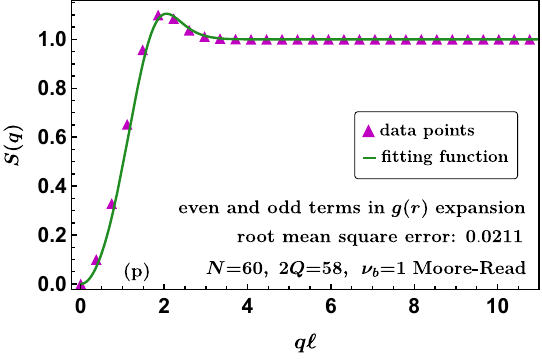}\\
         \includegraphics[width=0.24\columnwidth]{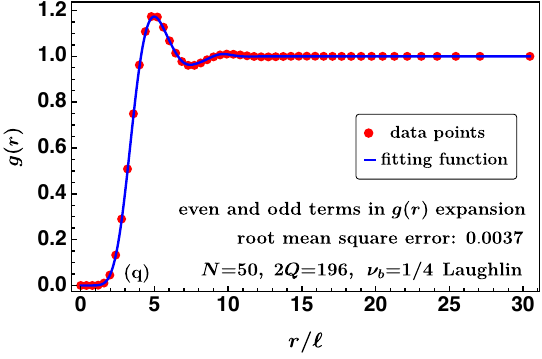}
        \includegraphics[width=0.24\columnwidth]{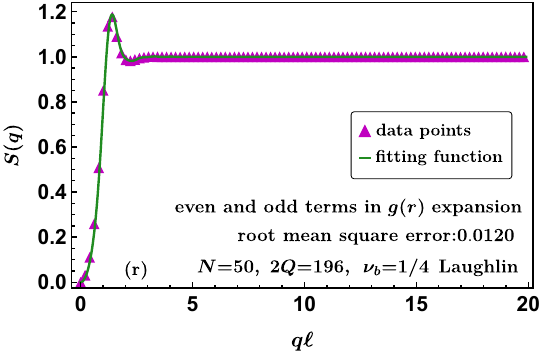}
        \includegraphics[width=0.24\columnwidth]{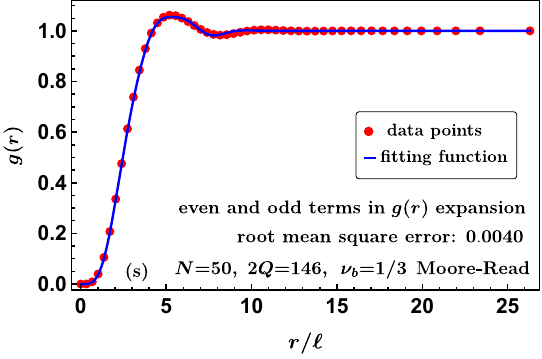}
        \includegraphics[width=0.24\columnwidth]{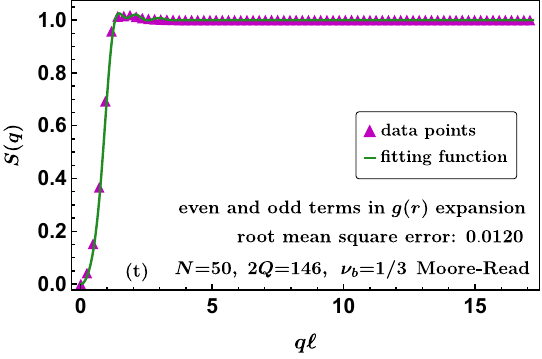}
          \caption{Comparison of the fitted and computed pair-correlation function $g(r)$ and the unprojected static structure factor $S(q)$ for many bosonic and fermionic fractional quantum Hall states. The fitting is done for the planar geometry while the actual computations are done on the spherical geometry. }
          \label{fig: comparision_fitted_computed_gr_Sq}
\end{figure*}
\end{widetext}

\begin{table}[htbp]
	\centering
	\begin{tabular}{|c|c|c|c|}
		\hline 
		$\nu$ 	&  State 						    & $s_{8}$ 		& $s_{10}$		\\ \hline
            $1/2$	& bosonic Laughlin       &   $0.01$			&    	$-0.03$			\\ \hline
            $1/3$	& fermionic Laughlin     &   	$-0.005$		&    $0.014$				\\ \hline
            $1/4$	& bosonic Laughlin       &   	$-0.191$		&    	$1.2$			\\ \hline
            $1/5$	& fermionic Laughlin     &   	$-30.86$		&    		$93.21$		\\ \hline \hline
		$1$	    & bosonic Moore-Read Pfaffian       &   $-0.023$			&    $0.034$				\\ \hline
            $1/2$	& fermionic Moore-Read Pfaffian     &   	$4.412$		&   $-23.536$ 				\\ \hline
            $1/3$	& bosonic Moore-Read Pfaffian       &   	$0.817$		&    $-2.707$				\\ \hline
	\end{tabular} 
	\caption{\label{tab: q8_expansion_Sq} The coefficient of the $q^{8}$ and $q^{10}$ terms in the long-wavelength expansion of the unprojected structure factor $S(q)$ for the Laughlin [see Eq.~\eqref{eq: Laughlin_S_q_expansion}] and Moore-Read Pfaffian [see Eq.~\eqref{eq: pfaffian_S_q_expansion}] states extracted from the numerical fitting procedure outlined in Appendix~\ref{app: fit_gr_Sq}.}
\end{table}

\bibliography{biblio_fqhe}

\end{document}